# Modified Causal Forest

Michael Lechner and Jana Mareckova[*]

Swiss Institute for
Empirical Economic Research

University of St.Gallen

This version: August 2022 (1st version: December 2018)

*Date this version has been printed:* **05 September 2022**

Comments are welcome

**Abstract:** Uncovering the heterogeneity of causal effects of policies and business decisions at various levels of granularity provides substantial value to decision makers. This paper develops estimation and inference procedures for multiple treatment models in a selection-on-observed-variables framework by modifying the Causal Forest approach (Wager and Athey, 2018) in several dimensions. The new estimators have desirable theoretical, computational, and practical properties for various aggregation levels of the causal effects. While an Empirical Monte Carlo study suggests that they outperform previously suggested estimators, an application to the evaluation of an active labour market programme shows their value for applied research.

**Correspondence to**: Michael Lechner or Jana Mareckova, Professors of Econometrics, Swiss Institute for Empirical Economic Research (SEW), University of St. Gallen, Switzerland, Michael.Lechner@unisg.ch, Jana.Mareckova@unisg.ch, www.sew.unisg.ch.

---

[*] Michael Lechner is also affiliated with CEPR, London, CESIfo, Munich, IAB, Nuremberg, IZA, Bonn, and RWI, Essen. This research project was part of the National Research Programme "Big Data" (NRP 75, Grant number 407540_166999) of the Swiss National Science Foundation (SNSF). Further information on the National Research Programme can be found at www.nrp75.ch or https://bigdata-dialog.ch/. We thank Michael Knaus and Anthony Strittmatter for many important comments and discussions on this and related topics and Michael Zimmert for pointing out an important computational detail in the multiple treatment version of the algorithm. Also, thanks to Jeff Smith for many important comments on various parts of the manuscript. This paper also greatly benefited from the data preparation done for previous papers using the same data. We are indebted to Martin Huber, Giovanni Mellace, Michael Knaus and Anthony Strittmatter for this. We also thank Jonathan Chassot, Daniel Goller, and Gabriel Okasa for help with optimizing the code used in the simulations. A previous version of the paper was presented at research seminars at the Universities of St. Gallen (2018) and Gent (2019), at Amazon, Seattle (2018), at the annual meetings of the Swiss Society of Economics and Statistics in Geneva (2019), the Econometrics Society in Manchester (2019), the International Association of Applied Econometrics in Nicosia (2019), the International Biometrics Society in Lausanne (2019), and EMbeDS in Pisa (2019). We thank participants for helpful comments and suggestions. The usual disclaimer applies.

# 1  Introduction

Although academia and the public celebrate the amazing predictive power of the new machine learning (ML) methods, many researchers experience some unease, simply because prediction does not imply causation. The ability to uncover causal relations is, however, at the core of most questions concerning the *effects* of policies, medical treatments, marketing campaigns, business decisions, etc. (see for example, Athey, 2017).

The rapidly expanding *causal* ML literature holds great promise for the improved estimation of causal effects by merging the statistics and econometrics literature on causality with the supervised ML literature focussing on prediction. On the one hand, the causality literature clarifies the conditions needed to identify and estimate causal effects. It also shows how to transform a counterfactual causal problem into specific prediction problems (e.g., Imbens and Wooldridge, 2009). On the other hand, the literature on ML provides tools that can be highly effective in solving prediction problems (e.g., Hastie, Tibshirani, and Friedman, 2009). Bringing those two literatures together can lead to more precise, less biased, and thus more reliable estimators of average causal effects. Furthermore, it is possible to uncover their heterogeneity systematically (for an overview, see Athey and Imbens, 2017).

In many important applications of causal methods in economics, as for example in evaluation studies of active labour market programmes (e.g., Card, Kluve, and Weber, 2018), identification is likely to be based on a selection-on-observed-variables assumption within a multiple-programme setting (for this so-called 'multiple treatment' case, see Imbens, 2000, and Lechner, 2001). In such studies, we seek to learn the effects of policies at different levels of granularity. For example, average treatment effects might serve as inputs into a cost-benefit analysis. In the context of programme evaluation these parameters will inform us about the average performance of the programmes for individuals in general or specific programme participants. On the other extreme, estimates of effects that capture many features of the individual units, i.e., they are (almost) unit-specific effects, not only allow us a better understanding of the causal mechanisms at work, but it also sheds some light on distributional aspects of the policy by, e.g., identifying groups who win and groups who lose. Furthermore, such disaggregated results will also hint at potential improvements that could be obtained by different allocations of, in this example, unemployed to different programmes. In programme evaluation, from these parameters we can potentially learn which types of unemployed win and who lose from specific programmes. However, the intermediate aggregation level, e.g., based on splitting the population in several easily understandable subgroups, is also of interest. One example is that such findings (which, e.g., could contrast effects for migrants to effects for non-migrants) are much easier to communicate to, and understood by decision makers. Of course, for all these estimated policy parameters, measures of estimation uncertainty are important to gauge the confidence one should have in the various point estimates. Furthermore, overall estimation costs should not be excessive.

The literature on estimating heterogeneous causal effects with the help of ML methods is expanding. Knaus, Lechner, and Strittmatter (2021, KLS21 henceforth) characterized this literature by three types of not necessarily exclusive approaches used. The first two approaches use standard tools of ML but modify the data to obtain causal effects instead of predictions. The options are either to modify the outcome (first proposed by Signorovitch, 2007) or to modify the covariates[1] (first proposed by Tian, Alizadeh, Gentles, and Tibshirani, 2014) in different ways. The third generic approach is to change the ML algorithm directly (early Regression Tree based examples are Su, Tsai, Wang, Nickerson, and Li, 2009, and Athey and Imbens, 2016). This is also the path taken in this paper.

This paper provides a new estimator that allows the researcher to obtain estimation and inference results for the above-mentioned different levels of aggregation at reasonable computational costs in one comprehensive estimation step. To be more precise, the Causal Forest estimators proposed by Wager and Athey (2018, WA18 henceforth) and Athey, Tibshirani, and Wager (2019, ATW19 henceforth) form the basis of the proposed methods. Their outcome-based version is extended in two important directions for which theoretical guarantees are provided.

---

[1] The terms *covariates* or *features* are used interchangeably below.

The first extension concerns the splitting rule used for growing the Causal Trees that form the Causal Forest of WA18. WA18 propose to base it on maximising estimated treatment effect heterogeneity. The rational is that this is an unbiased estimator of the MSE of the estimated causal effects. However, this is only true if selection bias is not relevant. Since this is usually not the case for the early splits of the tree, we propose an alternative splitting criterion (for the binary as well as the multiple treatment case). It is based on the observation that the estimation of a causal effect at the lowest level of granularity in a selection-of-observables setting corresponds to predicting the *difference* of two outcome regressions. The complication is that the observations used to estimate these regressions are observed in different subsamples defined by the observed treatment status. We propose to establish the necessary link by a matching step conducted prior to building the Causal Forest. Extensive simulation results show that there could be indeed considerable improvements when there is selection bias. As expected, in the case of experimental data the rule of WA18 and the new splitting criteria lead to similar results. The simulations also show that adding an additional component to the splitting criteria that favours splits leading to larger heterogeneity of treatment probabilities (propensity scores) across leaves may drastically reduce (finite-sample) selection bias.

The second modification addresses the desire to obtain point estimates and inference for many aggregation levels with one coherent estimation strategy. This decreases computational as well as monitoring costs and increases the internal consistency of the results, e.g. by insuring lower level effects correctly add-up to their higher level counterparts, compared to using different estimators for different aggregation levels. Therefore, this paper proposes a procedure that uses a single Causal Forest for estimation and inference for all these parameters. This procedure exploits the fact that predictions of Random Forests can be expressed explicitly as weighted sums of outcome variables. The weights, which are obtained for the lowest aggregation level, can be aggregated to obtain estimators and inference results at higher aggregation levels. The simulation results show that this approach leads to rather accurate inference for higher aggregation levels, and conservative inference for the lowest level. It also turned out that using in addition a method proposed by ATW19, local centering, which here is implemented by subtracting the estimated conditional-on-covariates but unconditional-on-treatment expectation of the outcome from the actual outcome, essentially removes the 'conservativeness' of the estimated standard errors while not changing the point estimates much. However, this 'one-shop' general weights-based estimator may come at the price of a potential efficiency loss due to the need for sample splitting, which is required for the implemented weights-based inference approach. However, again, the simulation results suggests that such a loss is small, if existent at all.

The weighted representation also allows us to theoretically show that the estimators are pointwise consistent and are asymptotically Gaussian at all levels of granularity. These results require that, as in WA18, the trees are deeply grown and satisfy the condition called honesty, i.e., that trees are grown using one subsample while the estimation uses a different subsample, to reduce bias. Here, honesty is guaranteed by working with two non-overlapping halves of data. The subsample for growing the trees stems from one half and the subsample for estimation stems from the other half. Subsampling is required only for the first subsample used for growing the trees. For the estimation part, it is possible to take the whole half of the data and not to subsample at all.[2] The rates of convergence of the bias and the variance mainly depend on the parameters of the subsampling process yielding the two subsamples. The distributional result relies on a central limit theorem for triangular arrays of weekly dependent random variables in Neumann (2013), exploiting the fact that the potential outcome estimators are weighted averages. These theoretical results extend the class of causal forest estimators estimating heterogeneous treatment effects with proven asymptotic properties.

---

[2] Note that this implementation of honesty is different than the one in WA18 where the sample is first subsampled and then split into halves for the double-sample trees, forcing the subsampling rate to be the same in both subsamples.

As a further contribution, the paper contains a large-scale simulation analysis of the properties of various 'one-shop' estimators addressing the behaviour of point estimators and their inference for various aggregation levels. The simulations come from an Empirical Monte Carlo Study (EMCS) that is based on an actual labour market programme evaluation. While KLS21 contains a very extensive EMCS-based analysis of point estimators, to the best of our knowledge so far there is no large-scale simulation evidence for inference procedures for disaggregated treatment effects estimated by causal ML algorithms.

In the next section, we introduce the related literature. In Section 3, we discuss the parameters of interest and their identification. Estimation and inference, including a summary of the asymptotic properties of the proposed estimators, are the topics of Section 4. Section 5 contains the Empirical Monte Carlo (EMCS) study. Finally, Section 6 presents an empirical application, and Section 7 concludes. Appendix A contains the proofs of the asymptotic properties of the estimators as well as implementational details. Appendix B presents specifics of the EMCS. Appendix C shows the results for several other DGPs relevant in observational studies to document the robustness and sensitivities of the results. Appendix D covers the simulation results for the case without selection bias (experiment). Free computer code is available for Python, R and Gauss.

## 2 Literature

There is a considerable and rapidly increasing literature related to the estimation of effect heterogeneity by ML methods in observational studies within a selection-on-observed-variables research design. Therefore, we will discuss the main papers only briefly and refer the reader to the much more in-depth and systematic discussion of KLS21.

We start this review with the methodological literature concerned with estimating heterogeneous causal or treatment effects by ML methods followed by applications in economics and comparative studies.[3] Contributions to this literature come from various fields, like epidemiology, econometrics, statistics, and informatics. Proposed estimators are based on regression trees (Su, Tsai, Wang, Nickerson, and Li, 2009; Athey and Imbens, 2016), Random Forests (Athey, Tibshirani, and Wager, 2019; Friedberg, Tibshirani, Athey, and Wager, 2020; Oprescu, Syrgkanis and Wu, 2019; Seibold, Zeileis, and Hothorn, 2018; Wager and Athey, 2018), bagging nearest neighbour estimators (Fan, Lv, and Wang, 2018), the least absolute shrinkage and selection operator (LASSO, Qian and Murphy, 2011; Tian, Alizadeh, Gentles, and Tibshirani, 2014), support vector machines (Imai and Ratkovic, 2013), boosting (Powers, Qian, Jung, Schuler, Shah, Hastie, and Tibshirani, 2018), neural networks (Ramachandra, 2018; Schwab, Linhardt, and Karlen, 2018; Shalit, Johansson and Sontag, 2017), and Bayesian ML methods (Hill, 2011; Wang and Rudin, 2015; Taddy, Gardner, Chen, and Draper, 2016). Finally, Chen, Tian, Cai, and Yu (2017), Künzel, Sekhon, Bickel, and Yu (2019), and Nie and Wager (2021) propose general estimation approaches that are not linked to any specific ML method.

---

[3] Flexible **C**onditional **A**verage **T**reatment **E**ffect (CATE) estimation has also been discussed using non-machine learning methods. These are usually multi-step procedures based on a first step estimation of the propensity score (and possibly the expectation of the outcome given treatment and confounders) and a second non- or semi-parametric step to obtain a low-dimensional CATE function. Finally, this function is used for predicting CATEs (or aggregated versions thereof) in and out-of-sample. For example, Xie, Brand, and B. Jann (2012) base their estimator on propensity score stratification and regression, while Abrevaya, Hsu, and Lieli (2015) use propensity score weighting, and Lee, Okui, and Whang (2017) use a doubly robust estimation approach.

While there are many proposed methods, in economics only a few studies have used these methods so far. Ascarza (2018) investigates retention campaigns for customers. Bertrand, Crépon, Marguerie, and Premand (2021) analyse active labour market programmes in a developing country. Davis and Heller (2017) investigate summer jobs in the US. Strittmatter (2018) reinvestigates a US welfare programme. Knaus, Lechner, and Strittmatter (2022) evaluate the heterogeneous effects of a Swiss job search programme for unemployed workers. Cockx, Lechner, and Bollens (2019) investigate several active labour market programmes in Flanders. Finally, Boller, Lechner, and Okasa (2021) analyse the effects of the level of sports activity on online dating. The last two papers use the same methodology as is suggested in this paper.

There are also a few simulation studies investigating the properties of the suggested methods. Of course, almost every methodological paper contains a simulation study, but these studies tend to be very specific. However, there appear to be at least three studies that compare a larger number of estimators: Two of them have an epidemiological background. Powers, Qian, Jung, Schuler, Shah, Hastie, and Tibshirani (2018) use data generating processes that are only to a limited extent informed by real data. The second study uses large medical databases to inform their simulation designs (Wendling, Callahan, Schuler, Shah, and Gallego, 2018). Since these data generating processes are very specific to biometrics, it is hard to draw strong lessons for many applications in the social sciences. The fourth study by KLS21 uses the same data and similar simulation designs as this paper. KLS21 investigate various Random Forest and LASSO based estimators for causal heterogeneity in a selection-on-observables setting. Generally, they conclude that the Forest based versions, in particular the Generalized Forest by Athey, Tibshirani, and Wager (2019) are among the best performing estimators if explicitly adjusted to take account of confounding. This adjustment is mainly done by a pre-estimation ML step that purges the outcomes from some of their dependence on the covariates by subtracting their estimated conditional-on-covariates mean (*local centering*).[4] Perhaps somewhat surprisingly, the estimators that predict the conditional mean of the outcomes among the treated and among the controls using standard Random Forest regression and subsequently take the difference of the two predictions perform often similar to the more sophisticated estimators explicitly optimized for causal estimation.

## 3 Causal framework

### 3.1 The potential outcome model

We use Rubin's (1974) potential outcome language to describe a multiple treatment model under unconfoundedness, selection-on-observables, or conditional independence (Imbens, 2000, Lechner, 2001). Let *D* denote the treatment that may take a known number of *M* different integer values from *0* to *M-1*. The (potential) outcome of interest that realises under treatment *d* is denoted by $Y^d$. For each observation, we observe only the particular potential outcome that is related to the treatment status the observation is observed to be in, $y_i = \sum_{d=0}^{M-1} \underline{1}(d_i = d) y_i^d$, where $\underline{1}(\cdot)$ denotes the indicator function, which equals one if its argument is true.[5] There are two groups of variables to condition on, $\tilde{X}$ and *Z*. $\tilde{X}$ contains those covariates needed to correct for selection bias (confounders), while *Z* contains variables that define (groups of) population members for which an average causal effect estimate is desired. For identification, $\tilde{X}$ and *Z* may be discrete, continuous, or both (for estimation, we consider discrete or discretized *Z* only). They may overlap in any way. In line with the ML literature, we call them 'features' from now on. Denote the union of the two groups of variables by $X$, $X = \{\tilde{X}, Z\}$, $\dim(X) = p$.[6]

---

[4] Oprescu, Syrgkanis, and Wu (2019) suggested a related, computationally more intensive adjustment procedure.

[5] If not obvious otherwise, capital letters denote random variables, and small letters their values. Small values subscripted by '*i*' denote the value of the respective variable of observation '*i*'.

Below, we investigate the following average causal effects:

$$IATE(m,l;x,\Delta) = E(Y^m - Y^l \mid X = x, D \in \Delta),$$

$$GATE(m,l;z,\Delta) = E(Y^m - Y^l \mid Z = z, D \in \Delta) = \int IATE(m,l;\tilde{x},z,\Delta) f_{\tilde{X}\mid Z=z, D\in\Delta}(\tilde{x}) d\tilde{x},$$

$$ATE(m,l;\Delta) = E(Y^m - Y^l \mid D \in \Delta) = \int IATE(m,l;x,\Delta) f_{X\mid D\in\Delta}(x) dx.$$

The **I**ndividualized **A**verage **T**reatment **E**ffects (IATEs), $IATE(m,l;x,\Delta)$ measure the mean impact of treatment $m$ compared to treatment $l$ for units with features $x$ that belong to treatment groups $\Delta$, where $\Delta$ denotes all treatments of interest.[7] The IATEs represent the causal parameters at the finest aggregation level of the features available. On the other extreme, the **A**verage **T**reatment **E**ffects (ATEs) represent the population averages. If $\Delta$ relates to the population with $D=m$, then this is the **A**verage **T**reatment **E**ffect on the **T**reated (ATET) for treatment $m$. However, for example, it might also relate to a combination of two or more treatment populations. The ATE and ATET are the classical parameters investigated in many econometric causal studies. The **G**roup **A**verage **T**reatment **E**ffect (GATE) parameters are in-between those two extremes with respect to their aggregation levels.[8] The IATEs and the GATEs are special cases of the so-called **C**onditional **A**verage **T**reatment **E**ffects (CATEs).

---

[6] To avoid complications, we assume $p$ to be finite (although it may be very large). In Section 4 we assume that the two groups fully overlap to avoid introducing new notation.

[7] Note that under the identifying assumption imposed in the next section, *IATE(m, l, x, Δ)* is the same for all treatment groups and does therefore not depend on *Δ*.

[8] Note that we presume that the analyst selects the variables *Z* prior to estimation. They are not assumed to be determined in a data driven way, e.g., by statistical variable selection procedures. However, the estimated IATE may be analysed by such methods to describe their dependence on certain features. See Section 6 for more details. Note that Abrevaya, Hsu, and Lieli (2015) and Lee, Okui, and Whang (2017) introduce similar aggregated parameters that depend on a reduced conditioning set and discuss inference in the specific settings of their papers.

## 3.2 Identifying assumptions

The classical set of assumptions made in an unconfoundedness setting consists of the following parts (see Imbens, 2000, Lechner 2001):[9]

$$\{Y^0,...,Y^m,...,Y^{M-1}\} \coprod D \mid X = x, \qquad \forall x \in \chi; \qquad (CIA)$$

$$0 < P(D = d \mid X = x) = p_d(x), \qquad \forall x \in \chi, \forall d \in \{0,...,M-1\}; \ (CS)$$

$$Y = \sum_{d=0}^{M-1} 1(D = d)Y^d; \qquad (Observation\ rule)$$

The conditional independence assumption (CIA) implies that there are no features other than $X$ that jointly influence treatment and potential outcomes (for the values of $X$ that are in the support of interest, $\chi$). The common support (CS) assumption stipulates that for each value in $\chi$, there must be the possibility to observe all treatments. The stable-unit-treatment-value assumption (SUTVA) implies that the observed value of the treatment and the outcome does not depend on the treatment allocation of the other population members (ruling out spillover and treatment scale effects). This is captured by the definition of the potential outcomes and the observation rule. Usually, to have an interesting interpretation of the effects, it is required that $X$ is not influenced by the treatment (exogeneity). In addition to these *identifying* assumptions, assume that a large random sample of size $N$ from the random variables $Y, D, X$, $\{y_i, d_i, x_i\}$, $i=1,...,N$, is available and that these random variables have at least first and second moments.[10]

If this set of assumption holds, then all IATEs are identified in the sense that they can be uniquely deduced from expectations of variables that have observable sample realisations (see Hurwicz, 1950):

---

[9] To simplify the notation, we take the strongest form of these assumptions. Some parameters are identified under weaker conditions as well (for details, see Lechner, 2001, or Imbens, 2000, 2004).

[10] The identification results will also hold under weights-based and dependent sampling (if the dependence is not too large and certain additional regularity conditions are imposed), but for simplicity we stick to the i.i.d. case. Second moments are not needed for identification, but are required for the theory and inference parts below.

$$\begin{aligned}
IATE(m,l;x,\Delta) &= E(Y^m - Y^l \mid X = x, D \in \Delta) \\
&= E(Y^m - Y^l \mid X = x) \\
&= E(Y^m \mid X = x, D = m) - E(Y^l \mid X = x, D = l) \\
&= E(Y \mid X = x, D = m) - E(Y \mid X = x, D = l) \\
&= IATE(m,l;x); \qquad \forall x \in \chi, \forall m \neq l \in \{0,...,M-1\}.
\end{aligned}$$

Note that the IATE does not depend on the conditioning treatment set, $\Delta$. Since the distributions used for aggregation, $f_{\tilde{X}\mid Z=z, D\in\Delta}(\tilde{x})$ and $f_{X\mid D\in\Delta}(x)$, relate to observed variables (X, D) only, they are identified as well (under standard regularity conditions). This in turn implies that the GATE and ATE parameters are identified (their dependence on $\Delta$ remains, if the distribution of the features depends on $\Delta$).

## 4 Estimation and inference

In this section, we discuss fundamental ideas, theoretical properties and inference of the proposed Causal Forest based estimators. The first subsection introduces the modified splitting criteria for the Causal Trees that form the Causal Forest. The following subsection briefly reviews the theoretical guarantees and properties such estimators have, followed by an implementation of weights-based approximate inference as a computationally convenient tool to conduct inference for all desired aggregation levels. Subsection 4 considers the aggregation steps required for the ATE and GATE parameters explicitly. The final subsection considers several issues related to the practical implementation of the estimator, including local centering.

### 4.1 IATE: Towards an MSE minimal estimator

Denoting the conditional expectations of Y given X in the subpopulation $D = d$ by $\mu_d(x) = E[Y \mid X = x, D = d]$ leads to the following expression of $IATE(m,l;x)$ as a difference of $\mu_m(x)$ and $\mu_l(x)$:

$$IATE(m,l;x) = \mu_m(x) - \mu_l(x); \qquad \forall x \in \chi, \forall m \neq l \in \{0,...,M-1\}.$$

This estimation task is different from standard ML problems because the two conditional expectations have to be estimated in different, treatment-specific subsamples.[11] Thus, the ML prediction of the difference cannot be directly validated in a holdout sample. This observation is indeed the starting point of the current causal ML literature. The papers then differ on how to tackle this issue.

An easy-to-implement estimator consists in estimating the two conditional expectations separately by standard ML tools, and taking a difference. Below, we denote this estimator as *Basic*, $\widehat{IATE}^{basic}(m,l;x) = \hat{\mu}_m^{basic}(x) - \hat{\mu}_l^{basic}(x)$. This approach has the disadvantage that standard ML methods attempt to maximise out-of-sample predictive power of the two estimators *separately*. More concretely, if a Random Forest is used, the difference of the predictions of the two different estimated forests (one estimated in subpopulation *m*, the other one estimated in subpopulation *l*) may suggest a variability of the IATE that is just estimation error due to (random) differences in the estimated forests. This problem can be particularly pronounced when the features are highly predictive for *Y*, but not for the IATEs. Another example is the case of very unequal treatment shares. As (standard) tree-building uses a stopping rule defined in terms of minimum leaf size, there will be, e.g., many more observations in treatment *m* compared to treatment *l,* even for similar values of *x*. Thus, the forest estimated for treatment *m* will be finer than the one estimated for treatment *l*. Again, this may lead to spurious effect heterogeneity. However, despite these methodological drawbacks, the large-scale EMCS of KLS21 finds that the Random Forest based *Basic* estimator may perform well compared to technically more sophisticated approaches, specifically when the IATEs are large and vary strongly with the features.[12]

---

[11] This is implied by the fact that the causal effect is a hypothetical construct that is per se unobservable. Thus, in the words of Athey and Imbens (2016), the 'ground truth' is unobservable in causal analysis.

[12] An additional drawback is that usually ML estimators tend to have a biased asymptotic distribution which makes inference difficult.

An alternative approach is to use the same trees in both subsamples in which $\mu_m(x)$ and $\mu_l(x)$ are estimated by $\hat{\mu}_m(x)$ and $\hat{\mu}_l(x)$, respectively. Of course, the key is then how to obtain a plausible splitting rule for this 'joint' forest (that leads to a correlation of $\hat{\mu}_m(x)$ and $\hat{\mu}_l(x)$). The dominant approach in the literature so far seems to consider the analogy to a classical Random Forest regression problem in which the 'ground truth', i.e., the individual treatment effect, would be observable. In this case, the tree estimates of $IATE(m,l;x)$ would be equal to the mean of the 'observed' individual treatment effects in each leaf. For such a case, some algebra reveals that minimising the mean squared expected error of the prediction and maximising the variance of the predicted treatment effects leads to the same sample splits. Therefore, Athey and Imbens (2016) suggest for their causal CARTs to split the parent leaf such as to maximise the heterogeneity of the estimated effects (subject to some adjustments for overfitting). This criterion is also used in one of the approaches of Wager and Athey (2018) and in Oprescu, Syrgkanis, and Wu (2019). However, in case of causal estimation, when the individual treatment effect is unobservable, there is no guarantee that the difference of the outcome means of treated and controls within all leaves equals the means of the true effects within all leaves. Without this condition, the maximisation of treatment effect heterogeneity is not equivalent to MSE minimisation of treatment effects prediction. The reason for the difference to the standard predictive case is due to potential selection bias. Intuitively, if selection bias does not matter, such an equality holds in expectation and this criterion should be a good approximation to minimizing the MSE of the estimated individual treatment effect. However, if selection bias is relevant (as it is likely to be, particularly in the early splits of the tree), then the quality of this approximation may be questionable.

An alternative approach, which is our first modification of the approach by WA18, is to derive a splitting rule that considers the mean square error of this estimation problem directly. For the comparison between two alternatives at a given value of $x$, we obtain:

$$MSE\left[\widehat{IATE}(m,l;x)\right] = E\left\{\left[\widehat{IATE}(m,l;x) - IATE(m,l;x)\right]^2\right\}$$

$$= E\left[\hat{\mu}_m(x) - \mu_m(x)\right]^2 + E\left[\hat{\mu}_l(x) - \mu_l(x)\right]^2 - 2E\left[\hat{\mu}_m(x) - \mu_m(x)\right]\left[\hat{\mu}_l(x) - \mu_l(x)\right]$$

$$= MSE\left[\hat{\mu}_m(x)\right] + MSE\left[\hat{\mu}_l(x)\right] - 2\underbrace{E\left[\hat{\mu}_m(x) - \mu_m(x)\right]\left[\hat{\mu}_l(x) - \mu_l(x)\right]}_{MCE[\hat{\mu}_m(x),\hat{\mu}_l(x)]}$$

$$= MSE\left[\hat{\mu}_m(x)\right] + MSE\left[\hat{\mu}_l(x)\right] - 2MCE\left[\hat{\mu}_m(x), \hat{\mu}_l(x)\right].$$

When there are more than two treatments, this is averaged over the treatments and over the observations used for forest building:

$$\overline{MSE} = \frac{1}{N}\sum_{i=1}^{N}\left\{\sum_{m=0}^{M-1}\sum_{l=m+1}^{M-1} MSE\left[\widehat{IATE}(m,l;x_i)\right]\right\}$$

$$= \frac{1}{N}\sum_{i=1}^{N}\left\{(M-1)\sum_{m=0}^{M-1} MSE\left[\hat{\mu}_m(x_i)\right] - 2\sum_{m=0}^{M-1}\sum_{l=m+1}^{M-1} MCE\left[\widehat{IATE}(m,l;x_i)\right]\right\}.$$

This derivation of the mean square error is instructive. It shows that the *Basic* estimator fails to take into account that estimation errors may be correlated, conditional on the features. Thus, it may be advantageous to tie the estimators together in a way such that the correlation of their estimation errors becomes positive (and errors cancel to some extent). The complication is that the **M**ean **C**orrelated **E**rror (MCE) is difficult to estimate.

For constructing estimators based on this criterion, the MSE of $\hat{\mu}_d(x)$ must be estimated. This is straightforward, as the MSEs of all $M$ functions $\hat{\mu}_d(x)$ can be computed in the respective treatment subsamples in the usual way. Denote by $N_{S_x}^d$ the number of observations with treatment value $d$ in a certain stratum (leaf) $S_x$, which is defined by the values of the features $x$. Then, the following estimator is a 'natural' choice:

$$\widehat{MSE}_{S_x}\left[\hat{\mu}_d(x)\right] = \frac{1}{N_{S_x}^d}\sum_{i=1}^{N}\underline{1}(x_i \in S_x)\underline{1}(d_i = d)\left[\hat{\mu}_d(x_i) - y_i\right]^2.$$

Note that the overall MSE in $S_x$ is the sum of the MSEs in the treatment specific subsamples of $S_x$, where each subsample receives the same weight (independent of the number of observations in that subsample), as implied by the above MSE formula for causal effect estimation.

To compute the correlation of the estimation errors, we need a proxy for cases when there are no observations with the same values of *x* in all treatment states (as is always true for continuous features). In this case, we propose using the closest neighbour available (which is denoted by $y_{(i,m)}$ below) instead.[13]

$$\widehat{MCE}(m,l;S_x) = \frac{1}{N_{S_x}^l + N_{S_x}^m} \sum_{i=1}^{N} \mathbf{1}(x_i \in S_x)\left[\underline{1}(d_i = m) + \underline{1}(d_i = l)\right]\left[\hat{\mu}_m(x_i) - \tilde{y}_{(i,m)}\right]\left[\hat{\mu}_l(x_i) - \tilde{y}_{(i,l)}\right],$$

$$\tilde{y}_{(i,m)} = \begin{cases} y_i & if \quad d_i = m \\ y_{(i,m)} & \quad d_i \neq m \end{cases}.$$

Please note that to ensure validity of this step, finding the closest neighbour would need to be done in every potential daughter node to estimate the relevant quantities for the corresponding leaves. To reduce computational time, closest neighbours are determined at the beginning based on the whole available sample for growing the trees.

The splitting rule that minimizes $\widehat{MSE}[\widehat{IATE}(m,l,x)]$ is motivated by maximising the predictive power of the estimator. Analysing the estimated $\widehat{MSE}\left[\widehat{IATE}(m,l;x)\right]$ reveals that its minimization favours splits maximizing the differences of all $\hat{\mu}_d$ between the daughter leaves. This part is similar to the maximization of treatment heterogeneity in Athey & Imbens (2016). Additionally, the splits also favour large differences between the potential outcomes $\hat{\mu}_m$ and $\hat{\mu}_l$ in the same leaf. However, in causal analysis inference is important. Thus, if the MSE-minimal estimator has a substantial bias, this is problematic. In causal studies, a substantial source of bias is a non-random allocation of treatment (*selection bias*). This is captured by the propensity score, $P(D = d \mid X = x)$, which thus has a certain role to play in many proposed estimators of IATEs. This is usually tackled by a first stage estimation of $P(D = d \mid X = x)$ and / or $E(Y \mid D = d, X = x)$ and treating it as a nuisance parameter in the Random Forest estimation of $\widehat{IATE}(m,l;x)$.[14] Here, we would like to avoid computer-time-consuming additional estimation steps, but still improve on the robustness of the estimator with respect to selection bias, in particular in smaller samples where the Random Forests may not be automatically fine enough to remove all selection biases.

---

[13] Closeness is based on a simplified Mahalanobis metric as in Abadie and Imbens (2006). This simplified version has the inverse of the variances of the features on the main diagonal. Off-diagonal elements are zero. The simplification avoids computational complications when inverting the variance-covariance matrix of potentially large-dimensional features at the cost of ignoring correlations between covariates.

These considerations lead us to suggest a further modification of the splitting rules. Denote by *leaf(x')* and *leaf(x'')* the values of the features in the daughter leaves resulting from splitting some parent leaf. We propose to add a penalty term to $\widehat{MSE}\left[\widehat{IATE}(m,l;x)\right]$ that penalizes possible splits where the treatment probabilities in the resulting daughter leaves are similar (splits leading to leaves with similar treatment shares will not be able to remove much selection bias, while if they are very different in this respect, they approximate differences in $P(D = d \mid X = x)$ well). In other words, the modified criterion prefers splits with high propensity score heterogeneity and puts explicit emphasis on tackling selection bias. In the simulations below, the following penalty function is added to the splitting criteria of (some) estimators considered:

$$penalty(x',x'') = \lambda \left\{ 1 - \frac{1}{M} \sum_{d=0}^{M-1} [P(D = d \mid X \in leaf(x')) - P(D = d \mid X \in leaf(x''))]^2 \right\}.$$

This penalty term is zero if the split leads to a perfect prediction of the probabilities in the daughter leaves (in which case, however, the common support condition will fail). It reaches its maximum value, $\lambda$, when all probabilities are equal. Thus, the algorithm prefers a split that is not only predictive for *Y* but also for *D*. We conjecture that this low-cost local estimate of the propensity score reduces selection bias in ways like classical regression or matching type estimators that condition on or weight with the propensity score in one way or another. Of course, the choice of the exact form of this penalty function is arbitrary. Furthermore, there is the issue of how to choose $\lambda$ (*without* expensive additional computations) which is taken up again in Section 4.5.

---

[14] See for example Oprescu, Syrgkanis, and Wu (2019). There is also a substantial literature on how to exploit so-called double-robustness properties when estimating the causal effects at higher aggregation level, see, e.g., Belloni, Chernozhukov, Fernández-Val, and Hansen (2017) and the references therein.

When there are more than two treatments, this algorithm can be implemented also in a different way. Based on the 'sample reduction properties' in Lechner (2001), one could obtain estimates of all parameters in pair-wise comparisons independent of each other. However, this may become computationally cumbersome when there are many treatments.

## 4.2 Properties

In this section we present the asymptotic properties of the **M**odified **C**ausal **F**orest estimator abstracting from local centering (discussed in Section 4.5 below).[15]

The plain-vanilla MCF procedure aggregates individual trees $T$ into a forest in the following way. The data points are sampled into the training set and honest set without replacement from non-overlapping halves of the full data set of sizes $N_1 = N_2 = \frac{N}{2}$. The sampling rates are $N_1^{\beta_1}$ and $N_2^{\beta_2}$ where $0 < \beta_1, \beta_2 \leq 1$ for the training and honest set respectively i.e., $s_1 \sim N_1^{\beta_1}$ for the training set and $s_2 \sim N_2^{\beta_2}$ for the honest set. Thus, subsample sizes $s_1$ and $s_2$ are proportional to *N*.[16] Denote the $N_1$ training observations as $R_{1,i} = (X_{1,i}, Y_{1,i}, D_{1,i})$ for $i = 1, ..., N_1$, and the $N_2$ honest observations as $R_{2,i} = (X_{2,i}, Y_{2,i}, D_{2,i})$ for $i = 1, ..., N_2$. The two sets collecting the potential training and honest data points $\vec{R}_1 = (R_{1,1},...,R_{1,N_1})$ and $\vec{R}_2 = (R_{2,1},...,R_{2,N_2})$ have zero overlap and the individual data come from the same distribution. Each tree of the form $T(m,l,x;\xi,\vec{R})$, where $\xi \sim \Xi$ is a source of auxiliary randomness in the tree building process (such as random choice of splitting variables) and $\vec{R} = \{\vec{R}_1, \vec{R}_2\}$, can be used to estimate $IATE(m,l;x)$. When not important for proofs, $\xi$ and/or $\vec{R}$ will be suppressed in the notation for better readability. A Random Forest is an average of trees trained over all possible size-$(s_1, s_2)$ subsamples of the training and honest data. In practice, the average is taken over $B$ trees:

$$F(m,l,x;\vec{R}) \approx \frac{1}{B}\sum_{b=1}^{B} T_b(m,l,x;\xi_b,\vec{R}_b),$$

where $\vec{R}_b = \{\vec{R}_{1,b}, \vec{R}_{2,b}\}$ collects $\vec{R}_{1,b} = \{R_{1,b1},...,R_{1,bs_1}\}$ and $\vec{R}_{2,b} = \{R_{2,b1},...,R_{2,bs_2}\}$ which are drawn without replacement from $\vec{R}_1$ and $\vec{R}_2$ respectively and $\xi_b$ is a random draw from $\Xi$. The $b$ subscript will be supressed when it will not lead to any confusion in the proofs. Like in WA18, the trees $T$ need to satisfy the following definitions.

---

[15] More details of the assumptions needed as well as the formal proofs are contained in Appendix A.1.

[16] If *N* is an odd number, one of the two subsamples will contain one observation more than the other one without any consequences. Note also that the dependence of the subsample sizes $s_1$ and $s_2$ on *N* is suppressed in most of the following notation.

**DEFINITION 1** A tree grown on a training sample $\{R_{1,b1},...,R_{1,bs_1}\}$ is *honest* if the tree does not use the responses $Y_{2,1},...,Y_{2,N_2}$ from the honest sample to place its splits.

Honesty is crucial for inference and bias bounds. The splitting and subsampling procedure described above guarantees that the trees are honest. The main difference to the honesty in WA18 is the reverse order of steps. WA18 first subsample and then split the data for their double-sample trees. Here, the data set is split first and then the training and honest sets are subsampled from the given split. This allows to better control bias and variance rates as the size of the training set will codetermine the size of the final leaves translating into bias and the size of the honest set will influence the variance rate.

**DEFINITION 2** A tree is a *random-split* tree if at every step of the tree-growing procedure, marginalizing over $\xi$, the probability that the next split occurs along the $u$-th feature is bounded below by $\pi/p$ for some $0 < \pi \leq 1$, for all $u = 1,...,p$.

To guarantee consistency, the final leaves must asymptotically shrink in all dimensions similarly as in Meinshausen (2006) and WA18. Definition 3 controls the shape of the leaves, and Definition 4 imposes *symmetry*, which is needed to derive the asymptotic results.

**DEFINITION 3** A tree predictor grown by recursive partitioning is $(\alpha,k)$-*regular* for some $\alpha > 0$ if (1) each split leaves at least a fraction $\alpha$ of the available training examples of each treatment on each side of the split, (2) the leaf containing $x$ has at least $k$ observations from each of the $M$ treatment groups for some $k \in \mathbb{N}$, and (3) the leaf containing $x$ has at least one treatment with less than $2k-1$ observations.

Regarding the role of the splitting rule on $(\alpha,k)$-regularity, the algorithm first determines splits that do not violate the regularity condition. For these, the splitting criterion is calculated and the split that achieves the minimum value of the objective function is chosen as the best one. By following this procedure, any influence of the splitting criterion (including the penalty) on the regularity of the final leaves can be ruled out.

**DEFINITION 4** A predictor is *symmetric* if the (possibly randomized) output of the predictor does not depend on the order in which the observations are indexed in the training and honest samples.

Given these concepts, we can state the main theorems guaranteeing consistency and asymptotic normality of the estimates. Further assumptions necessary for achieving this asymptotic distribution in the case of i.i.d. sampling are (i) Lipschitz continuity of first and second order moments of the outcome variable conditional on the features, (ii) using subsampling to obtain the training data for tree building (subsamples should increase with *N*, slower than *N*, but not too slow), and (iii) conditions on the features (independent, continuous with bounded support, *p* is fixed, i.e. low-dimensional).

The proposed estimators (see below) fulfil these conditions on tree building. However, some of these conditions are very specific and sometimes difficult to match (like covariates being independent which is a common assumption in this literature, e.g., WA18), or impossible to verify (like the regularity conditions on the conditional outcome expectations). Nevertheless, the simulation results show that the predictions from all Random/Causal Forest based estimators appear to be very close to be normally distributed even for the smallest sample size investigated (*N=1,000*).

In the following, the asymptotic properties of the MCF estimator are derived. All proofs are collected in the Appendix A.1. Each section in the appendix contains all the necessary proofs and intermediate results for the corresponding main theorem. One of the main differences to WA18 is that due to a different splitting and subsampling approach to achieve honesty, our proofs can utilize the weighted representation of the Causal Forest estimator. Meanwhile, the Causal Forest estimator in WA18 is an approximation of a U-statistic and their proofs are based on the corresponding Gaussian theory. The first result is the bound on the bias of the forest. The proof is similar to the proof in WA18. In the first step, we show that the leaves get small in volume as sample size $s_1$ gets large. In the second step, we show that the honest observations in the final leaf can be seen as a subset of nearest neighbours around the point $x$ and their expected distance and Lipschitz continuity help to bound the bias. Lemma 1 and its proof in Appendix A.1.1 give rates at which the Lebesgue measure of final leaves in a regular, random-split tree shrinks under the assumption of the features to be independent and uniformly distributed.

**THEOREM 1** Under the conditions of Lemma 1, suppose moreover that trees $T$ are honest and $E\left[Y^d \mid X = x\right]$ are Lipschitz continuous. Then, the bias of the Causal Forest at a given value of $x$ is bounded by

$$\left| E\left[\widehat{IATE}(m,l;x)\right] - IATE(m,l;x) \right| = O\left(s_1^{-\log(1-\alpha)/p\log(\alpha)}\right).$$

Note that the bias rate of the forest is the same as the bias rate of a single tree, as a forest prediction is a simple average of the tree predictions. The bias rate is mainly driven by the shrinkage of the Lebesgue measure of the leaf that is influenced by the choice of parameter $\alpha$. The upper bound in Theorem 1 resembles the bias rate of nearest neighbours regression estimators (NNRE) under Lipschitz continuity in a $p$ dimensional leaf that shrinks at $(1-\alpha)$ rate. That stems from the fact that the final leaves can be bounded by balls that shrink at the same rate as the final leaves as shown in the proof in the Appendix A.1.1. Note that this rate is rather conservative as it stems from bounding the shallowest leaf with a leaf that would always end up with a $(1-\alpha)$ share of observations at the same level of depth. The rate is faster than in WA18 as here we bound by the expected distance between the nearest neighbours and the point $x$ instead of the longest expected diameter of the leaf.

**THEOREM 2** Assume that there are i.i.d. data $(X_i, Y_i, D_i) \in [0,1]^p \times \mathbb{R} \times \{0,1,...,M-1\}$ collected in $\vec{R}$ and a given value of $x$. Moreover, features are independently and uniformly distributed $X_i \sim U\left([0,1]^p\right)$. Let $T$ be an honest, regular and symmetric random split tree. Further assume that $E\left[Y^d \mid X = x\right]$ and $E\left[(Y^d)^2 \mid X = x\right]$ are Lipschitz continuous and $Var\left(Y^d \mid X = x\right) > 0$. Then for $\beta_1 < \beta_2 < \frac{p+2}{p}\frac{\log(1-\alpha)}{\log(\alpha)}\beta_1$,

$$\frac{\widehat{IATE}(m,l;x) - IATE(m,l;x)}{\sqrt{Var(\widehat{IATE}(m,l;x))}} \to N(0,1).$$

The restrictions on the sampling rates require that $(2+p)\log(1-\alpha)/(p\log(\alpha)) > 1$. This means that $\alpha$ needs to be set closer and closer to 0.5 for larger $p$. This is a consequence of the curse of dimensionality as values of $\alpha$ closer to 0.5 make sure that the shallowest final leaves get tighter upper bounds making sure that the bias vanishes fast enough to 0. Additionally, the relationship between the subsampling rates further ensures that the final leaf does not end up with too many honest observations and the squared bias–variance ratio converges to 0, i.e., the estimator is consistent.

### 4.3 Inference

There are several suggestions in the literature on how to conduct inference and how to compute standard errors of Random Forest based predictions (e.g., Wager, Hastie, and Efron, 2014; Wager and Athey, 2018; and the references therein). Although these methods can be used to conduct inference on the IATE, it is yet unexplored how these methods could be readily generalized to take account of the aggregation steps needed for the GATE and ATE parameters.

Therefore, we suggest an alternative inference method useful for estimators that have a representation as weighted averages of the outcomes. This perspective is attractive for Random Forest based estimators (e.g., Athey, Tibshirani, and Wager, 2019) as they consist of trees that first stratify the data (when building a tree), and subsequently average over these strata (when building the forest). Thus, we exploit the weights-based representation explicitly for inference (see also Lechner, 2002, and Abadie and Imbens, 2006, for related approaches).

Let us start with a general weights-based estimator. Denote by $\hat{W}_i$ the weight (that is normalized such that all weights sum up to $N$) that the dependent variable $Y_i$ receives in the desired estimator, $\hat{\theta}$ (which could be one of the IATEs, or GATEs, or ATEs).

$$\hat{\theta} = \frac{1}{N}\sum_{i=1}^{N} \hat{W}_i Y_i; \qquad Var(\hat{\theta}) = Var\left(\frac{1}{N}\sum_{i=1}^{N} \hat{W}_i Y_i\right).$$

Next, we apply the law of total probability to the variance:[17]

$$Var\left(\frac{1}{N}\sum_{i=1}^{N}\hat{W}_i Y_i\right) = E_{\hat{W}} Var\left(\frac{1}{N}\sum_{i=1}^{N}\hat{W}_i Y_i \mid \hat{W}_1,...,\hat{W}_N\right) + Var_{\hat{W}} E\left(\frac{1}{N}\sum_{i=1}^{N}\hat{W}_i Y_i \mid \hat{W}_1,...,\hat{W}_N\right)$$

$$= E_{\hat{W}}\left(\frac{1}{N^2}\sum_{i=1}^{N}\hat{W}_i^2 Var(Y_i \mid \hat{W}_1,...,\hat{W}_N)\right)$$

$$+ Var_{\hat{W}}\left(\frac{1}{N}\sum_{i=1}^{N}\hat{W}_i E(Y_i \mid \hat{W}_1,...,\hat{W}_N)\right).$$

However, the large conditioning sets of $E(Y_i \mid \hat{W}_1,...,\hat{W}_N)$ and $Var(Y_i \mid \hat{W}_1,...,\hat{W}_N)$ make it impossible to estimate these terms precisely without further assumptions. The conditioning sets can be drastically reduced, though, if observation '$i$' is not used to build the forest,[18] and the data used for the computations of the conditional mean and variance are an i.i.d. sample. To see this, recall that Random Forest weights are computed as functions of $\vec{X}_1 = (x_1,...,x_{N_1})$ and $\vec{Y}_1 = (y_1,...,y_{N_1})$ in the training sample. These weights are then assigned to observation '$i$' based on the value $x_i$ only. Thus, the weights are functions of $x_i$ and the training data, $\hat{w}_i = \hat{w}(x_i, \vec{X}_1, \vec{Y}_1)$. If observation '$i$' does not belong to the training data and there is i.i.d. sampling, $Y_i$ and $\hat{W}_j = \hat{w}(x_j, \vec{X}_1, \vec{Y}_1)$ are independent. Thus, we obtain $E(Y_i \mid \hat{W}_1,...,\hat{W}_N) = E(Y_i \mid \hat{W}_i) = \mu_{Y|W}(\hat{W}_i)$ and $Var(Y_i \mid \hat{W}_1,...,\hat{W}_N) = Var(Y_i \mid \hat{W}_i) = \sigma^2_{Y|W}(\hat{W}_i)$. This leads to the following expression of the variance of the proposed estimators:[19]

$$Var\left(\frac{1}{N_2}\sum_{i=N_1+1}^{N}\hat{W}_i Y_i\right) = E_{\hat{W}}\left(\frac{1}{N_2^2}\sum_{i=N_1+1}^{N}\hat{W}_i^2 \sigma^2_{Y|\hat{W}}(\hat{W}_i)\right) + Var_{\hat{W}}\left(\frac{1}{N_2}\sum_{i=N_1+1}^{N}\hat{W}_i \mu_{Y|\hat{W}}(\hat{W}_i)\right).$$

---

[17] Letting $A$ and $B$ be two random variables, then $Var(A) = E_B Var(A \mid B) + Var_B E(A \mid B)$.

[18] Note that this condition of 'honesty' is based on the same sample split for all trees. Thus, it is different to the implementation of 'honesty' as in Wager and Athey (2018) which is based on continuously switching the role of observations used for tree building and effect estimation in their Causal Forest. The reason is that if latter splitting is used, each weight may still depend on many observations.

[19] Note that the weighting estimator uses only the honest data, therefore the weights here sum up to $N_2$.

The above expression suggests using the following estimator for the leading terms:

$$\widehat{Var(\hat{\theta})} = \frac{1}{N_2^2} \sum_{N_1+1=1}^{N} \hat{w}_i^2 \hat{\sigma}_{Y|\hat{W}}^2(\hat{w}_i) + \frac{1}{N_2(N_2-1)} \sum_{N_1+1=1}^{N} \left[ \hat{w}_i \hat{\mu}_{Y|\hat{W}}(\hat{w}_i) - \frac{1}{N_2} \sum_{N_1+1=1}^{N} \hat{w}_i \hat{\mu}_{Y|\hat{W}}(\hat{w}_i) \right]^2.$$

The conditional expectations and variances may be computed by standard non-parametric or ML methods, as this is a one-dimensional problem for which many well-established estimators exist. Bodory, Camponovo, Huber, and Lechner (2020) investigate *k*-nearest neighbour estimators to obtain estimates for these quantities. They found good results in a binary treatment setting for the ATET. The same method is used here.[20] As both, $\mu_{Y|\hat{W}}$ and $\sigma_{Y|\hat{W}}^2$, are bounded and the number of nearest neighbours in their estimation is chosen such that it grows slower than $N_2$ (as documented in the Appendix A.2), consistency of the conditional expectations is guaranteed, see e.g. Devroye, Györfi, Krzyzak, and Lugosi (1994). Additionally, for larger $N$ the non-zero weights shall concentrate at a close neighbourhood of the given point $x$ as the leaves are shrinking to 0. Finally, note that because of the weighting representation, we conjecture that this approach can also be used to account for, e.g., clustering (by aggregating weights within clusters), or to conduct joint tests of several linear hypotheses, such as that several groups have the same or no effect, leading to Wald-type statistics. It is, however, beyond the scope of this paper to analyse rigorously the exact statistical conditions needed for this estimator to lead to valid inference.

## 4.4 GATE and ATE

Estimates for GATEs and ATE are most easily obtained by averaging the IATEs in the respective subsamples defined by *z* (assuming discrete *Z*) and Δ. Although estimating ATEs and GATEs directly instead of aggregating IATEs may lead to more efficient and more robust estimators (e.g., Belloni, Chernozhukov, Fernández-Val, and Hansen, 2017), the computational burden would also be higher, in particular if the number of GATEs of interest is large, as is common in many empirical studies. Therefore, letting $\widehat{IATE(m,l;x)}$ be an estimator of $IATE(m,l;x)$, we suggest estimating the GATEs and ATEs as appropriate averages of $\widehat{IATE(m,l;x)}$s:

$$\widehat{GATE}(m,l;z,\Delta) = \frac{1}{N_2^{z,\Delta}} \sum_{i=N_1+1}^{N} \underline{1}(z_i = z, d_i \in \Delta)\widehat{IATE}(m,l;x_i)$$

$$= \frac{1}{N_2^{z,\Delta}} \sum_{i=N_1+1}^{N} \underline{1}(z_i = z, d_i \in \Delta) \frac{1}{N_2} \sum_{j=N_1+1}^{N} \hat{w}_j^{IATE(m,l;x_i)} y_j$$

$$= \frac{1}{N_2} \sum_{i=N_1+1}^{N} \left( \frac{1}{N_2^{z,\Delta}} \sum_{j=N_1+1}^{N} \underline{1}(z_j = z, d_j \in \Delta)\hat{w}_i^{IATE(m,l;x_j)} \right) y_i$$

$$= \frac{1}{N_2} \sum_{i=N_1+1}^{N} \hat{w}_i^{GATE(m,l;z,\Delta)} y_i;$$

$$\hat{w}_i^{GATE(m,l;z,\Delta)} = \frac{1}{N_2^{z,\Delta}} \sum_{j=N_1+1}^{N} \underline{1}(z_j = z, d_j \in \Delta)\hat{w}_i^{IATE(m,l;x_j)}; \qquad N_2^{z,\Delta} = \sum_{i=N_1+1}^{N} \underline{1}(z_i = z, d_i \in \Delta).$$

$$\widehat{ATE}(m,l;\Delta) = \frac{1}{N_2^{\Delta}} \sum_{i=N_1+1}^{N} \underline{1}(d_i \in \Delta)\widehat{IATE}(m,l;x_i)$$

$$= \frac{1}{N_2} \sum_{i=N_1+1}^{N} \hat{w}_i^{ATE(m,l;\Delta)} y_i;$$

$$\hat{w}_i^{ATE(m,l;z,\Delta)} = \frac{1}{N_2^{\Delta}} \sum_{j=N_1+1}^{N} \underline{1}(d_j \in \Delta)\hat{w}_j^{IATE(m,l;x_i)}; \qquad N_2^{\Delta} = \sum_{j=N_1+1}^{N} \underline{1}(d_i \in \Delta).$$

---

[20] They also found a considerable robustness on how exactly to compute the conditional means and variances. Note that since their results relate to aggregate treatment effect parameters, their generalisability to the level of IATE's is unclear.

From this expression, ATEs and GATEs have the same type of weights-based representation as the IATEs. Hence, asymptotic normality can be established in a similar way as for the IATEs, just using different weights.

**THEOREM 3** Let all assumptions from Theorem 2 hold. Then,

$$\frac{\widehat{ATE}(m,l) - ATE(m,l)}{\sqrt{Var(\widehat{ATE}(m,l))}} \to N(0,1).$$

**COROLLARY 2** Let all assumptions from Theorem 2 hold. Then,

$$\frac{\widehat{GATE}(m,l;z) - GATE(m,l;z)}{\sqrt{Var(\widehat{GATE}(m,l;z))}} \to N(0,1).$$

## 4.5 Implementation

Beyond the *Basic* estimator, we investigate seven different estimators (and some additional variants), all based on using the same single forest for all treatment subsamples and the same subsampling rule. They differ in their complexity. *OneF* ignores the MCE. It constructs the tree solely based on the sum of the treatment specific MSEs of estimating $\hat{\mu}_d(x)$. *OneF.MCE* estimates the MCE in a computationally not too expensive way by using nearest neighbours. *OneF.VarT* is the Causal Forest estimator of Wager and Athey (2018) based on maximising treatment effect heterogeneity. *OneF.MCE.Penalty* and *OneF.VarT.Penalty* use the same splitting criteria as *OneF.MCE* and *OneF.VarT* but add an additional penalty term to the splitting rule to reduce potential selection biases. *OneF.MCE.LCk* and *OneF.MCE.Penalty.LCk* recenter the outcome variables before using the *OneF.MCE* and the *OneF.MCE.Penalty* algorithms.

The main elements of the algorithms used for estimating the results in the simulation and application parts are the following:

1) Split the estimation sample randomly into two parts of equal size (sample A and sample B)
2) Estimate the trees that define the respective Random Forest in sample A.

   a. *Basic*: Estimate Random Forests for each treatment state in the subsamples defined by treatment state. The splitting rule consists of independently minimizing the mean squared prediction error within each subsample.

   b. *OneF*: Estimate the same forest for all treatment states jointly. The splitting rule is based on minimising the sum of the MSEs for all treatment state specific outcome predictions (MCEs are set to 0).

    c. *OneF.MCE*: Same as b), but before building the trees, for each observation in each treatment state, find a close 'neighbour' (same observation might be used as neighbour several times) in every other treatment state and save its outcome (to estimate MCE). The splitting rule is based on minimising the overall MSEs, taking account of all MCEs.

    d. *OneF.MCE.LCk*: Same as c) but with recentered outcomes (based on *K* folds, see Appendix A.3 for details).

    e. *OneF.VarT*: Same as b), but splitting is based on maximising estimated treatment effect heterogeneity.

    f. *OneF.MCE.Penalty*: Same as c) but a penalty term penalizing propensity score homogeneity is added.

    g. *OneF.MCE.Penalty.LCk*: Same as f) but with recentered outcomes.

    h. *OneF.VarT.Penalty*: Same as e) but a penalty term penalizing propensity score homogeneity is added.[21]

3) Apply the sample splits obtained in sample A to all subsamples (by treatment state) of sample B and take the mean of the outcomes in the respective leaf as the prediction that comes with this Forest.

4) Obtain the weights from the estimated Forest by counting how many times an observation in sample B is used to predict IATE for a particular value of *x*.

---

[21] Setting $\lambda$ to the square of the sum of the differences of the treatment means corresponds to the intuition used for *OneF.MCE.Penalty*. However, in the simulations below it appeared that such a value is far too small to reduce biases meaningfully when there is selectivity (available on request). Therefore, a value corresponding to 100 x that value is used below.

5) Aggregate the IATEs to GATEs by taking the average over observations in sample B that have the same value of *z* and treatment group Δ. Do the same aggregation with the weights to obtain the new weights valid for the GATEs.

6) Do the same steps as in 5) to obtain the ATEs, but average over all observations in treatment group Δ.

7) Compute weights-based standard errors as described above. Use the estimated standard errors together with the quantiles from the normal distribution to obtain critical values (p-values).

While Appendix A details the implementation further, at least four points merit some more discussion. The efficiency loss inherent in the two-sample approach could be avoided by cross-fitting, i.e. by repeating the estimation with exchanged roles of the two samples and averaging the two (or more) estimates.[22] However, in such a case it is unclear how to compute the weights-based inference for the averaged estimator as the two components of this average will be correlated.[23] A second issue concerns the fact of forming the neighbours by simplified Mahalanobis matching, which has the issue of being potentially large-dimensional. A lower dimensional alternative might be to estimate a prognostic score, $[\hat{\mu}_0(x),....,\hat{\mu}_{M-1}(x)]$, by ML methods and then use this score instead.[24] While this is a viable alternative, it requires again (costly) nuisance parameter estimation which we want to avoid with the suggested estimator. The third note concerns *K*-fold cross-fitting as suggested by ATW19 and implemented in the application of Athey and Wager (2019). Appendix A.3 details the algorithm performing the recentering that is modified (compared to ATW19) to consider the special structure of the two-sample approach used here to obtain weights-based inference.

---

[22] However, note that the simulations below comparing the two-sample estimator with a one-sample-with-honesty strategy suggest that the efficiency loss is minor, if existent at all.

[23] For such a cross-fitted estimator (e.g., computed as mean of the single estimators), conservative inference could be obtained by basing inference on normality with a variance taken as average over the single estimations.

[24] Note that propensity scores will not be helpful as the intended correction is not directly related to selection bias.

The fourth point concerns the penalty function. In the simulations below, setting λ as equal to *Var(Y)* works well. *Var(Y)* corresponds to the MSE when the effects are estimated by the sample mean without any splits. Thus, it provides some ad-hoc benchmark for plausible values of λ. In small-scale experiments with values smaller and larger than *Var(Y)* the MSE shows little sensitivity for values half as well as twice the size of *Var(Y)* (available on request). Generally, decreasing the penalty increases biases and reduces variances, et vice versa. As will be seen below in the simulations, biases are more likely to occur when selection is strong. Thus, if a priori knowledge about the importance of selectivity is available, then the researcher might increase (strong selectivity) or decrease (weak selectivity or experiment) the penalty term accordingly.

## 5   Simulation results of an Empirical Monte Carlo Study

### 5.1   Data base, simulation concept, and data generating processes

It is a general problem of simulation results that they depend on the design of the data generating process (DGP) chosen by the researcher, which might reduce their generalizability to specific empirical applications. While this may be innocuous if simulations are used to investigate specific theoretical properties of estimators or test statistics only (e.g., analysing what happens if the correlation of features increases), it is more problematic when simulations are used to investigate the suitability of estimators for specific applications. In this case, a simulation environment that closely mimics real applications is advantageous. In that way, the results generalize more easily to applications with a similar data structure.

Huber, Lechner, and Wunsch (2013) and Lechner and Wunsch (2013) proposed a specific data driven method for simulation analyses which they called Empirical Monte Carlo Study (EMCS). The main idea of an EMCS is to have a large real data set from which to draw random subsamples using much information from the real data. In the simulations below, we follow this approach (see Appendix B for the exact algorithm used).[25]

We base the simulations on Swiss social security data previously used to evaluate active labour market policies (Lechner et al. 2020). More precisely, as in Huber, Lechner, and Mellace (2017) for analysing a mediation framework, and in KLS21 for comparing different estimators of the IATE, we consider the effects of a job search programme on employment outcomes. This data is well suited for this type of analysis, as it is long and wide (about 95,000 observations, more than 50 base covariates). Furthermore, the programme is one for which the literature argues that a selection-on-observed-variable assumption is plausible when rich social security data are available (see, e.g., Gerfin and Lechner, 2002, for the Swiss case, and the survey by Card, Kluve, and Weber, 2018).[26]

The main steps are the following: Using the initial data, we estimate the propensity score, *p(x)*, using the same specification as KLS21. As such, *p(x)* depends on 77 features that enter a logit model estimated by maximum likelihood.[27] This estimated propensity score plays the role of the true selection process in the following steps. Next, the treated are removed from the data. Thus, for all remaining observations (approx. 84,000) we observe $Y^0$, the non-treatment outcome (measured as number of months in employment in the next 33 months after the pro-gramme starts), the features, and *p(x)*. This information is used to simulate the individual treatment effects (ITE), to compute the IATEs, and to compute $Y^1$ as the sum of $Y^0$ and the ITE. In the next step, we draw randomly a validation data set with 5,000 observations and remove it from the main data. From the remaining data, we draw random samples of size *N = 1,000, N = 4,000,* and *N = 8,000*, simulate a treatment status for each observation using the 'true' propensity score (shifted such that the expected treatment share is about 50%), and take the potential outcome that corresponds to the simulated treatment as the observed outcome, *Y*. These random samples are the training data for the algorithms, while their performance is measured out-of-sample on the 5,000 observations of the validation sample.

---

[25] Note the similarity of the concept of EMCS with the very recently proposed method of Synth Validation (Schuler, Jung, Tibshirani, Hastie, and Shah, 2017). Although the latter method is intended to select methods for specific datasets (in the ML spirit of comparing predictions with observed variables), both methods could be used for method selection as well as for method validation. Advani, Kitagawa, and Słoczyński (2019) point to the potential limits of generalizing results from such simulation exercises.

[26] Our implementation follows closely KLS21. Therefore, for the sake of brevity, we do not repeat their extensive documentation of all steps that lead to the final sample and their descriptions of the estimation sample. The reader interested in more details is referred to KLS21. The underlying data is available from FORS, Lausanne via this link: https://doi.org/10.23662/FORS-DS-1203-1 . The programmes used in the simulations are coded in Gauss 18.1.5 and 19.1.0.

[27] The number is larger than the number of base covariates due to the addition of some transformations of the base covariates, such as dummy variables.

What remains is the specification of the ITEs. Here we specify four different IATEs such that they reflect different correlations with the propensity score (as disentangling selectivity from effect heterogeneity is a key challenge for all estimators in this setting) and different strengths and variability of the effects. To be specific, the first specification sets the ITE to zero for all individuals. In the next two specifications, the IATE is a non-linear function of the propensity score given through a non-linear deterministic component $\xi(x)$:

$$\xi(x) = \sin\left(1.25 \, \pi \, \frac{p(x)}{\max_{i=1:N} p(x_i)}\right),$$

$$IATE(x) = \delta \frac{\xi(x) - \bar{\xi}}{SD(\xi)}; \qquad \bar{\xi} = \frac{1}{N}\sum_{i=1}^{N}\xi(x_i), \quad SD(\xi) = \sqrt{\frac{1}{N}\sum_{i=1}^{N}\left[\xi(x_i) - \bar{\xi}\right]^2}.$$

The parameter $\delta$ determines the variability of the IATE. In the simulations, we consider two values of $\alpha$ (i) $\delta = 2$ ('normal' heterogeneity); and (ii) $\delta = 8$ ('strong' heterogeneity). Due to the non-linear way in which the features enter the IATEs via the propensity score, it is a difficult task for every estimator not to confuse selection effects with heterogeneous treatment effects. Finally, we also consider an 'easier' case in which $\xi(x)$ depends linearly on the *insured earnings* of the unemployed. The latter is an officially defined pre-unemployment earnings measure used to compute unemployment benefits. Although this variable is related to the selection into programmes as well, the link to the IATEs is much weaker.

Adding two independent random components to the IATE leads to the ITE. The first random term is a (minus) Poisson (1) variate adjusted to have zero expectation. The second random term ensures that the ITE, and thus $Y^1$, keeps its character as an integer (month) in a way that the rounding 'error' is independent of the IATEs:

$$ITE(x) = IATE(x) + (1-u) + v;$$
$$u \sim Poisson(1);$$
$$v^* \sim Uniform[0,1];$$
$$v^{diff} = IATE(x) + u - floor(IATE(x) + u);$$
$$v = \underline{1}(v^* > v^{diff})(-v^{diff}) + \underline{1}(v^* \leq v^{diff})(1 - v^{diff}).$$

*Floor* denotes the integer part of *IATE(x)*.

The non-linear ('normal') IATE with $\delta = 2$ has a standard deviation of about 1.7 (*Corr(p(x), IATE)=0.73*), while the 'strong' ($\delta = 8$) IATE has a standard deviation of about 6.8 (*Corr(p(x), IATE)=0.85*). The standard deviation of the earnings related ('earnings') IATE is also about 1.7 (*Corr(p(x), IATE) = 0.24*). These numbers should be related to the standard deviation of the outcome of about 12.9. The IATEs are well predictable. For example, plain-vanilla Random Forests have out-of-sample $R^2$'s above 99% if IATEs are regressed on covariates. Naturally, the predictability of the ITE is smaller with out-of-sample $R^2$'s of about 76% ('normal'), 92% ('strong'), and 94% ('earnings') respectively.

These IATEs are also used to compute the ATE and the GATEs. The true ATE is taken as the average of the IATEs in the validation sample, while the true GATEs are taken as averages of the IATEs in the respective groups. There are two types of GATEs considered: The first type consists of two GATEs, one for men (56%) and one for women (44%). The second type of GATEs considers 32 yearly age categories (24-55). As younger individuals are more likely to become unemployed in Switzerland, the largest, i.e., youngest, age group has about three times as many observations as the smallest, i.e., oldest, age group.

In addition to varying the IATEs and the corresponding aggregate effects in four scenarios, we also vary the assignment process by considering random assignment of the treatment as in a randomized control trial (RCT) as well. Furthermore, we vary the sample size. Overall, this leads to 24 different DGPs in total (4 IATEs x 3 sample sizes x 2 selection processes). In addition to the estimators introduced in Section 4, we consider *Basic* in a version without the a priori sample splitting but with honest trees instead.[28] Finally, for the smaller sample, *N = 1,000*, estimation and inference is repeated on 1,000 independent random training samples (replications, *R*). Since computation time is a constraint and because the larger sample produces estimates with lower variability, only 250 replications are used for *N = 4,000*, and only 125 replications for *N=8,000*.

## 5.2 Results

### 5.2.1 General remarks

For each replication, we obtain 5,035 results in the validation sample, consisting of 5,000 estimates of different IATEs, 32 and 2 estimates of the GATEs, and 1 estimate of the ATE. For each of these parameters, we compute the usual measures for point estimators, like bias (col. (2) of the following table), standard deviation (col. (9)), mean squared error (MSE, col. (5)), skewness (col. 6), and kurtosis (col. 7). The Jarque-Bera statistic (JB, col. (8)) summarizes these third and fourth moments. These three measures are used to check whether the estimators are normally distributed (JB is $\chi^2(2)$ distributed when the estimators are normally distributed and shifts to the right when they are not; the 5% and 1%-critical values are 6 and 9.2, respectively).

---

[28] So that for each random subsample used for a particular tree the data is split randomly into two parts: Half the data is used to build the tree, and half the data is used to estimate the effects given the tree.

Finally, we compute the bias of the estimated standard errors (col. 10) and the coverage probability of the 90% confidence interval (col. 11).[29] For the MSE, we also compute measures of its variability across replications to assess the simulation error (in the footnote of the tables).

It is nonsensical to report these measures for all 5,035 parameters. While we report the measures for the ATE directly, we aggregate the measures for the IATEs and the GATEs by taking their average across groups (there are 2 respectively 32 groups for the GATE, and 5,000 groups for the IATEs). Note that due to the way the estimators are constructed, the bias of the ATE and the average biases of the IATEs are identical. Since this type of cancellation of biases for the IATEs is undesirable in a quality measure (as a negative bias is as undesirable as a positive one), we report their average absolute bias (col. 2). Finally, we report the standard deviation of all the true (col. 3) and estimated GATEs and IATEs (col. 4) to see how the estimators capture the cross-sectional variability of the true effect heterogeneity.

### 5.2.2 Discussion of detailed results

We begin the discussion with the main simulation results for the DGP with *N=8,000* and a 'normal' IATE ($\delta = 2$) presented in Table 1. Further simulation results are relegated to various appendices. Appendix C.1 contains the results for the two smaller sample sizes, while Appendix C.2 considers the other three specifications of the IATEs, and Appendix C.3 shows the results for the additional estimators discussed in Section 4.5. Finally, Appendix D contains the results for the experimental case.

Table 1 contains the results for the estimators *Basic*, *OneF.VarT*, and *OneF.MCE* and its locally centred version based on two folds, *OneF.MCE.LC-2*. Both MCE based estimators are also considered in their penalized form (*OneF.MCE.Pen, OneF.MCE.Pen.LC-2*). To see the impact of the (computationally more expensive) increase in folds for the locally centred estimator, we also consider a version based on 5-folds (*OneF.MCE.Pen.LC-5*).

---

[29] This is the share of replications for which the true value was included in the 90% confidence interval (computed with the point estimate and the variance estimate under a normality assumption). 90% instead of the more common 95% is used, because the number of replications may be too small (in particular for the larger sample) to estimate the more extreme tail probability precisely. Since all estimators are asymptotically normally distributed, the particular tail quantile should hardly affect the conclusions.

*Basic* is substantially biased, but it captures the cross-sectional variation of the heterogeneity well. Furthermore, *Basic* (as well as all other estimators considered) appears to be normally distributed. The estimated weights-based standard errors are somewhat too large. The main problem for inference is, however, the substantial bias leading to too small coverage probabilities for all effects, in particular for ATE and the GATEs.

Estimating only one forest instead of two as in *Basic* and using the splitting rule of WA18 (*OneF.VarT*) improves the MSE somewhat for IATEs (but increases them for ATE and GATEs), but biases get even larger. This leads to lower coverage probabilities than the already too low ones of *Basic*.

*Table 1: Simulation results for N=8,000, main DGP, and main estimators*

| | | | True & estimated effects | | | Estimation error of effects (averages) | | | | | Estimation of std. error | |
|---|---|---|---|---|---|---|---|---|---|---|---|---|
| | Groups | Est. | Avg. bias | X-sectional std. dev. | | MSE | Skewness | Kurtosis | JB-Stat. | Std. err. | Avg. bias | CovP (90) in % |
| | # | | | true | est. | | | | | | | |
| | (1) | | (2) | (3) | (4) | (5) | (6) | (7) | (8) | (9) | (10) | (11) |
| ATE | 1 | Basic | 1.25 | - | - | 1.75 | 0.0 | 2.9 | 0.1 | 0.43 | 0.04 | 15 |
| GATE | 2 | | 1.27 | - | - | 1.78 | -0.1 | 2.8 | 0.7 | 0.45 | 0.05 | 18 |
| GATE | 32 | | 1.21 | 0.17 | 0.14 | 1.81 | 0.0 | 2.8 | 1.6 | 0.56 | 0.08 | 38 |
| IATE | 5000 | | 1.38 | 1.72 | 1.42 | 4.89 | 0.0 | 2.9 | 1.9 | 1.45 | 0.22 | 78 |
| ATE | 1 | OneF. | 1.71 | - | - | 3.12 | 0.3 | 3.1 | 1.5 | 0.51 | 0.06 | 3 |
| GATE | 2 | VarT | 1.71 | - | - | 3.14 | 0.2 | 3.1 | 2.2 | 0.54 | 0.07 | 4 |
| GATE | 32 | | 1.68 | 0.17 | 0.11 | 3.07 | 0.2 | 3.0 | 1.0 | 0.58 | 0.10 | 7 |
| IATE | 5000 | | 1.73 | 1.72 | 1.33 | 4.64 | 0.0 | 3.0 | 2.4 | 1.50 | 0.50 | 71 |
| ATE | 1 | OneF. | 1.29 | - | - | 1.86 | -0.1 | 3.3 | 0.4 | 0.46 | 0.04 | 14 |
| GATE | 2 | MCE | 1.28 | - | - | 1.86 | -0.1 | 3.2 | 0.4 | 0.46 | 0.05 | 17 |
| GATE | 32 | | 1.27 | 0.17 | 0.08 | 1.87 | 0.0 | 3.1 | 1.1 | 0.50 | 0.14 | 34 |
| IATE | 5000 | | 1.34 | 1.72 | 0.88 | 3.92 | 0.0 | 2.9 | 2.0 | 1.00 | 0.46 | 75 |
| ATE | 1 | OneF. | 0.90 | - | - | 1.06 | 0.1 | 2.3 | 2.5 | 0.50 | 0.02 | 44 |
| GATE | 2 | MCE. | 0.90 | - | - | 1.06 | 0.1 | 2.5 | 1.9 | 0.51 | 0.02 | 48 |
| GATE | 32 | LC-2 | 0.88 | 0.17 | 0.06 | 1.10 | 0.1 | 2.5 | 1.9 | 0.56 | 0.03 | 55 |
| IATE | 5000 | | 1.12 | 1.72 | 0.70 | 3.36 | 0.0 | 3.0 | 2.3 | 1.09 | 0.08 | 70 |
| ATE | 1 | OneF. | 0.21 | - | - | 0.27 | 0.1 | 3.3 | 0.8 | 0.49 | 0.28 | 97 |
| GATE | 2 | MCE. | 0.20 | - | - | 0.28 | 0.1 | 3.2 | 0.7 | 0.49 | 0.29 | 98 |
| GATE | 32 | Pen | 0.20 | 0.17 | 0.16 | 0.36 | 0.1 | 3.2 | 1.6 | 0.56 | 0.31 | 97 |
| IATE | 5000 | | 0.32 | 1.72 | 1.72 | 2.29 | 0.1 | 3.0 | 2.7 | 1.44 | 0.78 | 98 |
| ATE | 1 | OneF. | 0.23 | - | - | 0.49 | -0.2 | 3.3 | 1.5 | 0.66 | 0.01 | 90 |
| GATE | 2 | MCE. | 0.24 | - | - | 0.50 | -0.2 | 3.3 | 1.5 | 0.67 | 0.01 | 90 |
| GATE | 32 | Pen | 0.23 | 0.16 | 0.15 | 0.53 | -0.2 | 3.2 | 1.7 | 0.69 | 0.02 | 89 |
| IATE | 5000 | LC-2 | 0.39 | 1.72 | 1.65 | 2.39 | 0.0 | 3.0 | 1.5 | 1.45 | 0.10 | 90 |
| ATE | 1 | OneF. | 0.21 | - | - | 0.35 | 0.3 | 3.1 | 1.5 | 0.55 | 0.02 | 86 |
| GATE | 2 | MCE. | 0.21 | - | - | 0.35 | 0.3 | 3.0 | 1.5 | 0.56 | 0.03 | 88 |
| GATE | 32 | Pen | 0.21 | 0.16 | 0.15 | 0.37 | 0.2 | 3.1 | 1.7 | 0.57 | 0.04 | 88 |
| IATE | 5000 | LC-5 | 0.37 | 1.72 | 1.57 | 1.83 | 0.1 | 2.9 | 2.0 | 1.27 | 0.07 | 90 |

Note: For GATE and IATE the *average bias* is the absolute value of the bias for the specific group (GATE) / observation (IATE) averaged over all groups / observation (each group / observation receives the same weight). *CovP (90%)* denotes the (average) probability that the true value is part of the 90% confidence interval. The simulation errors of the mean MSEs are around 0.08.

Using the MCE splitting rule proposed in this paper (*OneF.MCE*) leads to substantially lower MSEs than for *Basic* (for IATEs) and *OneF.VarT* (for all parameters). Part of this gain in MSEs comes from a reduction of biases (compared to *OneF.VarT*) and another part comes from a reduction of the variances of the IATEs. However, due to the still substantial bias, coverage probabilities are still too small to be useful for inference. Adding local centering to the MCE criterion (*OneF.MCE.LC-2)* reduces the bias for all parameters and improves the MSE. However, despite a rather accurate estimation of the standard errors, coverage probabilities are still too low, because the bias is still relevant.

Adding the penalty terms to the splitting rule drastically reduces the remaining biases for the MCE based estimators (same for *OneF.VarT*, see Appendix C.3) and leads to substantially smaller MSEs. The bias reductions come at the cost of some additional variance. Interestingly, once the penalty is used, local centering does not affect the bias much, but improves the estimation of the weights-based standard errors. It is a general feature of all simulations that for estimators not using local centering estimated standard errors particularly for the IATEs are always too large (leading to conservative inference). However, estimated standard errors for the locally centred estimators are fairly accurate. The comparison of the estimators using two-fold and five-fold local centering shows that the additional variance reduction coming from using more than two folds can be relevant. This leads to the conclusion that across all parameters, and considering point estimates and inference jointly, either *OneF.MCE.Pen* (if having conservative inference is no problem) or *OneF.MCE.Pen.LC-5* are the preferred estimators.

A similar conclusion also holds for the case of no effect heterogeneity (Table C.5) as well as for the case of earnings dependent heterogeneity (Table C.11). However, the gain of local centering for estimating the IATEs is less clear (in fact, the IATEs are best estimated by *OneF.MCE*; for ATE and the GATEs, the penalty function is important, though). The most challenging case for estimation is the one with strong effects strongly linked to the propensity score ($\alpha = 8$, Table C.8). In this case, the clear winner of all estimators is the uncentred, penalised, MCE based estimator (*OneF.MCE.Pen*). Thus, if one cannot rule out strong heterogeneity strongly linked to selection, *OneF.MCE.Pen* is the estimator of choice because it is the only one that performs well in all scenarios. The cost of this robustness is that for IATE estimated standard errors are too large leading to conservative inference. Otherwise, as mentioned before, its locally centred version may perform substantially better with respect to inference.

Considering the results for the smaller samples based on the same DGP as in Table 1 (Tables C.1 and C.2), we arrive at very similar conclusion favouring *One.MCE.Pen.LC-5*. Using the three sample sizes of 1,000, 4,000, and 8,000 allows to consider approximate convergence rates of the estimators (of course, based on only 3 data points). For example, for the ATE and *One.MCE.Pen.LC-5* the biases (standard errors) fall with sample size as 0.79 – 0.39 – 0.21 (1.34 – 0.73 – 0.55) which seems to be roughly in line with $\sqrt{N}$-convergence. The absolute biases (standard errors) of the IATE fall slower at 0.79 – 0.53 – 0.37 (1.96 – 1.45 – 1.27). A similar picture appears for the other specification of the IATEs. The results for the strong-effect-highly-correlated-with-selection case (Tables C.6 and C.7) point to the need for a large enough sample in such case. For *N = 1,000* none of the estimators performs well, while for *N=4,000* their performances improve, but even the best estimator for this DGP (*OneF.MCE.Pen*) has a bias large enough to lead to too small coverage probabilities. This problem disappears for *N=8,000* (Table C.8).

Considering now further estimators (for details see Appendix C.3), it turns out that *OneF* performs well in the case when there is no effect. This is expected, because in this case the dependence of the conditional expectation of the outcome on covariates is the same among treated and controls. In all other cases, one of the other estimators always dominates it. Adding a penalty term to *OneF.VarT* leading to *OneF.VarT.Pen* also reduces its bias, but this is not enough to become competitive with estimators performing well in these scenarios. Finally, comparing *Basic* in its one-sample (*Basic.OneSam*) and two-sample versions shows that the one-sample-with-honesty version might have a slightly lower MSE (if at all). However, its bias tends to be somewhat larger and, as expected, the estimated standard errors are too small. This indicates that the costs of sample splitting needed for inference seem to be low and unavoidable. However, this does not rule out the fact that estimators that are 'optimized' for either the ATE, GATE, or IATE may use the sample more efficiently.

The simulations with random assignment (see Appendix D) show the importance of the selection process for the differential performance of the estimators. The first finding for the experimental case is that the other estimators considered dominate *Basic* (except for the strong-effect DGP and the small sample). The second finding is in fact a non-finding in the sense that it is very difficult to rank the other single-forest estimators among themselves as they perform very similarly. This is as expected, as the splitting rule of WA18 seems to be well justified if there is no selectivity. Another positive aspect of this finding is that the penalty term does not inflict much harm when used in cases in which it is redundant (as there is no selection bias in these DGPs). Finally, since the biases of all estimators are small in the experimental case, coverage probabilities substantially improve.

The conclusion of the simulation exercise is the following: If there is some a priori knowledge that the propensity score and the effect heterogeneity are not too important and tied too closely together, then *OneF.MCE.Pen.LC-5* is the preferred estimator. It has low MSE and good coverage. However, *OneF.MCE.Pen* is more robust and works in this difficult case in medium sized samples. However, this robustness comes at the price of the inference being too conservative. If effects are very precisely estimated, because of their large size or of large samples, such costs may not matter much and may be outweighed by the additional robustness.

# 6  Application to the evaluation of a job search programme

This section shows how the estimator that came out best in the Empirical Monte Carlo study across all scenarios, i.e., *OneF.MCE.Penalty*, can be productively applied in empirical studies. To this extent, we rely on the data set that formed the basis of the Empirical Monte Carlo study above, but we use a more homogeneous subsample of men living in cantons where German is the dominant language (about 38,000 observations). As before, we investigate the effect of participating in a job search programme (share of participants is 8%) on months of employment accumulated over 9 months, as well as over 3 years, respectively. Based on a previous reanalysis of the effects of this programme in Knaus, Lechner, and Strittmatter (2022), we expect to find essentially zero effects with very limited heterogeneity over three years, but substantial heterogeneity over the first 9-month period, the so-called lock-in period. On top of this, we expect selective programme participation (see below). Thus, this is a challenging setting.[30]

The estimation with *OneF.MCE.Penalty* is based on 1,000 subsampling replications (50% subsampling share). The minimum leaf size $N^{min} = (13, 50)$ and the number of coefficients used for leaf splitting $M = (4, 10, 27)$ (out of 31 ordered and 8 unordered variables;[31] 6 variables have been deleted a priori as their unconditional correlation with both the outcome and programme participation is below 1%) are treated as tuning parameters. $N^{min} = 13$ and $M = 27$ are chosen by out-of-bag minimisation of the optimization criterion of this estimator.

The matching literature has shown that it may be important to ensure that common support holds post-estimation (e.g., Imbens, 2004). This issue has not yet been discussed in the context of ML methods of the type proposed here.[32] The key issue is that for every combination of values of *X* that is relevant for estimation, there should be enough treated and non-treated observations for reliable estimation of the effects. While this has been shown to be relevant for the (conventional) estimation of *ATE* and related aggregated parameters (e.g., Lechner and Strittmatter, 2019), it is even more important when estimating *IATE(x)*, which will be particularly sensitive to support violations close to their evaluation points. Here, we predict the propensity scores using the trees of the already estimated (outcome) forest in a first step. In a second step, all values of *X* related to very few predicted treated or controls are deleted. Here, we deleted all values of *X* with propensity scores outside the [5%, 95%]-interval. This led to discarding about 0.5% (9 months) and 1.1% (3 years) of the observations. In the discarded group, almost 80% of the observations have a native language that is different from any of the main Swiss official languages (German, French, and Italian). Clearly, this procedure can be improved which is however left to future work.

---

[30] This section is merely a demonstration of possible findings. Therefore, for the sake of brevity, it is written very densely. For more details on programmes, institutional details, and data, the reader is referred to the previous papers using this setting, e.g., Huber, Lechner, and Mellace (2017), Knaus, Lechner, and Strittmatter (2021, 2022) and the references therein.

[31] *k-means-clustering* is used to deal with unordered variables (as proposed in Chou, 1991, and Hastie, Tibshirani, and Friedman, 2009).

[32] For a general discussion on the implications of the overlap assumption in a high-dimensional covariate space, see D'Amour, Ding, Feller, Lei, and Sekhon (2021).

From the point of view of eliminating selection bias, any good estimator in this setting should ensure covariate balance. Table 2 therefore shows the means and standardized differences for selected covariates (mostly directly related to the pre-specified heterogeneity shown in Table 3 below) prior to the estimation. To see how the different forests related to the two respective outcome variables change the balancing, the same covariates are predicted using the estimated forests (and their implied weights; columns headed by 'post-estimation'). In both cases, the balancing improves considerably.

*Table 2: Pre- and post-estimation balancing of covariates*

|  | Pre-estimation | | | | Post-estimation | |
|---|---|---|---|---|---|---|
| **Features** | Mean $d=1$ | Mean $d=0$ | Difference | Stand. diff. (%) | Difference | |
| **Outcome: Months of employment in** | | | | | 9 months | 3 years |
| # of pre UE employment episodes (last 2 years) | 0.09 | 0.12 | -0.027 | 20 | -0.019 | -0.016 |
| Foreign native language | 0.24 | 0.32 | -0.08 | 19 | -0.05 | -0.05 |
| Employability (1, 2, 3) | 2.01 | 1.93 | 0.09 | 18 | 0.02 | 0.05 |
| Pre-UE monthly earnings (in CHF) | 5435 | 4899 | 537 | 26 | 50 | 26 |

Note: The standardized difference (stand. diff.) is defined as $Stand.diff = \dfrac{|\bar{x}^1 - \bar{x}^0|}{\sqrt{[Var(x^1)+Var(x^0)]/2}}$, where the superscripts 1 and 0 denote the subpopulations of treated and controls. Post-estimation results are obtained by treating the variables as outcomes and running them through the estimated forests. Since these forests differ for the 9 and the 36 months outcomes, the post-estimation results differ with respect to the outcome variable used to train the forests.

Tables 3 and 4 contain the results of the estimation for the ATE (upper panel), the various GATEs that appeared to be of a priori interest (in the middle of the table), as well as summary statistics for the IATEs (lower panel). The GATEs relate to effect heterogeneity in the number of unemployment episodes in the 2 years prior to the unemployment spell analysed (12 categories), the native language being a 'Swiss' one or a different one (2 cat.), the employability index (3 cat.), and the economic sector of the last occupation (16 cat.). As mentioned before, all effects have the interpretation of additional months of regular employment due to programme participation.

*Table 3: Average, group effects, and individualized effects for outcome variable* months of employment accumulated over 9 months

| Average potential outcomes | | | | Average effects | | |
|---|---|---|---|---|---|---|
| Treatment | | No treatment | | | | |
| $E(Y^1|X=x)$ | Std. error | $E(y^0|X=x)$ | Std. error | ATE | Std. error | p-value |
| 1.59 | 0.06 | 2.09 | 0.02 | -0.52 | 0.06 | 0.0 % |
| **Group average treatment effects (GATEs)** | | | | | | |
| Group variable *(Z)* | | GATE | Std. err. | p-value | Group variable *(Z)* | GATE | Std. err. |
| **Number of** | 0 | -0.52 | 0.06 | 93 % | Sector 1 | -0.56 | 0.07 |
| **employment** | 1 | -0.51 | 0.06 | 37 % | Sector 2 | -0.64 | 0.07 |
| **spells prior** | 2 | -0.50 | 0.06 | 93 % | Sector 3 | -0.51 | 0.06 |
| **to training** | 3 | -0.50 | 0.07 | 43 % | Sector 4 | -0.41 | 0.09 |
| | 4 | -0.50 | 0.07 | 55 % | Sector 5 | -0.44 | 0.07 |
| | 5 | -0.49 | 0.07 | 87 % | Sector 6 | -0.45 | 0.07 |
| | 6 | -0.49 | 0.08 | 39 % | Sector 7 | -0.50 | 0.07 |
| | 7 | -0.51 | 0.07 | 30 % | Sector 8 | -0.50 | 0.06 |
| | 8 | -0.48 | 0.08 | 59 % | Sector 9 | -0.53 | 0.06 |
| | 9 | -0.51 | 0.08 | 76 % | Sector 10 | -0.50 | 0.06 |
| | 10 | 0.53 | 0.10 | - | Sector 11 | -0.64 | 0.09 |
| **# of employment spells** joint test (p-value) | | | | 94 % | Sector 12 | -0.49 | 0.07 |
| **Swiss language** native | | -0.51 | 0.06 | 68 | Sector 13 | -0.54 | 0.07 |
| not native | | -0.53 | 0.06 | - | Sector 14 | -0.45 | 0.07 |
| **Swiss** joint test | | | | 68 | Sector 15 | -0.46 | 0.07 |
| **Employability** good | | -0.49 | 0.06 | 11 | Sector 16 | -0.47 | 0.07 |
| intermediate | | -0.52 | 0.06 | 52 | **Sector** joint test (p-value) | | 0.1 % |
| bad | | -0.54 | 0.07 | - | | | |
| **Employability** joint test (p-value) | | | | 17 % | | | |
| **Individualized average treatment effects (IATEs)** | | | | | | |
| Mean of IATE | Standard deviation of IATE | Share < 0 | Share > 0 | Average of estimated standard errors of IATEs | Share of IATEs significant at 5% level |
| -0.52 | 0.15 | 100 % | 0 % | 0.15 | 92 % |

Note: Treatment is participating in a job search programme. The p-values relate to hypothesis tests based on the (asymptotic) *t*- or Wald-type. For the ATE, it is the *t*-test that the effect is zero. For the GATEs, it is the *t*-test that the adjacent values are identical (therefore it is not given for the sectors that are not in any natural order) as well as the Wald test that all GATEs relating to these variables are identical. Therefore, this test has (# of groups-1) degrees of freedom.

For the average treatment effect, the first four columns show the mean of the potential outcomes as well as their estimated standard errors. The remaining columns show the average effects, their estimation error, and the p-value for the hypothesis that the average effects equal zero. The results imply that over nine months (3 years), these individuals work 1.6 (16.5) month if they participate in the programme and 2.1 (16.8) months if not. This negative short-run effect is statistically significant at conventional levels (despite inference likely being conservative), while the smaller 3-year effect is not.

The middle part of the tables contains the group-mean effects and their standard errors for the four different GATEs relating to the variables discussed above. The p-values shown, however, do not relate to the hypothesis that their effects are zero but to the hypothesis that the differences of the effects minus the ones in the next row are zero. Such a test contains relevant information only if the heterogeneity variable is ordered. Since this is not true for the economic sectors, the p-values are not shown for those GATEs. Finally, for each GATE the last row shows the p-value of a Wald-test (again, based on the weighted representation of the estimator, which allows uncovering the full variance-covariance matrix of the estimated GATEs by computing new weights for testing based on differences of the weights used for estimation) with degrees of freedom equal to the number of different categories for the particular GATE for the hypothesis that all GATEs relating to this variable are equal.

No significant heterogeneity appears for the 3-year outcome, neither in a statistical nor in a substantive sense. However, for the 9-month employment outcome there appears to be heterogeneity with respect to the economic sector. Here, the Wald test clearly rejects homogeneity and the various negative effects differ substantially.[33]

While the estimated GATEs are presented in total since they are not based on too many different groups, this is impossible for the estimated IATEs. Thus, for a better description we may want to condense their information further. The first part of this exercise is reported in the lower panels of Table 3 and 4. The first two columns of these tables show their average and their standard deviation. Next, there is the share of IATEs with positive and negative values. Finally, we report the average of their estimated standard errors and the share of IATEs that are significantly different from zero (at the 5% level). For the 9-month outcome, all effects are negative and more than 90% significantly so (in a statistical sense). Their standard deviation across individuals is about 4 days. For the 3-year outcome, their standard deviation is larger, but this is necessarily so because the potential outcomes are about 10 x larger. About two-thirds of the IATEs are smaller than zero, and one third are larger. However, only 2% of them are statistically significantly different from zero.

---

[33] Given the space constraints of this paper, we refrain from analysing and interpreting this heterogeneity further.

Table 4: *Average, group effects, and individualized effect for outcome variable* months of employment accumulated over 36 months

| Average potential outcomes | | | | Average effects | | |
|---|---|---|---|---|---|---|
| Treatment | | No treatment | | | | |
| $E(Y^1|X=x)$ | Std. err. | $E(Y^0|X=x)$ | Std. err. | ATE | Std. err. | p-value |
| 16.50 | 0.33 | 16.79 | 0.099 | -0.30 | 0.34 | 39 % |
| **Group average treatment effects (GATEs)** | | | | | | |
| Group variable (Z) | | GATE | Std. err. | p-value | Group variable (Z) | GATE | Std. err. |
| Number of | 0 | -0.55 | 0.37 | 12 % | Sector 1 | 0.19 | 0.47 |
| employment | 1 | -0.19 | 0.36 | 12 % | Sector 2 | -0.62 | 0.42 |
| spells prior | 2 | 0.04 | 0.40 | 16 % | Sector 3 | -0.30 | 0.37 |
| to training | 3 | 0.16 | 0.44 | 43 % | Sector 4 | 0.77 | 0.67 |
| | 4 | 0.22 | 0.50 | 9 % | Sector 5 | 0.04 | 0.42 |
| | 5 | 0.34 | 0.55 | 96 % | Sector 6 | -0.13 | 0.40 |
| | 6 | 0.34 | 0.50 | 37 % | Sector 7 | -0.27 | 0.46 |
| | 7 | 0.15 | 0.50 | 65 % | Sector 8 | -0.25 | 0.39 |
| | 8 | 0.27 | 0.55 | 92 % | Sector 9 | -0.44 | 0.44 |
| | 9 | 0.24 | 0.60 | 37 % | Sector 10 | -0.32 | 0.37 |
| | 10 | -0.16 | 0.60 | - | Sector 11 | -0.45 | 0.58 |
| # of employment spells joint test (p-value) | | | | 94 % | Sector 12 | -0.49 | 0.07 |
| Swiss language native | | -0.22 | 0.40 | 66% | Sector 13 | -0.42 | 0.37 |
| not native | | -0.33 | 0.36 | - | Sector 14 | -0.17 | 0.44 |
| Swiss joint test | | | | 66% | Sector 15 | -0.46 | 0.07 |
| Employability good | | -0.09 | 0.38 | 81 % | Sector 16 | -0.19 | 0.40 |
| intermediate | | -0.33 | 0.35 | 35 % | Sector joint test (p-value) | | 42 % |
| bad | | -0.44 | 0.38 | 25 % | | | |
| Employability joint test (p-value) | | | | 34 % | | | |
| **Individualized average treatment effects (IATEs)** | | | | | | |
| Mean of IATE | Standard deviation of IATE | Share < 0 | Share > 0 | Average of estimated standard errors of IATEs | | Share of IATEs significant at 5% level |
| -0.30 | 0.86 | 64 % | 36 % | 1.10 | | 0.02 % |

Note: Treatment is participating in a job search programme. The p-values relate to hypothesis tests based on the (asymptotic) t- or Wald-type. For the ATE, it is the *t*-test that the effect is zero. For the GATEs, it is the *t*-test that the adjacent values are identical (therefore it is not given for the sectors that are not in any natural order) as well as the Wald test that all GATEs relating to these variables are identical. Therefore, this test has (# of groups-1) degrees of freedom.

The final step consists in a post-estimation analysis of the IATEs that *describes* the correlation of the estimated IATEs with exogenous variables. In principle, all standard descriptive tools could be used for this exercise. In a multivariate analysis, there is the difficulty that the covariate space is very large. Thus, we use a dimension reduction tool, here the LASSO, and perform OLS estimation with the covariates that had a non-zero coefficient in the LASSO estimation (this is also called Post-LASSO; for theoretical properties of the Post-LASSO and its advantages over LASSO see, e.g., Belloni, Chernozhukov and Wei, 2016).[34] Since model selection (LASSO) and coefficient estimation are performed on two random, non-overlapping subsamples (50% each), the OLS standard errors remain valid given the selected model.

*Table 5: Selected coefficients of post-lasso OLS estimation of IATE*

| Variable | Coefficient | Standard error | t-value | p-value | Uncond. correlation (if > 10%) |
|---|---|---|---|---|---|
| **Outcome: Months of employment in first 6 months** | | | | | |
| \|Age case worker – age UE\| in years | -0.002 | 0.001 | -2.0 | 4.7 % | -24 % |
| \|Age case worker – age UE\| < 5 | not | selected | - | - | - |
| Age | 0.45 | 0.22 | 2.0 | 4.6 % | 27 % |
| # E spells | 0.10 | 0.05 | 2.0 | 4.6 % | - |
| Time in employment before (share) | -0.10 | 0.05 | -2.0 | 4.5 % | -16 % |
| Earnings before | -0.04 | 0.02 | -1.9 | 5.2 % | - |
| Employability | -0.004 | 0.002 | 1.7 | 9.5 % | - |
| **Outcome: Months of employment in first 3 years** | | | | | |
| \|Age case worker – age UE\| in years | not | selected | - | - | 11 % |
| \|Age case worker – age UE\| < 5 | -0.12 | 0.03 | -4.1 | < 1 % | - |
| Age | -0.01 | 0.001 | -7.8 | < 1 % | -20 % |
| # E spells | 0.12 | 0.01 | 14 | <1 % | 26 % |
| Time in employment before (share) | -1.20 | 0.06 | -21 | <1 % | -32 % |
| Earnings before | -0.008 | 0.0006 | -14 | <1 % | - |
| Employability | -0.17 | 0.03 | -7.3 | <1 % | - |

Note: OLS regression. Dependent variable is the estimated IATE. Independent variables are those with non-zero coefficients in the LASSO estimation. OLS and LASSO were estimated on randomly splitted subsamples (to facilitate the use of ordinary, robust OLS standard errors). The penalty term of the LASSO estimation was chosen by minimization of the 10-fold cross-validated mean square prediction error and the one-standard error rule. $R^2$ = 71% (9 months), 20% (3 years). *N*=8990. A constant term and further variables capturing caseworker, local and individual characteristics are included as well.

Table 5 shows a selection of variables only, to serve as examples. Here, heterogeneity appears for both outcome variables, but related to variables not specified in the GATEs, like the age difference of the caseworker and the unemployed, age, number of employment spells, pre-unemployment time in employment, and earnings, as well as the employability according to the judgement of the caseworker. While these are highly significant predictors for the 3-year outcome, their 9-month effects are significant at approximately the 5% level.

---
[34] This approach is very similar in spirit to Zhao, Small, and Ertefaie (2022).

In summary, this section gave a first indication about the usefulness of the new methods in applications and provides some suggestions on how to deal with some practical issues (i.e. common support, balancing of covariates, describing the IATEs) that arise when applying such ML tools in a causal framework.[35]

# 7 Conclusion

In this paper, we develop new estimators to estimate aggregate as well as heterogeneous treatment effects for multiple treatments in a selection-on-observed-variables setting and provide theoretical guarantees. We compare them to existing estimation approaches in an empirically informed simulation study, and apply the best performing estimator to an empirical programme evaluation study. They are an extension of the Causal Forest approach proposed by Wager and Athey (2018).

The new estimators deviate from the original Causal Forest in that a new splitting criterion is proposed to build the trees that form the forest. The new splitting criterion has two components: First, we approximate the mean square error of the causal effects by combining the causal problem with the identification strategy directly. Second, a penalty term is added that favours splits that are heterogeneous with respect to the implied treatment propensity to reduce selection bias directly. It turned out that both changes worked as intended and that they can improve the performance of the Causal Forest approach in observational studies substantially.

---

[35] An application of this estimation approach to the evaluation of parts of the Flemish active labour market policy with many additional results is provided by Cockx, Lechner, and Bollens (2019).

An additional advantage of the Causal Forest approach is that the common estimators have a representation as weighted means of the outcome variables. This makes them particularly amendable to using a common inference framework for estimation effects at various aggregation levels, like for the average treatment effect (ATE), the group treatment effect (GATE), or the individualized treatment effect (IATE). The advantage is that ATE and GATEs (and their inference) can be obtained directly from aggregating the estimated IATEs without the need for different additional ML estimations at the various aggregation levels of interest.

These changes together with the good performance of the proposed methods in the simulation study and the application make this new estimator (even more) attractive to use in empirical studies. However, several issues remain unresolved and deserve further inquiry in the future: One open issue is how to choose the value of the penalty term in an optimal way at low computational cost. Another topic is to explore alternative methods to estimate the mean correlated errors, which is the main innovation in the new splitting rule. Three other issues concern the proposed inference methods: First, it would be desirable to have a more solid theoretical justification why the inference methods work well, e.g., by deriving explicitly the regularity conditions that the weights must fulfil to lead to a consistent standard error estimator. Second, this paper explored just one way to implement this weights-based inference. It seems worthwhile to investigate alternatives, also with the goal of tackling the issue that the current (approximate) standard errors for the IATEs seem to be too large without local centering. Third, it would be useful to understand better the relation of the other tuning parameters that come with these types of Random or Causal Forests (e.g., number of coefficients to be randomly chosen, minimum leaf size, subsampling size) to the quality of inference (and point estimation).

A continuously updated and extended Python module (*mcf*) is available on PyPI (https://pypi.org/project/mcf/), R code is available on request from Jana.Mareckova@unisg.ch,

and (legacy) Gauss code can be downloaded from www.michael-lechner.eu/statistical-software.

# Appendix A *(Online)*: More details on estimation and inference

Here, we present detailed proofs and some more details of the Causal Forest algorithms used in the EMCS.

## Appendix A.1: Theoretical Proofs

The following notation as in Wager and Athey (2018, WA18) is used for the asymptotic scaling: $f(s) \gtrsim g(s)$ means that $\liminf_{s \to \infty} f(s)/g(s) \geq 1$, and $f(s) \lesssim g(s)$ means that $\liminf_{s \to \infty} f(s)/g(s) \leq 1$. Further, $f(s) = \Omega(g(s))$ means that $\liminf_{s \to \infty} |f(s)|/g(s) > 0$, i.e., that $|f(s)|$ is bounded below by $g(s)$ asymptotically.

*Appendix A.1.1: Proofs for Theorem 1*

In this section, we collect all necessary results for the bias bound. In the first step, we show that the volume of the tree shrinks with larger subsample size $s_1$. We focus on the volume instead of the diameter in comparison to WA18 as the volume is important to determine the expected value of honest samples in the leaf as the subsampling rates in the training and honest set may differ.

**LEMMA 1 (based on Lemma 1 in Wager and Athey (2018))** Let $S(x)$ be a final leaf containing the point $x$ in a regular, random-split tree according to the definitions above and let $\lambda(S(x))$ be its Lebesgue measure. Suppose that $X_1, ..., X_{s_1} \sim U\left([0,1]^p\right)$ independently. Then for $\alpha \leq 0.5$, the expected value of the Lebesgue measure of the final leaf has the following bounds

$$E[\lambda(S(x))] = O\left(s_1^{-\log(1-\alpha)/\log(\alpha)}\right),$$
$$E[\lambda(S(x))] = \Omega\left(s_1^{-1}\right).$$

**Proof:** Let $c(x)$ denote number of splits leading to the leaf $S(x)$ and $s_{1,d}$ be the number of observations treated with treatment $d$ in the training subsample. By Wager and Walther (2015), in particular using their Lemma 12, Lemma 13 and Corollary 14, with high probability and simultaneously for all but last $O(\log(\log s_1))$ parent nodes above $S(x)$, the number of training examples in the node divided by $s_1$ is within a factor $1+o(1)$ of the Lebesgue measure of the node. Therefore, for large enough $s_1$ with probability greater than $1-1/s_1$ it holds that

$$\lambda(S(x)) \leq (1-\alpha+o(1))^{c(x)}.$$

To further evaluate the upper bound for $\lambda(S(x))$, the smallest number of splits that could lead to a leaf $S(x)$ needs to be determined. Let $s_{\min} = \min_d s_{1,d}$ denote the smallest treatment in the training subsample. By regularity, the following holds $s_{\min}\alpha^{c(x)} \leq 2k-1$. Since $s_{\min} \gtrsim s_1\varepsilon$, then $c(x) \geq \log((2k-1)/(s_1\varepsilon))/\log(\alpha)$ for large $s_1$ and $\lambda(S(x))$ is bounded by

$$\lambda(S(x)) = O\left((1-\alpha)^{\log((2k-1)/(s_1\varepsilon))/\log(\alpha)}\right) = O\left(s_1^{-\log(1-\alpha)/\log(\alpha)}\right).$$

This also translates into the upper bound for $E[\lambda(S(x))]$.

Let $|S(x)|$ and $|S(x,d)|$ denote the number of observations in leaf $S(x)$ and the number of observations treated with $d$ in leaf $S(x)$, respectively. Regarding the lower bound, by regularity $|S(x,d)| \geq k$ for all treatments. It can be shown that the probability of $|S(x)| \geq kM$ when $\lambda(S(x)) \leq kM/s_1$ decays at least at a rate of $1/s_1$ for uniformly distributed covariates. Therefore, $kM/s_1 \leq \lambda(S(x))$ with probability $1-O(1/s_1)$ yielding $E[\lambda(S(x))] = \Omega(s_1^{-1})$. To guarantee that the upper bound is above the lower bound, $\alpha \leq 1/2$. ∎

**THEOREM 1** Under the conditions of Lemma 1, suppose moreover that trees $T$ are honest and $E\left[Y^d \mid X = x\right]$ are Lipschitz continuous. Then, the bias of the Causal Forest at a given value of $x$ is bounded by

$$\left|E\left[\widehat{IATE}(m,l;x)\right] - IATE(m,l;x)\right| = O\left(s_1^{-\log(1-\alpha)/p\log(\alpha)}\right).$$

**Proof:** As forest is an average of trees, the rate of the bias of a tree is also the rate of the bias of the forest. Let $T_b(d;x)$ denote a tree estimate of a potential outcome of treatment $d$ at point $x$ of a form

$$T_b(d;x) = \sum_{j=1}^{s_2} \hat{W}_{bj,N,b}(D_{bj}, X_{bj}, d, x) Y_{bj},$$

where we use the following notation for the weights:

$$\hat{W}_{bj,N,b}(D_{bj}, X_{bj}, d, x) = \begin{cases} |S_b(d,x)|^{-1} & \text{if } X_{bj} \in S_b(x) \text{ and } D_{bj} = d \\ 0 & \text{else} \end{cases}.$$

The tree estimator of the treatment effect is $T(m,l;x) = T(m;x) - T(l;x)$. As the treatment effect estimator is a difference of two potential outcome estimators, the absolute bias can be bounded by the rate of a bias of the potential outcome estimator as

$$\left|E[T(m,l;x)] - IATE(m,l;x)\right| \leq \sum_{d \in \{m,l\}} \left|E[T(d;x)] - E\left[Y^d \mid X = x\right]\right|.$$

Therefore, in the following the bias of the tree estimator for the potential outcome will be analysed.

The next observation is that if $E\left[Y^d \mid X = x\right]$ is Lipschitz, then $E\left[Y \mid X = x, D = d\right]$ is also Lipschitz as the two expectations coincide under the CIA, CS and observation rule. By using Jensen's inequality and Lipschitz continuity, the absolute bias can be bounded by

$$\left|E[T_b(d;x)] - E[Y^d|X=x]\right| = \left|E\left[\frac{1}{|S(d,x)|}\sum_{i\in S(d,x)} Y_i\right] - E[Y^d|X=x]\right|$$

$$= \left|E\left[E\left[\frac{1}{|S(d,x)|}\sum_{i\in S(d,x)} Y_i \middle| S(d,x)\right]\right] - E[Y^d|X=x]\right|$$

$$= \left|E\left[\frac{1}{|S(d,x)|}\sum_{i\in S(d,x)} E[Y_i|S(d,x)]\right] - E[Y^d|X=x]\right|$$

$$\leq E\left[\frac{1}{|S(d,x)|}\sum_{i\in S(d,x)} \left|E[Y_i|S(d,x)] - E[Y^d|X=x]\right|\right]$$

$$= E\left[\frac{1}{|S(d,x)|}\sum_{i\in S(d,x)} \left|E[Y_i|S(d,x)] - E[Y_i|X_i=x, D_i=d]\right|\right]$$

$$\leq E\left[\frac{1}{|S(d,x)|}\sum_{i\in S(d,x)} C_d \|X_i - x\|\right].$$

As the $\alpha$ regularity would yield a very loose bound on the expected distance, we take a different approach here based on the nearest neighbours (NN) of the point $x$ that also contain all the observations in the final leaf. Each final leaf can be bounded by a ball with the centre at $x$ and radius equal to the longest segment of the leaf which we denote as $diam(S(x))$. Then the number of treated observations in the ball, denoted as $|B(S(x),d)|$, follows a binomial distribution for $s_2$ observations and the success probability being the Lebesgue measure of the ball that can be seen as a function of the $diam(S(x))$ since the features are independent and uniformly distributed. At the same time the number of observations in the final leaf $|S(x,d)|$ follows also a binomial distribution for $s_2$ observations and the success probability being the Lebesgue measure of the final leaf that can be seen as a function of the $diam(S(x))$ and the angles to the vertices from the origin point using a high-dimensional polar system. As both random variables depend on the $diam(S(x))$, we can conclude that $|B(S(x),d)| \leq O(|S(x,d)|)$ with a constant larger than 1. As the ball contains the final leaf, the following inequality holds

$$\sum_{i\in S(d,x)} \|X_i - x\| \leq \sum_{i\in B(S(x),d)} \|X_i - x\|.$$

When we randomly split all data points in the honest subsample with treatment $d$ into $|B(S(x),d)|+1$ segments, the first $|B(S(x),d)|$ segments will have a length $s_{2,d}/|B(S(x),d)|$. Denote $\tilde{X}_j^x$ as the first nearest neighbour in the $j^{th}$ segment. Then,

$$\sum_{i \in B(S(x),d)} \|X_i - x\| \leq \sum_{j=1}^{|B(S(x),d)|} \|\tilde{X}_j^x - x\|.$$

Therefore, the absolute bias can be further bounded by

$$\left| E[T_b(d,x)] - E[Y^d | X = x] \right| \leq C_d E\left[ \frac{1}{|S(d,x)|} \sum_{j=1}^{|B(S(x),d)|} \|\tilde{X}_j^x - x\| \right]$$

$$= C_d E\left[ E\left[ \frac{1}{|S(d,x)|} \sum_{j=1}^{|B(S(x),d)|} \|\tilde{X}_j^x - x\| \, \big| \, |S(d,x)| \right] \right]$$

$$= C_d E\left[ E\left[ \frac{|B(S(x),d)|}{|S(d,x)|} \|\tilde{X}_1^x - x\| \, \big| \, |S(d,x)| \right] \right]$$

$$\leq C_{d,B} E\left[ E\left[ \|\tilde{X}_1^x - x\| \, \big| \, |S(d,x)| \right] \right]$$

$$= C_{d,B} E\left[ \|X_{(1, \lfloor s_{2,d}/|B(S(x),d)| \rfloor)} - x\| \, \big| \, |S(d,x)| \right],$$

where $X_{(1,N)}$ denotes the first nearest neighbour among $N$ observations and $C_{d,b}$ collects the Lipschitz constant and the constant from the ratio of the observations in the ball and the final leaf. For a fixed $|S(d,x)|$ in a regular, random split tree, we can use results in Györfi, Kohler, Krzyzak and Walk (2002) for nearest neighbour (NN) estimators that also use the expected distance of the first neighbours in their Theorem 6.2 and Lemma 6.4 yielding the bound

$$C_{d,B} E\left[ \|X_{(1, \lfloor s_{2,d}/|B(S(x),d)| \rfloor)} - x\| \, \big| \, |S(d,x)| \right] \leq \tilde{c}_d E\left[ \left( \frac{|S(d,x)|}{s_{2,d}} \right)^{\frac{1}{p}} \right],$$

where $\tilde{c}_d$ collects the constant $C_{d,B}$ and all constants that emerge in the NN proof. The upper bound for the last expectation can be then found by applying Jensen's inequality,

$$E\left[E\left[\left(\frac{|S(d,x)|}{s_{2,d}}\right)^{\frac{1}{p}}\middle|\lambda(S(x))\right]\right] \leq E\left[\left(\frac{E\left[|S(d,x)|\lambda(S(x))\right]}{s_{2,d}}\right)^{\frac{1}{p}}\right]$$

$$\leq \tilde{c}_\varepsilon E\left[(\lambda(S(x)))^{\frac{1}{p}}\right] \leq \tilde{c}_\varepsilon \left(E[\lambda(S(x))]\right)^{\frac{1}{p}}$$

where $\tilde{c}_\varepsilon$ collects constants stemming from the common support assumption. Using the results from Lemma 1, the result at the tree level is

$$\left|E[T_b(d,x)] - E[Y^d | X = x]\right| = O\left(s_1^{-\log(1-\alpha)/p\log(\alpha)}\right).$$

The final constant is a function of the Lipschitz constant, constant from the ratio of the observations in the ball and the final leaf, common support parameter $\varepsilon$ and $k$, the regularity parameter controlling the number of the observations in the final leaf. As the forest is average of trees, the result above holds also for the forest estimate. ∎

*Appendix A.1.2: Proofs for Theorem 2*

The asymptotic normality proofs build on a central limit theorem for weakly dependent random variables of a form $A_{i,N_2} = \hat{W}_i Y_i - E[\hat{W}_i Y_i]$, introduced in Neumann (2013). In this section, we prove that $A_{i,N_2}$ satisfy the conditions of the CLT yielding the first necessary result. As not all $Y_i$ are unbiased estimators of $\mu_d(x)$, the weighted average is not an unbiased estimator. Therefore, in the second step it is necessary to show that the ratio of bias and variance converges to zero as the sample gets larger and the final leaf shrinks asymptotically.

For the analysis of the variance of the *IATE* forest estimator, we make the following observation.

**COROLLARY 1** For any forest estimator $F$ that averages $B$ tree estimators $T$, the rate of the forest variance $Var(F)$ is bounded from above and below by the rate of the individual tree variance $Var(T)$.

**Proof:** The variance of a forest in a simplified notation is

$$Var(F) = \frac{1}{B^2}\left[\sum_{b=1}^{B} Var(T_b) + \sum_{b=1}^{B}\sum_{b'\neq b} Cov(T_b, T_{b'})\right].$$

As a forest is an average of trees, the worst upper bound of the variance of the forest is the variance of an individual tree. Lemma 3 shows the upper bound for the tree estimate that converges to zero for $\beta_2 > \beta_1$.

Since any covariance has to go to zero as fast as the variance, the lower bound of the rate of the variance is also determined by the variance of an individual tree. The lower bound is scaled by $1/B$. ∎

In the following, we therefore focus on deriving the properties of the tree weights.

**LEMMA 2** Let $\hat{W}_{i,b} := \hat{W}_{bi,N,b}(D_{bi}, X_{bi}, d, x)$. Suppose that the assumptions from Lemma 1 hold and the tree is symmetric. Moreover $\beta_2 > \beta_1/2$. Then, the moments of the tree weights have the following rates and bounds:

a) $E\left[\hat{W}_{i,b}\right] \sim \dfrac{1}{N^{\beta_2}}$,

b) $E\left[\hat{W}_{i,b}^2\right] = \Omega\left(N^{\frac{\log(1-\alpha)}{\log(\alpha)}\beta_1 - 2\beta_2}\right)$ and $E\left[\hat{W}_{i,b}^2\right] = O\left(N^{\beta_1 - 2\beta_2}\right)$,

c) $Var(\hat{W}_{i,b})$ has the same bounds as $E\left[\hat{W}_{i,b}^2\right]$,

d) $\left|Cov\left(\hat{W}_{i,b}, \hat{W}_{j,b}\right)\right| = \Omega\left(N^{\frac{\log(1-\alpha)}{\log(\alpha)}\beta_1 - 3\beta_2}\right)$ and $\left|Cov\left(\hat{W}_{i,b}, \hat{W}_{j,b}\right)\right| = O\left(N^{\beta_1 - 3\beta_2}\right)$.

**Proof:**

a) Due to symmetry, the expected value of the first moment of the tree weights can be expressed as:

$$E\left[\hat{W}_{i,b}\right] = E\left[E\left[\hat{W}_{i,b}\,\big|\,|S(d,x)|\right]\right]$$

$$= E\left[\frac{s_2 - |S(d,x)|}{s_2} \cdot 0 + \frac{|S(d,x)|}{s_2}\frac{1}{|S(d,x)|}\right]$$

$$= \frac{1}{s_2} \sim \frac{1}{N^{\beta_2}}.$$

b) Let $\tilde{p}_d(S(x)) = s_2 p_d(S(x))/s_{2,d}$ where $p_d(S(x))$ is the propensity score of getting treatment $d$ when on leaf $S(x)$. Due to the common support assumption $\tilde{p}_d(S(x))$ must lie in an interval $(\varepsilon/(1-\varepsilon),(1-\varepsilon)/\varepsilon)$. Thus, under symmetry, the expected value of the second moment of the tree weights can be expressed as:

$$E\left[\hat{W}_{i,b}^2\right] = E\left[E\left[\hat{W}_{i,b}^2\,\big|\,|S(d,x)|\right]\right]$$

$$= E\left[\frac{s_2 - |S(d,x)|}{s_2} \cdot 0 + \frac{|S(d,x)|}{s_2}\frac{1}{|S(d,x)|^2}\right]$$

$$= \frac{1}{s_2}E\left[\frac{1}{|S(d,x)|}\right] = \frac{1}{s_2}E\left[\frac{1-(1-\lambda(S(x))\tilde{p}_d(S(x)))^{s_{2,d}}}{s_{2,d}\lambda(S(x))\tilde{p}_d(S(x))}\right],$$

where the last equality uses the fact that $|S(d,x)|$ is a positive binomial random variable. Since $E\left[1/|S(d,x)|\right] > 0$, $E\left[1/s_{2,d}\lambda(S(x))\tilde{p}_d(S(x))\right]$ determines the lower bound. The lower bound can then be derived as follows using Jensen's inequality and results in Lemma 1:

$$E\left[\hat{W}_{i,b}^2\right] = \Omega\left(\frac{1}{s_2 s_{2,d} E[\lambda(S(x))]}\right) = \Omega\left(N^{-2\beta_2 + \frac{\log(1-\alpha)}{\log(\alpha)}\beta_1}\right).$$

The upper bound can be similarly derived for $\beta_2 > \beta_1/2$ as

$$E\left[\frac{1}{|S(x,d)|}\right] = E\left[\frac{1-(1-\lambda(S(x))\tilde{p}_d(S(x)))^{s_{2,d}}}{s_{2,d}\lambda(S(x))\tilde{p}_d(S(x))}\right]$$

$$= E\left[\frac{1-(1-\lambda(S(x))\tilde{p}_d(S(x)))^{s_{2,d}}}{s_{2,d}\lambda(S(x))\tilde{p}_d(S(x))}\bigg|\lambda(S(x))\leq kM/s_1\right]\Pr(\lambda(S(x))\leq kM/s_1)$$

$$+E\left[\frac{1-(1-\lambda(S(x))\tilde{p}_d(S(x)))^{s_{2,d}}}{s_{2,d}\lambda(S(x))\tilde{p}_d(S(x))}\bigg|\lambda(S(x))>kM/s_1\right]\Pr(\lambda(S(x))>kM/s_1)$$

$$\leq 1\cdot O(1/s_1)+O(s_1/s_{2,d})\cdot 1 = O\left(1/N^{\beta_2-\beta_1}\right).$$

The two results yield

$$E\left[\hat{W}_{i,b}^2\right] = \Omega\left(N^{-2\beta_2+\frac{\log(1-\alpha)}{\log(\alpha)}\beta_1}\right),$$

$$E\left[\hat{W}_{i,b}^2\right] = O\left(N^{-2\beta_2+\beta_1}\right).$$

Both rates have constants that depend on $\varepsilon$, the common support parameter and regularity parameter $k$. The upper bound also depends on the number of treatments as the number of treatments influences the smallest expected Lebesgue measure of the leaf.

c) Since $E\left[\hat{W}_{i,b}^2\right] \to 0$ for $\beta_2 > \beta_1/2$, the bounds for $Var(\hat{W}_{i,b})$ are the same as for $E\left[\hat{W}_{i,b}^2\right]$.

d) The covariance can be expressed as

$$Cov(\hat{W}_{i,b},\hat{W}_{j,b}) = Cov\left(\hat{W}_{i,b}, 1-\sum_{k\neq j}\hat{W}_{k,b}\right) = -Var(\hat{W}_{i,b}) - \sum_{k\neq i,j}Cov(\hat{W}_{i,b},\hat{W}_{k,b})$$

yielding

$$Cov(\hat{W}_{i,b},\hat{W}_{j,b}) = -\frac{Var(\hat{W}_{i,b})}{s_2-1}.$$

The bounds for covariance are

$$\left|Cov\left(\hat{W}_{i,b},\hat{W}_{j,b}\right)\right|=\Omega\left(N^{-3\beta_2+\frac{\log(1-\alpha)}{\log(\alpha)}\beta_1}\right),$$

$$\left|Cov\left(\hat{W}_{i,b},\hat{W}_{j,b}\right)\right|=O\left(N^{-3\beta_2+\beta_1}\right).$$

Note that

$$\sum_{k\neq j}Cov(\hat{W}_{i,b},\hat{W}_{k,b})=\frac{Var(\hat{W}_{i,b})}{s_2-1}.\blacksquare$$

**LEMMA 3** Suppose that the tree conditions from Lemma 2 hold i.e., we build regular, random-split, symmetric trees. Additionally, the tree is honest and $E\left[Y^d\mid X=x\right]$ is Lipschitz continuous. Moreover, assume that $E\left[\left(Y^d\right)^2\mid X=x\right]$ is also Lipschitz continuous and $Var\left(Y^d\mid X=x\right)>0$. Further, the sampling rates satisfy $\beta_2>\beta_1$. Then the tree variance has the following rates

$$Var\left(\sum_{i=1}^{s_2}\hat{W}_{i,b}Y_i\right)=O\left(N^{\beta_1-\beta_2}\right),$$

$$Var\left(\sum_{i=1}^{s_2}\hat{W}_{i,b}Y_i\right)=\Omega\left(N^{\frac{\log(1-\alpha)}{\log(\alpha)}\beta_1-\beta_2}\right).$$

**Proof:** The variance of a tree estimate of $\mu_d(x)$ can be decomposed as

$$Var\left(\sum_{i=1}^{s_2}\hat{W}_{i,b}Y_i\right)=\sum_{i=1}^{s_2}Var\left(\hat{W}_{i,b}Y_i\right)+\sum_{i=1}^{s_2}\sum_{j\neq i}Cov\left(\hat{W}_{i,b}Y_i,\hat{W}_{j,b}Y_j\right).$$

The upper bound for $Var\left(\hat{W}_{i,b}Y_i\right)$ is

$$\begin{aligned}
Var\left(\hat{W}_{i,b}Y_i\right) &= E\left[\hat{W}_{i,b}^2 Y_i^2\right] - E^2\left[\hat{W}_{i,b}Y_i\right] \\
&\leq E\left[\hat{W}_{i,b}^2 Y_i^2\right] = E\left[E\left[\hat{W}_{i,b}^2 \mid X_i\right]E\left[Y_i^2 \mid X_i\right]\right] \\
&\leq E\left[E\left[\hat{W}_{i,b}^2 \mid X_i\right]E\left[\sup_{d,x} E\left[(Y_i)^2 \mid X_i = x, D_i = d\right]\right]\right] \\
&= E\left[E\left[\hat{W}_{i,b}^2 \mid X_i\right]E\left[\sup_{d,x} E\left[(Y_i^d)^2 \mid X_i = x\right]\right]\right] \\
&\leq E\left[\hat{W}_{i,b}^2\right]C_2 = O\left(N^{-2\beta_2 + \beta_1}\right),
\end{aligned}$$

using the identifying assumptions and Lipschitz continuity of $E\left[(Y_i^d)^2 \mid X_i\right]$ on a bounded covariate space. This yields that

$$\sum_{i=1}^{s_2} Var\left(\hat{W}_{i,b}Y_i\right) \leq s_2 E\left[\hat{W}_{i,b}^2\right]C_2 = O\left(N^{\beta_1 - \beta_2}\right).$$

Note that this upper bound converges to zero for $\beta_2 > \beta_1$. In order to derive the upper bound for the covariance part, we rewrite the covariance as

$$\begin{aligned}
\left|Cov(\hat{W}_{i,b}Y_i, \hat{W}_{j,b}Y_j)\right| &= \left|Cov\left(\hat{W}_{i,b}Y_i, \left(1 - \sum_{k \neq j}\hat{W}_{k,b}\right)Y_j\right)\right| \\
&= \left|-\sum_{k \neq j} Cov\left(\hat{W}_{i,b}Y_i, \hat{W}_{k,b}Y_j\right)\right| \\
&= \left|\sum_{k \neq j} Cov\left(\hat{W}_{i,b}Y_i, \hat{W}_{k,b}Y_j\right)\right|.
\end{aligned}$$

Next, we derive the upper bound of $\sum_{k \neq j} Cov(\hat{W}_{i,b}Y_i, \hat{W}_{k,b}Y_j)$

$$\begin{aligned}
\sum_{k \neq j} Cov(\hat{W}_{i,b}Y_i, \hat{w}_{k,b}Y_j) &= \sum_{k \neq j} E[\hat{W}_{i,b}Y_i \hat{W}_{k,b}Y_j] - E[\hat{W}_{i,b}Y_i] E[\hat{W}_{k,b}Y_j] \\
&= \sum_{k \neq j} E[\hat{W}_{i,b}Y_i \hat{W}_{k,b}] E[Y_j] - E[\hat{W}_{i,b}Y_i] E[\hat{W}_{k,b}] E[Y_j] \\
&= \sum_{k \neq j} \left( E[\hat{W}_{i,b}Y_i \hat{W}_{k,b}] - E[\hat{W}_{i,b}Y_i] E[\hat{W}_{k,b}] \right) E[Y_j] \\
&= \sum_{k \neq j} \Big( E\big[ E[\hat{W}_{i,b}\hat{W}_{k,b}|X_i,X_k] E[Y_i|X_i] \big] \\
&\quad - E\big[ E[\hat{W}_{i,b}|X_i] E[Y_i|X_i] \big] E[\hat{W}_{k,b}] \Big) E[Y_j] \\
&= \sum_{k \neq j} \Big( E\big[ ( E[\hat{W}_{i,b}\hat{W}_{k,b}|X_i,X_k] - E[\hat{W}_{i,b}|X_i] E[\hat{W}_{k,b}] ) E[Y_i|X_i] \big] \Big) E[Y_j] \\
&\leq \sum_{k \neq j} \Big( E[\hat{W}_{i,b}\hat{W}_{k,b}] - E[\hat{W}_{i,b}] E[\hat{W}_{k,b}] \Big) C_1^2 \\
&= \sum_{k \neq j} Cov(\hat{W}_{i,b}, \hat{W}_{k,b}) C_1^2 = \frac{Var(\hat{W}_{i,b})}{s_2 - 1} C_1^2,
\end{aligned}$$

where the inequality uses the same supremum argument as the proof for the variance, the fact that the expected potential outcomes are Lipschitz continuous on a bounded covariate space, i.e., $E[Y_i^d|X_i] \in [-C_1, C_1]$ and $E[Y_i^d] \in [-C_1, C_1]$ for some positive constant $C_1$, and the fact that the final sum of covariances is positive and therefore the product certainly bounds the original sum from above. This yields an upper bound for

$$\sum_{i=1}^{s_2} \sum_{j \neq i} \left| Cov(\hat{W}_{i,b}Y_i, \hat{W}_{j,b}Y_j) \right| \leq s_2(s_2-1) \frac{Var(\hat{W}_{i,b})}{s_2 - 1} C_1^2 $$
$$= O(N^{\beta_1 - \beta_2}).$$

The overall upper bound for the tree variance is

$$Var\left( \sum_{i=1}^{s_2} \hat{W}_{i,b} Y_i \right) = O(N^{\beta_1 - \beta_2}).$$

The constant depends on $\varepsilon$, the common support parameter, regularity parameter $k$, number of treatments $M$ and a constant related to Lipschitz continuity of the outcome variable.

Due to non-negativity of the variance, it is enough to find the lower bound for $E\left[\hat{W}_{i,b}^2 Y_i^2\right]$ to analyze the lower bound of $Var\left(\hat{W}_{i,b} Y_i\right)$.

$$E\left[\hat{W}_{i,b}^2 Y_i^2\right] = E\left[E\left[\hat{W}_{i,b}^2 | X_i\right] E\left[Y_i^2 | X_i\right]\right]$$
$$\geq E\left[E\left[\hat{W}_{i,b}^2 | X_i\right] E\left[\inf_{d,x} E\left[Y_i^2 | X_i = x, D_i = d\right]\right]\right]$$
$$= E\left[E\left[\hat{W}_{i,b}^2 | X_i\right]\right] E\left[\inf_{d,x} E\left[\left(Y_i^d\right)^2 | X_i = x\right]\right]$$
$$= E\left[\hat{W}_{i,b}^2\right] E\left[\inf_{d,x} E\left[\left(Y_i^d\right)^2 | X_i = x\right]\right]$$
$$= \Omega\left(N^{\frac{\log(1-\alpha)}{\log(\alpha)}\beta_1 - 2\beta_2}\right).$$

Therefore, $\sum_{i=1}^{s_2} Var\left(\hat{W}_{i,b} Y_i\right) = \Omega\left(N^{\frac{\log(1-\alpha)}{\log(\alpha)}\beta_1 - \beta_2}\right)$.

The lower bound for the covariance bound can be derived with help of the triangular inequality

$$\left|Cov(\hat{W}_{i,b} Y_i, \hat{W}_{j,b} Y_j)\right| = \left|E\left[\hat{W}_{i,b} Y_i \hat{W}_{j,b} Y_j\right] - E\left[\hat{W}_{i,b} Y_i\right] E\left[\hat{W}_{j,b} Y_j\right]\right|$$
$$\geq \left\|E\left[\hat{W}_{i,b} Y_i \hat{W}_{j,b} Y_j\right]\right| - \left|E\left[\hat{W}_{i,b} Y_i\right] E\left[\hat{W}_{j,b} Y_j\right]\right\|$$

and finding lower bounds for the individual elements:

$$\left|E\left[\hat{W}_{i,b} Y_i \hat{W}_{j,b} Y_j\right]\right| = \left|E\left[E\left[\hat{W}_{i,b} Y_i \hat{w} \hat{W}_{j,b} Y_j | \lambda(S(x))\right]\right]\right|$$
$$= \left|E\left[E\left[1/|S(x,d)|^2 Y_i Y_j | \lambda(S(x)), X_i, X_j \in S(x)\right] \Pr\left(X_i, X_j \in S(x) | \lambda(S(x))\right)\right]\right|$$
$$\geq \frac{1}{(s_{2,d})^2} E\left[\inf_{d,x} E\left[|Y_i Y_j| \Big| D_i, D_j = d, X_i, X_j = x\right] (\lambda(S(x)))^2\right]$$
$$= \Omega\left(N^{-2\beta_2 - 2\beta_1}\right)$$

$$\left|E\left[\hat{W}_{i,b}Y_i\right]\right| = \left|E\left[E\left[\hat{W}_{i,b}Y_i \mid \lambda(S(x))\right]\right]\right|$$

$$= \left|E\left[E\left[\hat{W}_{i,b}Y_i \mid X_i \in S(x), \lambda(S(x))\right]\Pr\left(X_i \in S(x) \mid \lambda(S(x))\right)\right]\right|$$

$$\geq \frac{1}{s_{2,d}} E\left[\inf_{d,x} E\left[|Y_i| \mid D_i = d, X_i = x\right]\lambda(S(x))\right]$$

$$= \Omega\left(N^{-\beta_2 - \beta_1}\right)$$

Since $\left|E\left[\hat{W}_{i,b}Y_i\hat{W}_{j,b}Y_j\right]\right| = \Omega\left(N^{-2\beta_2 - 2\beta_1}\right)$, $\left|E\left[\hat{W}_{i,b}Y_i\right]\right|\left|E\left[\hat{W}_{j,b}Y_j\right]\right| = \Omega\left(N^{-2\beta_2 - 2\beta_1}\right)$ and

$\left|E\left[\hat{W}_{i,b}Y_i\right]\right|\left|E\left[\hat{W}_{j,b}Y_j\right]\right| = O\left(N^{-2\beta_2}\right)$, the final lower bound is

$$\left|Cov(\hat{W}_{i,b}Y_i, \hat{W}_{j,b}Y_j)\right| = \Omega\left(N^{-2\beta_2 - 2\beta_1}\right).$$

Putting these results together yields a lower bound for covariance:

$$\sum_{i=1}^{s_2}\sum_{j \neq i}\left|Cov\left(\hat{W}_{i,b}Y_i, \hat{W}_{j,b}Y_j\right)\right| = \Omega\left(N^{-2\beta_1}\right).$$

As the lower bound for the covariance exceeds the lower bound for the variance, the lower bound for the tree variance is determined by the one converging slower to zero i.e.

$$Var\left(\sum_{i=1}^{s_2}\hat{W}_{i,b}Y_i\right) = \Omega\left(N^{\frac{\log(1-\alpha)}{\log(\alpha)}\beta_1 - \beta_2}\right).$$

The constant depends on $\varepsilon$, the common support parameter, regularity parameter $k$ and a constant related to Lipschitz continuity of the outcome variable. ∎

**LEMMA 4** Let the assumptions from Lemma 3 hold and $A_{i,N_2} := \hat{W}_{i,N}Y_i - E\left[\hat{W}_{i,N}Y_i\right]$, where $\hat{W}_{i,N}$ are the forest weights. Then $(A_{i,N_2})_{i=1,\ldots,N_2}$ is a triangular array satisfying:

a) $E[A_{i,N_2}] = 0$,

b) $\sum_{i=1}^{N_2} E[A_{i,N_2}^2] < \infty$ for all $N$ and $i$,

c) $\sigma_N^2 := Var(A_{1,N_2} + \ldots + A_{N_2,N_2}) \xrightarrow[N \to \infty]{} 0$,

d) $\sum_{i=1}^{N_2} E\left[A_{i,N_2}^2 \mathbf{1}\left(|A_{i,N_2}| > \tilde{\varepsilon}\right)\right] \xrightarrow{N \to \infty} 0$ for all $\tilde{\varepsilon} > 0$,

e) There is a summable sequence $(\pi_r)_{r \in \mathbb{N}}$ such that for all $u \in \mathbb{N}$ and all indices $1 \leq s_1 < s_2 < ... < s_u < s_u + r = t_1 \leq t_2 \leq N_2$, the following upper bounds for covariances hold true for all measurable functions $g : \mathbb{R}^u \to \mathbb{R}$ with $\|g\|_\infty = \sup_{x \in \mathbb{R}^u} |g(x)| \leq 1$:

$$\left|\text{cov}(g(A_{s_1,N_2},...,A_{s_u,N_2})A_{s_u,N_2}, A_{t_1,N_2})\right| \leq \left(E[A_{s_u,N_2}^2] + E[A_{t_1,N_2}^2] + N_2^{-1}\right)\pi_r$$

and

$$\left|\text{cov}(g(A_{s_1,N_2},...,A_{s_u,N_2}), A_{t_1,N_2}, A_{t_2,N_2})\right| \leq \left(E[A_{t_1,N_2}^2] + E[A_{t_2,N_2}^2] + N_2^{-1}\right)\pi_r.$$

**Proof:**

a) $E\left[\hat{W}_{i,N} y_i - E\left[\hat{W}_{i,N} y_i\right]\right] = 0$.

b) $E\left[A_{i,N_2}^2\right] = \text{Var}\left(\hat{W}_{i,N} Y_i\right) = \left(E\left[\hat{W}_{i,N}^2 Y_i^2\right] - E^2\left[\hat{W}_{i,N} Y_i\right]\right)$

Deriving upper bounds for $E[\hat{W}_{i,N} Y_i]$ and $E[\hat{W}_{i,N}^2 Y_i^2]$, we use that $\hat{W}_{i,N}$ and $Y_i$ are independent conditional on $X_i$ and $E\left[Y^d \mid X = x\right]$ and $E\left[(Y^d)^2 \mid X = x\right]$ are Lipschitz continuous and therefore can be bounded from above by $C_1 < \infty$ and $C_2 < \infty$ respectively on a bounded covariate space. The first and second moment of the forest weights are

$$E\left[\hat{W}_{i,N}\right] = \frac{N_2 - s_2}{N_2} \cdot 0 + \frac{s_2}{N_2} E\left[\frac{1}{B}\sum_{b=1}^{B} \hat{W}_{i,N,b}\right] = \frac{s_2}{N_2} \frac{1}{B} B \frac{1}{s_2} = \frac{1}{N_2} \sim N^{-1}$$

and

$$E\left[\hat{W}_{i,N}^2\right] = \frac{N_2 - S_2}{N_2} \cdot 0 + \frac{S_2}{N_2} E\left[\left(\frac{1}{B}\sum_{b=1}^{B}\hat{W}_{i,N,b}\right)^2\right]$$

$$= \frac{S_2}{N_2}\frac{1}{B^2}\left[\sum_{b=1}^{B} E\left[\hat{W}_{i,N,b}^2\right] + \sum_{b=1}^{B}\sum_{b'\neq b} E\left[\hat{W}_{i,N,b}\hat{W}_{i,N,b'}\right]\right]$$

$$= \frac{S_2}{N_2}\frac{1}{B^2}\left[B E\left[\hat{W}_{i,N,b}^2\right] + B(B-1)\left(Cov(\hat{W}_{i,N,b},\hat{W}_{i,N,b'}) + \left(E\left[\hat{W}_{i,N,b}\right]\right)^2\right)\right]$$

$$\leq \frac{S_2}{N_2}\frac{1}{B^2}\left[B E\left[\hat{W}_{i,N,b}^2\right] + B(B-1)\left(Var(\hat{W}_{i,N,b}) + \left(E\left[\hat{W}_{i,N,b}\right]\right)^2\right)\right]$$

$$\leq \frac{S_2}{N_2}\frac{1}{B^2}\left[B^2 E\left[\hat{W}_{i,N,b}^2\right]\right] = O\left(N^{-1-\beta_2+\beta_1}\right).$$

These results can be used to further bound the following expectations:

$$E\left[\hat{W}_{i,N}Y_i\right] = E\left[E\left[\hat{W}_{i,N}|X_i\right]E\left[Y_i|X_i\right]\right] = O(N^{-1}),$$

$$E[\hat{W}_{i,N}^2 Y_i^2] = E\left[E\left[\hat{W}_{i,N}^2|X_i\right]E\left[Y_i^2|X_i\right]\right] = O\left(N^{-1-\beta_2+\beta_1}\right).$$

It follows that

$$\sum_{i=1}^{N_2} E[A_{i,N_2}^2] = O\left(N^{-\beta_2+\beta_1}\right) < \infty.$$

c) $$Var(A_{1,N_2} + ... + A_{N_2,N_2}) = Var\left(\sum_{i=1}^{N_2}\hat{W}_{i,N}Y_i - \sum_{i=1}^{N_2}E\left[\hat{W}_{i,N}Y_i\right]\right) = Var\left(\sum_{i=1}^{N_2}\hat{W}_{i,N}Y_i\right)$$

$$= \sum_{i=1}^{N_2}Var\left(\hat{W}_{i,N}Y_i\right) + \sum_{i=1}^{N_2}\sum_{j\neq i}Cov\left(\hat{W}_{i,N}Y_i,\hat{W}_{j,N}Y_j\right)$$

$$\leq \sum_{i=1}^{N_2}Var\left(\hat{W}_{i,N}Y_i\right) + \sum_{i=1}^{N_2}\sum_{j\neq i}\left|Cov\left(\hat{W}_{i,N}Y_i,\hat{W}_{j,N}Y_j\right)\right|$$

Using the results from b), the first element is bounded at rate $O\left(N^{-\beta_2+\beta_1}\right)$ and by a similar logic as in the tree case the sum of all covariances has to decay to zero as fast as the sum of variances. These results yield that $Var(A_{1,N_2} + ... + A_{N_2,N_2}) \xrightarrow[N\to\infty]{} 0$.

d) Due to monotonicity and the result in b):

$$\sum_{i=1}^{N_2} E\left[A_{i,N_2}^2 \mathbf{1}\left(\left|A_{i,N_2}\right| > \tilde{\varepsilon}\right)\right] \le \sum_{i=1}^{N_2} E\left[A_{i,N_2}^2\right] \to 0 \text{ for all } \tilde{\varepsilon} > 0.$$

e) Due to exchangeability which implies strict stationarity and the result in c), it is possible to interpret $\left(A_{i,N_2}\right)_{i=1,\ldots,N_2}$ as a $\rho$-mixing process. Since every $\phi$-mixing process is also $\rho$-mixing, then along the Lemma 20.1 in Billingsley (1968) for $\phi$-mixing processes, we can also bound the covariances of a $\rho$-mixing process as follows:[36]

$$\left|Cov\left(g\left(A_{s_1,N_2},\ldots,A_{s_u,N_2}\right)A_{s_u,N_2},A_{t_1,N_2}\right)\right|$$
$$\le 2\sqrt{\phi_{t_1-s_u}}\sqrt{E\left[g^2\left(A_{s_1,N_2},\ldots,A_{s_u,N_2}\right)A_{s_u,N_2}^2\right]}\sqrt{E\left[A_{t_1,N_2}^2\right]}$$

and

$$\left|Cov(g(A_{s_1,N_2},\ldots,A_{s_u,N_2}),A_{t_1,N_2}A_{t_2,N_2})\right| \le 2\phi_{t_1-s_u}E\left[\left|A_{t_1,N_2}A_{t_2,N_2}\right|\right] \le 2\phi_{t_1-s_u}E\left[A_{t_1,N_2}^2\right]$$

where the last inequality follows from the stationarity of the process. The two results can be further bounded

$$\left|Cov(g(A_{s_1,N_2},\ldots,A_{s_u,N_2})A_{s_u,N_2},A_{t_1,N_2}\right| \le 2\sqrt{\phi_{t_1-s_u}}\sqrt{E\left[A_{s_u,N_2}^2\right]}\sqrt{E\left[A_{t_1,N_2}^2\right]}$$
$$\le \sqrt{\phi_{t_1-s_u}}\left(E\left[A_{s_u,N_2}^2\right]+E\left[A_{t_1,N_2}^2\right]\right)$$
$$\le \sqrt{\phi_{t_1-s_u}}\left(E\left[A_{s_u,N_2}^2\right]+E\left[A_{t_1,N_2}^2\right]+N_2^{-1}\right)$$

by property of the function $g()$ and inequality of arithmetic and geometric mean and

$$\left|Cov(g(A_{s_1,N_2},\ldots,A_{s_u,N_2}),A_{t_1,N_2}A_{t_2,N_2}\right| \le \sqrt{\phi_{t_1-s_u}}\left(E\left[A_{t_1,N_2}^2\right]+E\left[A_{t_2,N_2}^2\right]+N_2^{-1}\right)$$

by stationarity and the fact that the mixing coefficients are smaller or equal to 1. Then the weak dependence conditions are fulfilled for $\pi_r = \sqrt{\phi_r}$. ∎

---

[36] Bradley (1986) showed that the coefficients of the two processes fulfil the following inequality: $\rho_r = 2\sqrt{\phi_r}$.

**LEMMA 5** Define $A_{i,N_2}$ as in Lemma 4, then $\widehat{IATE}(m,l;x) - E\left[\widehat{IATE}(m,l;x)\right] = \sum_{i=1}^{N_2} A_{i,N_2}$ and

$$\frac{\widehat{IATE}(m,l;x) - E\left[\widehat{IATE}(m,l;x)\right]}{\sqrt{Var(\widehat{IATE}(m,l;x))}} \to N(0,1).$$

**Proof:** The normality proof for triangular arrays satisfying the conditions in Lemma 4 is proven in Neumann (2013) and can be applied on estimation of potential outcomes $\mu_m(x)$ and $\mu_l(x)$. As those are estimated on an honest data set, the difference of the two quantities will also follow a normal distribution. ∎

**THEOREM 2** Assume that there are i.i.d. data $(X_i, Y_i, D_i) \in [0,1]^p \times \mathbb{R} \times \{0,1,...,M-1\}$ collected in $\vec{R}$ and a given value of $x$. Moreover, features are independently and uniformly distributed $X_i \sim U\left([0,1]^p\right)$. Let $T$ be an honest, regular and symmetric random split tree. Further assume that $E\left[Y^d \mid X=x\right]$ and $E\left[\left(Y^d\right)^2 \mid X=x\right]$ are Lipschitz continuous and $Var\left(Y^d \mid X=x\right) > 0$. Then for $\beta_1 < \beta_2 < \frac{p+2}{p}\frac{\log(1-\alpha)}{\log(\alpha)}\beta_1$,

$$\frac{\widehat{IATE}(m,l;x) - IATE(m,l;x)}{\sqrt{Var(\widehat{IATE}(m,l;x))}} \to N(0,1).$$

**Proof:** Given the result in Lemma 5, it remains to show that

$$\frac{E\left[\widehat{IATE}(m,l;x)\right] - IATE(m,l;x)}{\sqrt{Var(\widehat{IATE}(m,l;x))}} \to 0.$$

The final result will follow then from Slutsky's lemma. By Theorem 1, we have

$$\left|E\left[\widehat{IATE}(m,l;x)\right] - IATE(m,l;x)\right| = O\left(s_1^{-\log(1-\alpha)/p\log(\alpha)}\right).$$

From Corollary 1, we get

$$Var(\widehat{IATE}(m,l;x)) = \Omega\left(N^{\frac{\log(1-\alpha)}{\log(\alpha)}\beta_1 - \beta_2}\right).$$

It follows that

$$\frac{\left(E\left[\widehat{IATE}(m,l;x)\right] - IATE(m,l;x)\right)}{\sqrt{Var(\widehat{IATE}(m,l;x))}} = O\left(N^{-\left(\frac{1}{p}+\frac{1}{2}\right)\frac{\log(1-\alpha)}{\log(\alpha)}\beta_1 + \frac{\beta_2}{2}}\right).$$

The ratio converges to zero when

$$\frac{p}{2+p}\beta_2 < \frac{\log(1-\alpha)}{\log(\alpha)}\beta_1. \blacksquare$$

*Appendix A.1.3: Proofs for Theorem 3*

**THEOREM 3** Let all assumptions from Theorem 2 hold and define $\widehat{ATE}(m,l)$ as an average of all corresponding $\widehat{IATE}(m,l;x)$. Then,

$$\frac{\widehat{ATE}(m,l) - ATE(m,l)}{\sqrt{Var(\widehat{ATE}(m,l))}} \to N(0,1).$$

**Proof:** Using the CLT for triangular arrays of weakly dependent random variables requires to check that all requirements in Lemma 4 hold for $A_{i,N_2} := \hat{W}_i^{ATE(m,l)} Y_i - E\left[\hat{W}_i^{ATE(m,l)} Y_i\right]$. The proof uses the observation that the rates for the *ATE* cannot be worse than for the *IATE* and, therefore, the conditions in Lemma 4 will be satisfied for the *ATE* weights. Due to the pointwise convergence at different points $x$, the convergence rate is affected.

a) $E\left[A_{i,N_2}\right] = 0$ holds trivially.

b) $\sum_{i=1}^{N_2} E\left[A_{i,N_2}^2\right] = \sum_{i=1}^{N_2} Var\left(\hat{W}_i^{ATE(m,l)} Y_i\right)$. The upper bound on the individual variances is the upper bound of $E\left[\left(\hat{W}_i^{ATE(m,l)}\right)^2 Y_i^2\right]$:

$$E\left[\left(\hat{W}_i^{ATE(m,l)}\right)^2 Y_i^2\right] = E\left[\left(\frac{1}{N_2}\sum_{j=1}^{N_2}\hat{W}_i^{IATE(m,l,x_j)}\right)^2 Y_i^2\right] =$$

$$\leq \frac{1}{(N_2)^2} E\left[(N_2)^2 E\left[\left(\hat{W}_i^{IATE(m,l,x_j)}\right)^2 \Big| X_i\right] E\left[Y_i^2 | X_i\right]\right]$$

$$= O\left(N^{-1-\beta_2+\beta_1}\right).$$

The second requirement is also satisfied as $\sum_{i=1}^{N_2} Var\left(\hat{W}_i^{ATE(m,l)} Y_i\right) = O\left(N^{-\beta_2+\beta_1}\right) < \infty$.

c) $Var(A_{1,N_2} + ... + A_{N_2,N_2}) = Var\left(\sum_{i=1}^{N_2}\hat{W}_{i,N}^{ATE(m,l)}Y_i - \sum_{i=1}^{N_2}E\left[\hat{W}_{i,N}^{ATE(m,l)}Y_i\right]\right) = Var\left(\sum_{i=1}^{N_2}\hat{W}_{i,N}^{ATE(m,l)}Y_i\right)$

$$\leq \sum_{i=1}^{N} Var\left(\hat{W}_{i,N}^{ATE(m,l)}Y_i\right) + \sum_{i=1}^{N}\sum_{j\neq i}\left|Cov\left(\hat{W}_{i,N}^{ATE(m,l)}Y_i, \hat{W}_{j,N}^{ATE(m,l)}Y_j\right)\right|$$

Using the results from b), the first element is bounded at rate $O\left(N^{-\beta_2+\beta_1}\right)$ and by a similar logic as in the tree case the sum of all covariances must decay to zero as fast as the sum of the variances. These results yield that $Var(A_{1,N_2} + ... + A_{N_2,N_2}) \xrightarrow{N\to\infty} 0$.

d) Due to monotonicity and result in b):

$$\sum_{i=1}^{N_2} E\left[A_{i,N_2}^2 \mathbf{1}\left(|A_{i,N_2}| > \tilde{\varepsilon}\right)\right] \leq \sum_{i=1}^{N_2} E\left[A_{i,N_2}^2\right] \to 0 \text{ for all } \tilde{\varepsilon} > 0.$$

e) The proof follows the same logic as in Lemma 4.

With this, all requirements for the CLT for triangular arrays of weakly dependent random variables hold, so that we get

$$\frac{\widehat{ATE}(m,l) - ATE(m,l)}{\sqrt{Var(\widehat{ATE}(m,l))}} \to N(0,1).$$

Note that $E\left[\sum_{i=1}^{N_2}\hat{W}_i^{APO(d)}Y_i\right] = E\left[Y^d\right]$ due to exchangeability and the fact that the weights sum to 1. Therefore, we could have applied the CLT directly to the quantity of interest. The next corollary would follow a similar proof. ∎

## Appendix A.2: Tuning parameters

In the simulations, the number of randomly chosen variables to form the next split in a tree ($V$) is either chosen from a *min(58,1+Poisson(5))* process or from a *min(58,1+Poisson(38))* process (58 variables in total). The means of the two Poisson processes are considered as the only tuning parameters. They are chosen based on the out-of-bag value of the objective function of the particular estimator. The motivation for using a random number of features is to foster the independence of the trees that appear in the Random Forest.[37]

The minimum leaf size in each tree equals 5. The number of trees contained in any forest equals 1000. Trees are formed on random subsamples drawn without replacement (subsampling) with a sample size of 50% of the size of sample A.

The nonparametric regressions that enter the estimation of the standard errors are based on *k*-NN estimation with number of neighbours equal to *2 sqrt(N)*.

## Appendix A.3: Local centering

Recentering is implemented in the following way:

1) Estimate the trees that define the Random Forest for $E(Y \mid X = x)$ in sample A.

2) Recentering of outcomes in sample A: Split sample A randomly into *K* equally sized parts, *A-1* to *A-K*. Use the outcomes in the union of the *K-1*-folds *A-1* to *A-(K-1)* to obtain the Random Forest predictions given the forest estimated in step 1). Use these predictions to predict $E(Y \mid X = x)$ in fold *A-K*. Do this for all-possible combinations of folds (cross-fitting as used in k-fold cross-validation). Subtract the predictions from the actual values of *Y*.

3) Redo step 2 in sample B using the estimated forests of sample A.

Concerning the specifics of this algorithm, there are a couple of points worth mentioning.

---

[37] This has also been suggested by Denil, Matheson, and de Freitas (2014) for regression forests.

First, in order to avoid overfitting, the outcomes of observation 'i' are not used to predict itself. Therefore, the implementation is chosen similar to cross-validation.

Second, weights-based inference requires avoiding a dependence of the weights in sample B on outcomes of sample A. However, since recentering uses outcome variables independent of the treatment state, this could induce a correlation between the recentered outcomes in different treatment states. This finite sample correlation will be ignored here (as in Athey, Tibshirani, and Wager, 2019).

Third, the number of folds is a tuning parameter that influences the noise added to the recentered outcome by subtracting an estimated quantity. The simulation results indicate that the computationally most attractive choice of *K=2* may be too small in medium sized samples and that a somewhat larger number of folds may be needed to avoid much additional noise to the estimators.

# Appendix B *(Online)*: Protocol of EMCS

The Empirical Monte Carlo study follows almost exactly the one used in KLS21. The main differences are in how the effects are generated, as explained in the main body of the text. In this appendix, we repeat their protocol for completeness, with some small adjustments.

1) Take the full sample and estimate the propensity score, $p^f(x)$. We use the specification of Huber, Lechner, and Mellace (2017) which is based on a logit model.

2) Remove all treated from the dataset and keep only the $N^{nt}$ non-treated observations ($Y^0$).

3) Specify the true individual treatment effect (ITE). Add them to $Y^0$ to obtain $Y^1$ for all observations.

4) Remove $N^v=5000$ observations from the main sample to form the validation sample.

5) Calculate the true GATEs and ATE by aggregating the true IATEs in the validation sample.

6) Modify $p^f(x)$ by increasing the constant term in the index function used such that $p^f(x)$ equals approximately 50% and compute its value for all $N$ units.

7) Draw random samples with $N=1,000$, $N=4,000,$ and $N=8,000$ from the ($N^{nt}$ - $N^v$) observations.

8) Assign a treatment state based on the outcome of a draw in the Bernoulli distribution with this modified probability as parameter.

9) Depending on the assigned treatment state, use the value $Y^0$ or $Y^1$ as observable outcome variable $Y$.

10) Use these $N$ observations as training sample to compute all effects.

11) Predict all effects in the validation sample.

12) Predict the quality measures in the validation sample for each parameter to estimate.

13) Repeat steps 7 to 12 1,000 ($N=1,000$), 250 ($N=4,000$), or 125 ($N=8,000$) times.

14) Calculate quality measures by aggregating over all estimated effects.

# Appendix C *(Online)*: Further simulation results for the case with selection bias

The tables in this appendix follow exactly the structure of those in Section 5 of the main body of the text.

## Appendix C.1: Main specification with smaller sample sizes

This appendix contains the results of the main specification for the smaller samples of *N=1,000* and *N=4,000*.

*Table C.1: Simulation results for N=1,000, main DGP, and main estimators*

| | Groups # | Est. | True & estimated effects | | | Estimation error of effects (averages) | | | | | Estimation of std. error | |
|---|---|---|---|---|---|---|---|---|---|---|---|---|
| | | | Avg. bias | X-sectional std. dev. | | MSE | Skewness | Kurtosis | JB-Stat. | Std. err. | Avg. bias | CovP (90) in % |
| | | | | true | est. | | | | | | | |
| | (1) | | (2) | (3) | (4) | (5) | (6) | (7) | (8) | (9) | (10) | (11) |
| ATE | 1 | Basic | 1.90 | - | - | 4.92 | -0.1 | 2.7 | 5.4 | 1.14 | 0.08 | 53 |
| GATE | 2 | | 1.90 | - | - | 4.96 | -0.1 | 2.7 | 4.7 | 1.16 | 0.09 | 54 |
| GATE | 32 | | 1.87 | 0.17 | 0.09 | 4.98 | -0.1 | 2.8 | 4.2 | 1.22 | 0.13 | 60 |
| IATE | 5000 | | 1.91 | 1.72 | 0.96 | 8.47 | -0.1 | 3.1 | 3.4 | 1.86 | 0.46 | 79 |
| ATE | 1 | OneF. | 2.09 | - | - | 5.64 | -0.1 | 3.0 | 0.7 | 1.14 | 0.11 | 48 |
| GATE | 2 | VarT | 2.09 | - | - | 5.67 | -0.1 | 3.0 | 0.7 | 1.15 | 0.12 | 50 |
| GATE | 32 | | 2.06 | 0.17 | 0.07 | 5.60 | -0.1 | 3.0 | 0.8 | 1.16 | 0.15 | 53 |
| IATE | 5000 | | 2.09 | 1.72 | 0.75 | 8.01 | 0.0 | 3.0 | 3.0 | 1.55 | 0.54 | 74 |
| ATE | 1 | OneF. | 1.97 | - | - | 5.28 | -0.1 | 3.1 | 1.6 | 1.18 | 0.04 | 50 |
| GATE | 2 | MCE | 1.97 | - | - | 5.31 | -0.1 | 3.1 | 2.2 | 1.19 | 0.06 | 52 |
| GATE | 32 | | 1.94 | 0.17 | 0.06 | 5.32 | -0.1 | 3.0 | 2.0 | 1.24 | 0.11 | 56 |
| IATE | 5000 | | 1.99 | 1.72 | 0.55 | 8.43 | 0.0 | 3.0 | 2.0 | 1.63 | 0.56 | 75 |
| ATE | 1 | OneF. | 1.39 | - | - | 3.56 | 0.0 | 3.1 | 0.2 | 1.28 | 0.11 | 76 |
| GATE | 2 | MCE. | 1.38 | - | - | 3.60 | 0.0 | 3.1 | 0.3 | 1.30 | 0.11 | 77 |
| GATE | 32 | LC-2 | 1.36 | 0.17 | 0.04 | 3.64 | 0.0 | 3.1 | 0.4 | 1.33 | 0.12 | 78 |
| IATE | 5000 | | 1.52 | 1.72 | 0.43 | 7.02 | 0.0 | 3.1 | 2.1 | 1.75 | 0.20 | 76 |
| ATE | 1 | OneF. | 1.11 | - | - | 3.35 | -0.1 | 2.9 | 2.6 | 1.46 | 0.10 | 81 |
| GATE | 2 | MCE. | 1.11 | - | - | 3.36 | -0.1 | 3.1 | 3.1 | 1.46 | 0.11 | 82 |
| GATE | 32 | Pen | 1.10 | 0.17 | 0.13 | 3.37 | -0.1 | 2.9 | 2.7 | 1.47 | 0.13 | 83 |
| IATE | 5000 | | 1.12 | 1.72 | 1.50 | 5.87 | -0.2 | 3.1 | 2.0 | 2.05 | 0.56 | 90 |
| ATE | 1 | OneF. | 0.97 | - | - | 3.42 | -0.1 | 3.1 | 1.6 | 1.57 | 0.03 | 84 |
| GATE | 2 | MCE. | 0.97 | - | - | 3.44 | -0.1 | 3.1 | 1.7 | 1.58 | 0.03 | 84 |
| GATE | 32 | Pen | 0.95 | 0.17 | 0.10 | 3.46 | -0.1 | 3.1 | 1.7 | 1.60 | 0.03 | 84 |
| IATE | 5000 | LC-2 | 1.00 | 1.72 | 1.24 | 6.52 | -0.1 | 3.2 | 9.5 | 2.25 | 0.12 | 86 |
| ATE | 1 | OneF. | 0.79 | - | - | 2.41 | -0.1 | 2.8 | 1.3 | 1.34 | 0.06 | 85 |
| GATE | 2 | MCE. | 0.79 | - | - | 2.43 | -0.1 | 2.8 | 1.3 | 1.34 | 0.06 | 85 |
| GATE | 32 | Pen | 0.78 | 0.17 | 0.10 | 2.44 | -0.1 | 2.8 | 1.4 | 1.35 | 0.06 | 85 |
| IATE | 5000 | LC-5 | 0.89 | 1.72 | 1.20 | 5.03 | -0.1 | 3.1 | 2.0 | 1.96 | 0.08 | 86 |

Note: For GATE and IATE the *average bias* is the absolute value of the bias for the specific group (GATE) / observation (IATE) averaged over all groups / observation (each group / observation receives the same weight). *CovP (90%)* denotes the (average) probability that the true value is part of the 90% confidence interval. The simulation errors of the mean MSEs are around 0.1.

*Table C.2: Simulation results for N=4,000, main DGP, and main estimators*

| | | | True & estimated effects | | | Estimation error of effects (averages) | | | | | Estimation of std. error | |
|---|---|---|---|---|---|---|---|---|---|---|---|---|
| | Groups # | Est. | Avg. bias | X-sectional std. dev. | | MSE | Skew ness | Kurt- osis | JB- Stat. | Std. err. | Avg. bias | CovP (90) in % |
| | | | | true | est. | | | | | | | |
| | (1) | | (2) | (3) | (4) | (5) | (6) | (7) | (8) | (9) | (10) | (11) |
| ATE | 1 | Basic | 1.39 | - | - | 2.30 | 0.3 | 3.0 | 2.9 | 0.62 | 0.02 | 31 |
| GATE | 2 | | 1.39 | - | - | 2.35 | 0.2 | 3.1 | 2.4 | 0.64 | 0.03 | 35 |
| GATE | 32 | | 1.35 | 0.17 | 0.12 | 2.40 | 0.2 | 2.9 | 3.3 | 0.74 | 0.06 | 48 |
| IATE | 5000 | | 1.48 | 1.72 | 1.30 | 5.71 | 0.0 | 3.0 | 2.3 | 1.58 | 0.24 | 78 |
| ATE | 1 | OneF. | 1.78 | - | - | 3.57 | -0.2 | 3.0 | 1.0 | 0.63 | 0.05 | 17 |
| GATE | 2 | VarT | 1.78 | - | - | 3.58 | -0.1 | 3.0 | 1.4 | 0.65 | 0.06 | 18 |
| GATE | 32 | | 1.75 | 0.17 | 0.09 | 3.52 | -0.1 | 3.0 | 1.0 | 0.66 | 0.08 | 22 |
| IATE | 5000 | | 1.79 | 1.72 | 1.13 | 5.31 | -0.1 | 3.1 | 5.7 | 1.14 | 0.50 | 73 |
| ATE | 1 | OneF. | 1.46 | - | - | 2.50 | -0.6 | 3.7 | 20.1 | 0.61 | 0.04 | 22 |
| GATE | 2 | MCE | 1.45 | - | - | 2.50 | -0.6 | 3.6 | 17.4 | 0.63 | 0.06 | 27 |
| GATE | 32 | | 1.43 | 0.17 | 0.07 | 2.53 | -0.5 | 3.4 | 11.9 | 0.67 | 0.12 | 40 |
| IATE | 5000 | | 1.50 | 1.72 | 0.79 | 4.87 | -0.1 | 3.0 | 3.2 | 1.16 | 0.50 | 76 |
| ATE | 1 | OneF. | 1.05 | - | - | 1.54 | -0.2 | 2.9 | 1.9 | 0.71 | 0.04 | 56 |
| GATE | 2 | MCE. | 1.05 | - | - | 1.56 | -0.2 | 2.9 | 1.6 | 0.73 | 0.05 | 59 |
| GATE | 32 | LC-2 | 1.02 | 0.17 | 0.05 | 1.59 | 0.2 | 2.9 | 2.1 | 0.78 | 0.05 | 62 |
| IATE | 5000 | | 1.24 | 1.72 | 0.63 | 4.14 | 0.0 | 3.0 | 3.6 | 1.36 | 0.13 | 72 |
| ATE | 1 | OneF. | 0.39 | - | - | 0.89 | -0.1 | 2.9 | 0.9 | 0.86 | 0.10 | 90 |
| GATE | 2 | MCE. | 0.39 | - | - | 0.90 | -0.1 | 2.9 | 0.8 | 0.87 | 0.11 | 90 |
| GATE | 32 | Pen | 0.38 | 0.17 | 0.16 | 0.93 | -0.1 | 2.8 | 1.1 | 0.89 | 0.14 | 91 |
| IATE | 5000 | | 0.51 | 1.72 | 1.89 | 2.83 | -0.1 | 3.0 | 2.3 | 1.53 | 0.53 | 94 |
| ATE | 1 | OneF. | 0.39 | - | - | 1.02 | 0.1 | 3.0 | 0.1 | 0.90 | -0.03 | 86 |
| GATE | 2 | MCE. | 0.39 | - | - | 1.04 | 0.1 | 3.0 | 0.2 | 0.91 | -0.03 | 86 |
| GATE | 32 | Pen | 0.39 | 0.17 | 0.14 | 1.07 | 0.0 | 2.9 | 0.3 | 0.94 | -0.02 | 87 |
| IATE | 5000 | LC-2 | 0.51 | 1.72 | 1.59 | 3.38 | 0.0 | 3.1 | 2.9 | 1.80 | 0.10 | 89 |
| ATE | 1 | OneF. | 0.41 | - | - | 0.70 | -0.2 | 3.1 | 1.5 | 0.73 | 0.04 | 87 |
| GATE | 2 | MCE. | 0.41 | - | - | 0.71 | -0.2 | 3.1 | 1.5 | 0.74 | 0.05 | 87 |
| GATE | 32 | Pen | 0.40 | 0.17 | 0.13 | 0.75 | -0.2 | 3.1 | 1.9 | 0.76 | 0.05 | 87 |
| IATE | 5000 | LC-5 | 0.53 | 1.72 | 1.44 | 2.57 | -0.1 | 3.0 | 3.5 | 1.45 | 0.09 | 89 |

Note: For GATE and IATE the *average bias* is the absolute value of the bias for the specific group (GATE) / observation (IATE) averaged over all groups / observation (each group / observation receives the same weight). *CovP (90%)* denotes the (average) probability that the true value is part of the 90% confidence interval. The simulation errors of the mean MSEs are around 0.1.

## Appendix C.2: Alternative specifications of IATEs for main estimators

In this section, we show the results for alternative specifications of the individualized effects for all sample sizes.

*Appendix C.2.1: No individual effects (ITE = 0)*

*Table C.3: Simulation results for N=1,000, no effect, and main estimators*

| | | | True & estimated effects | | | Estimation error of effects (averages) | | | | | Estimation of std. error | |
|---|---|---|---|---|---|---|---|---|---|---|---|---|
| | Groups | Est. | Avg. bias | X-sectional std. dev. | | MSE | Skewness | Kurtosis | JB-Stat. | Std. err. | Avg. bias | CovP (90) in % |
| | # | | | true | est. | | | | | | | |
| | (1) | | (2) | (3) | (4) | (5) | (6) | (7) | (8) | (9) | (10) | (11) |
| ATE | 1 | Basic | 1.13 | - | - | 2.64 | 0.1 | 3.3 | 3-4 | 1.17 | 0.04 | 77 |
| GATE | 2 | | 1.13 | - | - | 2.71 | 0.1 | 3.2 | 3.7 | 1.19 | 0.05 | 78 |
| GATE | 32 | | 1.09 | 0 | 0.12 | 2.78 | 0.1 | 3.3 | 3.9 | 1.25 | 0.08 | 80 |
| IATE | 5000 | | 1.14 | 0 | 0.60 | 5.11 | 0.0 | 3.1 | 2.4 | 1.86 | 0.42 | 90 |
| ATE | 1 | OneF. | 1.23 | - | - | 2.80 | -0.1 | 3.0 | 0.8 | 1.12 | 0.11 | 76 |
| GATE | 2 | VarT | 1.24 | - | - | 2.84 | -0.1 | 3.0 | 0.9 | 1.14 | 0.13 | 76 |
| GATE | 32 | | 1.23 | 0 | 0.04 | 2.85 | -0.1 | 3.0 | 0.8 | 1.15 | 0.15 | 78 |
| IATE | 5000 | | 1.24 | 0 | 0.39 | 3.98 | 0.0 | 3.0 | 1.9 | 1.51 | 0.53 | 91 |
| ATE | 1 | OneF. | 1.11 | - | - | 2.53 | -0.1 | 2.9 | 2.0 | 1.14 | 0.06 | 77 |
| GATE | 2 | MCE | 1.11 | - | - | 2.58 | -0.1 | 2.9 | 2.1 | 1.15 | 0.08 | 78 |
| GATE | 32 | | 1.10 | 0 | 0.06 | 2.64 | -0.1 | 2.9 | 1.7 | 1.20 | 0.13 | 81 |
| IATE | 5000 | | 1.11 | 0 | 0.33 | 3.87 | 0.0 | 2.9 | 2.0 | 1.59 | 0.56 | 93 |
| ATE | 1 | OneF. | 0.81 | - | - | 2.31 | 0.1 | 3.0 | 1.0 | 1.29 | 0.08 | 85 |
| GATE | 2 | MCE. | 0.81 | - | - | 2.36 | 0.1 | 3.0 | 1.2 | 1.31 | 0.09 | 85 |
| GATE | 32 | LC-2 | 0.80 | 0 | 0.04 | 2.44 | 0.1 | 3.0 | 1.2 | 1.34 | 0.09 | 86 |
| IATE | 5000 | | 0.81 | 0 | 0.23 | 3.79 | 0.0 | 3.1 | 1.6 | 1.76 | 0.19 | 90 |
| ATE | 1 | OneF. | 0.63 | - | - | 2.26 | -0.1 | 3.1 | 0.9 | 1.37 | 0.16 | 89 |
| GATE | 2 | MCE. | 0.63 | - | - | 2.29 | -0.1 | 3.1 | 1.0 | 1.38 | 0.17 | 90 |
| GATE | 32 | Pen | 0.63 | 0 | 0.06 | 2.34 | -0.1 | 3.1 | 0.9 | 1.39 | 0.18 | 90 |
| IATE | 5000 | | 0.88 | 0 | 0.73 | 4.70 | -0.1 | 3.2 | 9.5 | 1.93 | 0.63 | 93 |
| ATE | 1 | OneF. | 0.50 | - | - | 2.59 | -0.1 | 3.1 | 1.8 | 1.53 | 0.05 | 89 |
| GATE | 2 | MCE. | 0.51 | - | - | 2.62 | -0.1 | 3.1 | 1.7 | 1.54 | 0.05 | 89 |
| GATE | 32 | Pen | 0.51 | 0 | 0.04 | 2.68 | -0.1 | 3.1 | 1.5 | 1.56 | 0.06 | 89 |
| IATE | 5000 | LC-2 | 0.64 | 0 | 0.50 | 5.43 | -0.1 | 3.3 | 6.3 | 2.34 | 0.14 | 90 |
| ATE | 1 | OneF. | 0.41 | - | - | 1.95 | -0.2 | 3.1 | 10.0 | 1.34 | 0.05 | 90 |
| GATE | 2 | MCE. | 0.42 | - | - | 1.98 | -0.2 | 3.1 | 9.8 | 1.34 | 0.05 | 90 |
| GATE | 32 | Pen | 0.42 | 0 | 0.03 | 2.01 | -0.2 | 3.1 | 9.4 | 1.35 | 0.06 | 90 |
| IATE | 5000 | LC-5 | 0.58 | 0 | 0.48 | 4.03 | -0.2 | 3.2 | 10.0 | 1.90 | 0.13 | 90 |

Note: For GATE and IATE the *average bias* is the absolute value of the bias for the specific group (GATE) / observation (IATE) averaged over all groups / observation (each group / observation receives the same weight). *CovP (90%)* denotes the (average) probability that the true value is part of the 90% confidence interval. The simulation errors of the mean MSEs are around 0.15.

*Table C.4: Simulation results for N=4,000, no effect, and main estimators*

| | | | True & estimated effects | | | Estimation error of effects (averages) | | | | | Estimation of std. error | |
|---|---|---|---|---|---|---|---|---|---|---|---|---|
| | Groups | Est. | Avg. bias | X-sectional std. dev. | | MSE | Skewness | Kurtosis | JB-Stat. | Std. err. | Avg. bias | CovP (90) in % |
| | # | | | true | est. | | | | | | | |
| | (1) | | (2) | (3) | (4) | (5) | (6) | (7) | (8) | (9) | (10) | (11) |
| **ATE** | *1* | Basic | 0.84 | - | - | 1.09 | 0.0 | 2.8 | 0.2 | 0.63 | 0.00 | 60 |
| **GATE** | *2* | | 0.85 | - | - | 1.15 | 0.0 | 2.8 | 0.9 | 0.65 | 0.01 | 63 |
| **GATE** | *32* | | 0.78 | 0 | 0.17 | 1.18 | -0.1 | 2.9 | 1.1 | 0.73 | 0.06 | 73 |
| **IATE** | *5000* | | 1.00 | 0 | 0.86 | 3.84 | 0.0 | 3.0 | 2.4 | 1.54 | 0.24 | 87 |
| **ATE** | *1* | OneF. | 1.05 | - | - | 1.43 | 0.1 | 2.7 | 1.5 | 0.58 | 0.09 | 55 |
| **GATE** | *2* | VarT | 1.05 | - | - | 1.47 | 0.1 | 2.7 | 1.6 | 0.60 | 0.11 | 58 |
| **GATE** | *32* | | 1.03 | 0 | 0.06 | 1.44 | 0.1 | 2.8 | 1.4 | 0.61 | 0.12 | 61 |
| **IATE** | *5000* | | 1.06 | 0 | 0.57 | 2.60 | 0.0 | 2.9 | 1.7 | 1.08 | 0.53 | 89 |
| **ATE** | *1* | OneF. | 0.86 | - | - | 1.06 | -0.1 | 2.8 | 0.6 | 0.57 | 0.05 | 59 |
| **GATE** | *2* | MCE | 0.87 | - | - | 1.08 | -0.1 | 2.8 | 0.7 | 0.58 | 0.07 | 63 |
| **GATE** | *32* | | 0.84 | 0 | 0.08 | 1.13 | -0.1 | 2.9 | 1.1 | 0.64 | 0.14 | 75 |
| **IATE** | *5000* | | 0.87 | 0 | 0.43 | 2.18 | 0.0 | 3.0 | 2.4 | 1.12 | 0.52 | 94 |
| **ATE** | *1* | OneF. | 0.72 | - | - | 0.88 | 0.1 | 2.9 | 0.6 | 0.60 | 0.09 | 76 |
| **GATE** | *2* | MCE. | 0.72 | - | - | 0.91 | 0.1 | 2.9 | 0.7 | 0.62 | 0.09 | 76 |
| **GATE** | *32* | LC-2 | 0.71 | 0 | 0.05 | 0.95 | 0.1 | 3.0 | 1.5 | 0.67 | 0.10 | 79 |
| **IATE** | *5000* | | 0.72 | 0 | 0.29 | 1.95 | 0.0 | 3.0 | 2.2 | 1.16 | 0.18 | 89 |
| **ATE** | *1* | OneF. | 0.25 | - | - | 0.66 | -0.1 | 2.8 | 1.0 | 0.77 | 0.20 | 94 |
| **GATE** | *2* | MCE. | 0.25 | - | - | 0.67 | -0.1 | 2.8 | 0.9 | 0.78 | 0.20 | 95 |
| **GATE** | *32* | Pen | 0.26 | 0 | 0.05 | 0.70 | -0.1 | 2.9 | 1.6 | 0.80 | 0.22 | 95 |
| **IATE** | *5000* | | 0.62 | 0 | 0.59 | 2.42 | 0.0 | 2.9 | 1.6 | 1.41 | 0.66 | 96 |
| **ATE** | *1* | OneF. | 0.38 | - | - | 0.84 | 0.1 | 2.6 | 1.4 | 0.84 | 0.06 | 89 |
| **GATE** | *2* | MCE. | 0.39 | - | - | 0.86 | 0.1 | 2.7 | 1.4 | 0.85 | 0.06 | 89 |
| **GATE** | *32* | Pen | 0.38 | 0 | 0.04 | 0.89 | 0.0 | 2.7 | 1.4 | 0.86 | 0.07 | 89 |
| **IATE** | *5000* | LC-2 | 0.56 | 0 | 0.47 | 2.87 | 0.0 | 3.0 | 2.8 | 1.58 | 0.21 | 91 |
| **ATE** | *1* | OneF. | 0.17 | - | - | 0.55 | -0.4 | 3.5 | 10.2 | 0.73 | 0.06 | 91 |
| **GATE** | *2* | MCE. | 0.17 | - | - | 0.56 | -0.4 | 3.4 | 8.9 | 0.73 | 0.06 | 91 |
| **GATE** | *32* | Pen | 0.18 | 0 | 0.03 | 0.60 | -0.4 | 3.5 | 9.8 | 0.76 | 0.06 | 92 |
| **IATE** | *5000* | LC-5 | 0.36 | 0 | 0.34 | 2.22 | -0.1 | 3.1 | 2.3 | 1.43 | 0.12 | 91 |

Note: For GATE and IATE the *average bias* is the absolute value of the bias for the specific group (GATE) / observation (IATE) averaged over all groups / observation (each group / observation receives the same weight). *CovP (90%)* denotes the (average) probability that the true value is part of the 90% confidence interval. The simulation errors of the mean MSEs are around 0.1.

*Table C.5: Simulation results for N=8,000, no effect, and main estimators*

|  |  |  | True & estimated effects | | | Estimation error of effects (averages) | | | | | Estimation of std. error | |
|---|---|---|---|---|---|---|---|---|---|---|---|---|
|  | Groups # | Est. | Avg. bias | X-sectional std. dev. | | MSE | Skew-ness | Kurt-osis | JB-Stat. | Std. err. | Avg. bias | CovP (90) in % |
|  |  |  |  | true | est. |  |  |  |  |  |  |  |
|  | (1) |  | (2) | (3) | (4) | (5) | (6) | (7) | (8) | (9) | (10) | (11) |
| ATE | 1 | Basic | 0.77 | - | - | 0.77 | 0.1 | 2.9 | 0.4 | 0.40 | 0.05 | 46 |
| GATE | 2 |  | 0.79 | - | - | 0.81 | 0.2 | 2.8 | 1.1 | 0.42 | 0.06 | 52 |
| GATE | 32 |  | 0.70 | 0 | 0.25 | 0.85 | 0.0 | 2.7 | 1.6 | 0.54 | 0.10 | 69 |
| IATE | 5000 |  | 0.99 | 0 | 0.91 | 3.39 | 0.0 | 2.9 | 1.7 | 1.39 | 0.25 | 86 |
| ATE | 1 | OneF. | 1.06 | - | - | 1.28 | -0.1 | 2.8 | 0.3 | 0.41 | 0.09 | 28 |
| GATE | 2 | VarT | 1.06 | - | - | 1.30 | -0.1 | 2.9 | 0.4 | 0.43 | 0.54 | 34 |
| GATE | 32 |  | 1.04 | 0 | 0.07 | 1.29 | -0.1 | 2.8 | 0.6 | 0.45 | 0.58 | 40 |
| IATE | 5000 |  | 1.09 | 0 | 0.64 | 2.48 | 0.0 | 3.0 | 1.9 | 0.97 | 1.47 | 86 |
| ATE | 1 | OneF. | 0.77 | - | - | 0.79 | -0.3 | 2.9 | 1.7 | 0.43 | 0.02 | 44 |
| GATE | 2 | MCE | 0.78 | - | - | 0.81 | -0.2 | 2.8 | 2.8 | 0.45 | 0.03 | 50 |
| GATE | 32 |  | 0.75 | 0 | 0.09 | 0.84 | -0.2 | 2.8 | 2.8 | 0.52 | 0.11 | 66 |
| IATE | 5000 |  | 0.79 | 0 | 0.42 | 1.74 | -0.1 | 3.0 | 3.0 | 0.98 | 0.48 | 94 |
| ATE | 1 | OneF. | 0.53 | - | - | 0.48 | -0.3 | 3.0 | 2.0 | 0.44 | 0.05 | 71 |
| GATE | 2 | MCE. | 0.54 | - | - | 0.50 | -0.3 | 2.9 | 1.9 | 0.45 | 0.05 | 74 |
| GATE | 32 | LC-2 | 0.52 | 0 | 0.06 | 0.54 | -0.2 | 2.9 | 1.7 | 0.51 | 0.06 | 76 |
| IATE | 5000 |  | 0.55 | 0 | 0.32 | 1.40 | 0.0 | 2.9 | 1.0 | 1.00 | 0.13 | 89 |
| ATE | 1 | OneF. | 0.09 | - | - | 0.42 | -0.3 | 4.5 | 12.4 | 0.64 | 0.14 | 95 |
| GATE | 2 | MCE. | 0.09 | - | - | 0.43 | -0.3 | 4.3 | 11.0 | 0.65 | 0.14 | 95 |
| GATE | 32 | Pen | 0.08 | 0 | 0.03 | 0.47 | -0.2 | 4.1 | 10.4 | 0.68 | 0.20 | 95 |
| IATE | 5000 |  | 0.34 | 0 | 0.36 | 2.40 | -0.1 | 3.0 | 2.3 | 1.49 | 0.77 | 98 |
| ATE | 1 | OneF. | 0.08 | - | - | 0.49 | -0.2 | 3.4 | 1.5 | 0.68 | -0.01 | 89 |
| GATE | 2 | MCE. | 0.09 | - | - | 0.49 | -0.2 | 3.4 | 1.3 | 0.69 | -0.01 | 89 |
| GATE | 32 | Pen | 0.09 | 0 | 0.02 | 0.52 | -0.2 | 3.3 | 1.6 | 0.72 | 0.00 | 90 |
| IATE | 5000 | LC-2 | 0.32 | 0 | 0.34 | 2.13 | 0.0 | 3.0 | 2.5 | 1.57 | 0.17 | 92 |
| ATE | 1 | OneF. | 0.08 | - | - | 0.34 | -0.4 | 3.9 | 7.5 | 0.58 | 0.02 | 90 |
| GATE | 2 | MCE. | 0.08 | - | - | 0.35 | -0.4 | 3.8 | 6.6 | 0.58 | 0.02 | 91 |
| GATE | 32 | Pen | 0.09 | 0 | 0.03 | 0.38 | -0.3 | 3.8 | 6.2 | 0.61 | 0.02 | 90 |
| IATE | 5000 | LC-5 | 0.28 | 0 | 0.29 | 1.72 | 0.0 | 3.1 | 1.9 | 1.27 | 0.10 | 91 |

Note: For GATE and IATE the *average bias* is the absolute value of the bias for the specific group (GATE) / observation (IATE) averaged over all groups / observation (each group / observation receives the same weight). *CovP (90%)* denotes the (average) probability that the true value is part of the 90% confidence interval. The simulation errors of the mean MSEs are around 0.1.

*Appendix C.2.2: Strong individual effects (α = 8)*

*Table C.6: Simulation results for N=1,000, strong effect, and main estimators*

|  | Groups # | Est. | True & estimated effects | | | Estimation error of effects (averages) | | | | | Estimation of std. error | |
|---|---|---|---|---|---|---|---|---|---|---|---|---|
|  |  |  | Avg. bias | X-sectional std. dev. | | MSE | Skewness | Kurtosis | JB-Stat. | Std. err. | Avg. bias | CovP (90) in % |
|  |  |  |  | true | est. |  |  |  |  |  |  |  |
|  | (1) |  | (2) | (3) | (4) | (5) | (6) | (7) | (8) | (9) | (10) | (11) |
| ATE | 1 | Basic | 3.22 | - | - | 12.01 | 0.1 | 2.7 | 4.0 | 1.26 | 0.01 | 20 |
| GATE | 2 |  | 3.22 | - | - | 11.99 | 0.1 | 2.7 | 3.8 | 1.28 | 0.03 | 22 |
| GATE | 32 |  | 3.18 | 0.66 | 0.39 | 11.98 | 0.1 | 2.8 | 2.5 | 1.34 | 0.06 | 27 |
| IATE | 5000 |  | 3.33 | 6.87 | 3.91 | 28.63 | 0.0 | 2.9 | 2.9 | 2.24 | 0.35 | 65 |
| ATE | 1 | OneF. | 6.47 | - | - | 15.62 | -0.1 | 2.9 | 2.8 | 1.46 | -0.08 | 18 |
| GATE | 2 | VarT | 3.66 | - | - | 15.53 | -0.1 | 2.9 | 2.5 | 1.47 | -0.07 | 19 |
| GATE | 32 |  | 3.61 | 0.66 | 0.25 | 15.39 | -0.1 | 2.9 | 2.4 | 1.48 | -0.02 | 22 |
| IATE | 5000 |  | 4.52 | 6.87 | 2.81 | 36.76 | -0.1 | 2.9 | 6.7 | 2.13 | 0.23 | 51 |
| ATE | 1 | OneF. | 3.48 | - | - | 14.09 | -0.2 | 3.1 | 1.4 | 1.40 | -0.01 | 20 |
| GATE | 2 | MCE | 3.46 | - | - | 13.99 | -0.2 | 3.2 | 1.4 | 1.41 | 0.01 | 21 |
| GATE | 32 |  | 3.42 | 0.66 | 0.25 | 13.94 | -0.2 | 3.1 | 1.4 | 1.42 | 0.05 | 25 |
| IATE | 5000 |  | 4.26 | 6.87 | 2.56 | 37.01 | -0.1 | 3.0 | 2.0 | 2.04 | 0.35 | 57 |
| ATE | 1 | OneF. | 2.67 | - | - | 9.09 | 0.0 | 3.1 | 1.0 | 1.39 | 0.06 | 42 |
| GATE | 2 | MCE. | 2.65 | - | - | 9.03 | 0.0 | 3.1 | 1.3 | 1.41 | 0.07 | 44 |
| GATE | 32 | LC-2 | 2.59 | 0.66 | 0.13 | 9.08 | 0.0 | 3.2 | 1.7 | 1.44 | 0.08 | 47 |
| IATE | 5000 |  | 4.95 | 6.87 | 1.41 | 42.16 | 0.0 | 3.2 | 7.0 | 1.94 | 0.12 | 42 |
| ATE | 1 | OneF. | 2.29 | - | - | 8.01 | 0.0 | 2.8 | 1.8 | 1.66 | -0.03 | 60 |
| GATE | 2 | MCE. | 2.28 | - | - | 7.97 | 0.0 | 2.8 | 1.7 | 1.66 | -0.02 | 60 |
| GATE | 32 | Pen | 2.24 | 0.66 | 0.41 | 7.91 | 0.0 | 2.8 | 2.0 | 1.67 | 0.00 | 62 |
| IATE | 5000 |  | 2.96 | 6.87 | 4.42 | 19.76 | 0.1 | 2.9 | 9.6 | 2.47 | 0.30 | 72 |
| ATE | 1 | OneF. | 1.87 | - | - | 6.26 | -0.2 | 3.5 | 16 | 1.67 | -0.04 | 66 |
| GATE | 2 | MCE. | 1.85 | - | - | 6.23 | -0.2 | 3.5 | 15 | 1.67 | -0.04 | 67 |
| GATE | 32 | Pen | 1.80 | 0.66 | 0.29 | 6.25 | -0.2 | 3.5 | 14 | 1.69 | -0.03 | 68 |
| IATE | 5000 | LC-2 | 3.62 | 6.87 | 3.18 | 25.67 | -0.2 | 3.3 | 37 | 2.52 | -0.14 | 56 |
| ATE | 1 | OneF. | 1.78 | - | - | 5.08 | -0.2 | 3.3 | 11 | 1.38 | 0.00 | 62 |
| GATE | 2 | MCE. | 1.76 | - | - | 5.02 | -0.2 | 3.3 | 11 | 1.39 | 0.00 | 62 |
| GATE | 32 | Pen | 1.71 | 0.66 | 0.27 | 5.04 | -0.2 | 3.3 | 10 | 1.40 | 0.01 | 64 |
| IATE | 5000 | LC-5 | 3.68 | 6.87 | 3.03 | 24.58 | -0.2 | 3.2 | 37 | 2.00 | -0.15 | 49 |

Note: For GATE and IATE the *average bias* is the absolute value of the bias for the specific group (GATE) / observation (IATE) averaged over all groups / observation (each group / observation receives the same weight). *CovP (90%)* denotes the (average) probability that the true value is part of the 90% confidence interval. The simulation errors of the mean MSEs are around 0.15 for ATE/GATE and 0.3 for IATE.

*Table C.7: Simulation results for N=4,000, strong effect, and main estimators*

| | | | True & estimated effects | | | Estimation error of effects (averages) | | | | | Estimation of std. error | |
|---|---|---|---|---|---|---|---|---|---|---|---|---|
| | Groups | Est. | Avg. bias | X-sectional std. dev. | | MSE | Skewness | Kurtosis | JB-Stat. | Std. err. | Avg. bias | CovP (90) in % |
| | # | | | true | est. | | | | | | | |
| | (1) | | (2) | (3) | (4) | (5) | (6) | (7) | (8) | (9) | (10) | (11) |
| ATE | 1 | Basic | 2.19 | - | - | 5.21 | 0.1 | 3.0 | 0.8 | 0.64 | 0.04 | 5 |
| GATE | 2 | | 2.20 | - | - | 5.28 | 0.1 | 3.0 | 1.3 | 0.66 | 0.05 | 5 |
| GATE | 32 | | 2.16 | 0.66 | 0.52 | 5.28 | 0.1 | 3.2 | 2.4 | 0.74 | 0.08 | 15 |
| IATE | 5000 | | 2.38 | 6.87 | 5.12 | 14.78 | 0.0 | 3.0 | 2.4 | 1.69 | 0.30 | 67 |
| ATE | 1 | OneF. | 2.69 | - | - | 8.29 | -0.1 | 2.7 | 1.4 | 1.01 | -0.23 | 10 |
| GATE | 2 | VarT | 2.68 | - | - | 8.23 | -0.1 | 2.7 | 1.3 | 1.01 | -0.21 | 10 |
| GATE | 32 | | 2.64 | 0.66 | 0.38 | 8.13 | -0.1 | 2.7 | 1.5 | 1.03 | -0.17 | 14 |
| IATE | 5000 | | 3.38 | 6.87 | 4.49 | 19.78 | -0.1 | 2.9 | 3.5 | 1.74 | 0.06 | 51 |
| ATE | 1 | OneF. | 1.91 | - | - | 4.24 | 0.5 | 3.2 | 8.9 | 0.77 | 0.08 | 28 |
| GATE | 2 | MCE | 1.90 | - | - | 4.23 | 0.4 | 3.3 | 10.1 | 0.78 | 0.09 | 29 |
| GATE | 32 | | 1.87 | 0.66 | 0.44 | 4.21 | 0.4 | 3.2 | 8.3 | 0.80 | 0.13 | 37 |
| IATE | 5000 | | 2.38 | 6.87 | 4.49 | 13.54 | 0.2 | 3.3 | 7.3 | 1.45 | 0.42 | 68 |
| ATE | 1 | OneF. | 1.92 | - | - | 4.44 | -0.2 | 3.2 | 2.8 | 0.85 | -0.07 | 25 |
| GATE | 2 | MCE. | 1.91 | - | - | 4.41 | -0.2 | 3.2 | 2.6 | 0.86 | -0.06 | 27 |
| GATE | 32 | LC-2 | 1.87 | 0.66 | 0.26 | 4.46 | -0.2 | 3.1 | 2.5 | 0.89 | -0.05 | 31 |
| IATE | 5000 | | 3.69 | 6.87 | 2.79 | 24.46 | 0.0 | 3.2 | 5.0 | 1.06 | -0.11 | 41 |
| ATE | 1 | OneF. | 0.94 | - | - | 1.56 | 0.3 | 3.7 | 10.0 | 0.82 | 0.13 | 80 |
| GATE | 2 | MCE. | 0.95 | - | - | 1.56 | 0.3 | 3.7 | 9.7 | 0.83 | 0.14 | 80 |
| GATE | 32 | Pen | 0.92 | 0.66 | 0.55 | 1.58 | 0.3 | 3.6 | 8.5 | 0.84 | 0.17 | 82 |
| IATE | 5000 | | 1.50 | 6.87 | 5.74 | 6.39 | 0.1 | 3.1 | 4.8 | 1.59 | 0.47 | 85 |
| ATE | 1 | OneF. | 1.21 | - | - | 2.60 | -0.4 | 3.1 | 7.8 | 1.05 | -0.17 | 53 |
| GATE | 2 | MCE. | 1.20 | - | - | 2.58 | -0.4 | 3.1 | 7.5 | 1.06 | -0.16 | 56 |
| GATE | 32 | Pen | 1.18 | 0.66 | 0.41 | 2.62 | -0.4 | 3.1 | 7.8 | 1.08 | -0.15 | 58 |
| IATE | 5000 | LC-2 | 2.58 | 6.87 | 4.36 | 14.03 | -0.2 | 3.0 | 7.6 | 2.03 | -0.25 | 57 |
| ATE | 1 | OneF. | 1.24 | - | - | 2.21 | -0.5 | 3.9 | 20.6 | 0.74 | -0.08 | 45 |
| GATE | 2 | MCE. | 1.23 | - | - | 2.20 | -0.5 | 3.9 | 18.3 | 0.75 | -0.07 | 46 |
| GATE | 32 | Pen | 1.19 | 0.66 | 0.37 | 2.22 | -0.5 | 3.9 | 19.9 | 0.77 | -0.07 | 50 |
| IATE | 5000 | LC-5 | 2.76 | 6.87 | 4.17 | 13.92 | -0.1 | 3.3 | 11.8 | 1.48 | -0.21 | 47 |

Note: For GATE and IATE the *average bias* is the absolute value of the bias for the specific group (GATE) / observation (IATE) averaged over all groups / observation (each group / observation receives the same weight). *CovP (90%)* denotes the (average) probability that the true value is part of the 90% confidence interval. The simulation errors of the mean MSEs are around 0.1 for ATE/GATE and 0.15 for IATE.

*Table C.8: Simulation results for N=8,000, strong effect, and main estimators*

|  |  |  | True & estimated effects | | | Estimation error of effects (averages) | | | | | Estimation of std. error | |
|---|---|---|---|---|---|---|---|---|---|---|---|---|
|  | Groups | Est. | Avg. bias | X-sectional std. dev. | | MSE | Skew ness | Kurt- osis | JB- Stat. | Std. err. | Avg. bias | CovP (90) in % |
|  | # |  |  | true | est. |  |  |  |  |  |  |  |
|  | (1) |  | (2) | (3) | (4) | (5) | (6) | (7) | (8) | (9) | (10) | (11) |
| ATE | 1 | Basic | 1.82 | - | - | 3.52 | -0.6 | 3.4 | 7.2 | 0.44 | 0.05 | 2 |
| GATE | 2 |  | 1.84 | - | - | 3.58 | -0.5 | 3.6 | 10.0 | 0.46 | 0.06 | 3 |
| GATE | 32 |  | 1.82 | 0.66 | 0.60 | 3.66 | -0.3 | 2.9 | 2.8 | 0.55 | 0.11 | 12 |
| IATE | 5000 |  | 2.07 | 6.87 | 5.58 | 11.06 | 0.0 | 2.9 | 2.0 | 1.52 | 0.26 | 68 |
| ATE | 1 | OneF. | 0.91 | - | - | 1.80 | 0.2 | 2.1 | 5.0 | 0.99 | -0.26 | 64 |
| GATE | 2 | VarT | 0.91 | - | - | 1.79 | 0.2 | 2.1 | 4.8 | 0.99 | -0.25 | 66 |
| GATE | 32 |  | 0.88 | 0.66 | 0.55 | 1.80 | 0.1 | 2.2 | 4.5 | 1.00 | -0.22 | 67 |
| IATE | 5000 |  | 1.39 | 6.87 | 5.90 | 6.94 | 0.1 | 2.7 | 4.1 | 1.86 | 0.12 | 82 |
| ATE | 1 | OneF. | 1.39 | - | - | 2.20 | -0.1 | 2.6 | 1.2 | 0.52 | 0.12 | 28 |
| GATE | 2 | MCE | 1.39 | - | - | 2.20 | -0.2 | 2.7 | 1.7 | 0.53 | 0.13 | 30 |
| GATE | 32 |  | 1.35 | 0.66 | 0.49 | 2.21 | -0.1 | 2.6 | 1.3 | 0.57 | 0.17 | 39 |
| IATE | 5000 |  | 1.86 | 6.87 | 5.05 | 8.76 | 0.0 | 2.9 | 1.7 | 1.23 | 0.40 | 70 |
| ATE | 1 | OneF. | 1.67 | - | - | 3.19 | -0.1 | 3.1 | 0.3 | 0.63 | -0.04 | 14 |
| GATE | 2 | MCE. | 1.66 | - | - | 3.16 | -0.1 | 3.0 | 0.5 | 0.64 | -0.04 | 15 |
| GATE | 32 | LC-2 | 1.62 | 0.66 | 0.33 | 3.19 | -0.1 | 3.0 | 0.7 | 0.66 | -0.03 | 24 |
| IATE | 5000 |  | 3.16 | 6.87 | 3.46 | 18.23 | -0.1 | 2.8 | 2.6 | 1.47 | -0.15 | 43 |
| ATE | 1 | OneF. | 0.52 | - | - | 0.57 | 0.4 | 3.0 | 3.3 | 0.55 | 0.17 | 89 |
| GATE | 2 | MCE. | 0.52 | - | - | 0.58 | 0.4 | 2.9 | 3.1 | 0.56 | 0.17 | 88 |
| GATE | 32 | Pen | 0.50 | 0.66 | 0.58 | 0.62 | 0.3 | 3.0 | 2.6 | 0.59 | 0.20 | 89 |
| IATE | 5000 |  | 1.01 | 6.87 | 6.14 | 4.17 | 0.1 | 3.1 | 2.0 | 1.40 | 0.48 | 90 |
| ATE | 1 | OneF. | 0.88 | - | - | 1.28 | -0.5 | 3.1 | 3.4 | 0.71 | -0.05 | 59 |
| GATE | 2 | MCE. | 0.87 | - | - | 1.27 | -0.4 | 3.1 | 2.9 | 0.72 | -0.05 | 60 |
| GATE | 32 | Pen | 0.85 | 0.66 | 0.48 | 1.32 | -0.4 | 3.1 | 3.8 | 0.75 | -0.05 | 63 |
| IATE | 5000 | LC-2 | 1.96 | 6.87 | 5.03 | 8.99 | 0.1 | 2.9 | 3.1 | 1.73 | -0.16 | 64 |
| ATE | 1 | OneF. | 0.97 | - | - | 1.30 | -0.3 | 3.4 | 3.0 | 0.60 | -0.07 | 38 |
| GATE | 2 | MCE. | 0.95 | - | - | 1.29 | -0.3 | 3.5 | 4.3 | 0.61 | -0.06 | 44 |
| GATE | 32 | Pen | 0.92 | 0.66 | 0.40 | 1.31 | -0.3 | 3.4 | 3.2 | 0.62 | -0.06 | 50 |
| IATE | 5000 | LC-5 | 2.55 | 6.87 | 4.33 | 11.6 | -0.1 | 3.1 | 5.8 | 1.46 | -0.18 | 44 |

Note: For GATE and IATE the *average bias* is the absolute value of the bias for the specific group (GATE) / observation (IATE) averaged over all groups / observation (each group / observation receives the same weight). *CovP (90%)* denotes the (average) probability that the true value is part of the 90% confidence interval. The simulation errors of the mean MSEs are around 0.1 (GATEs) -0.4 (IATEs).

*Appendix C.2.3: Earnings dependent individual effects*

The following tables contain the results for the IATEs that are dependent on insured earnings. A major difference to the other (non-zero) IATEs is that the earnings related IATEs have a much lower correlation with the propensity score. Thus, it should be 'easier' for the estimators to disentangle effect heterogeneity from selection bias.

*Table C.9: Simulation results for N=1,000, earnings dependent effect, and main estimators*

|  | Groups | Est. | True & estimated effects | | | Estimation error of effects (averages) | | | | | Estimation of std. error | |
|---|---|---|---|---|---|---|---|---|---|---|---|---|
|  |  |  | Avg. bias | X-sectional std. dev. | | MSE | Skew­ness | Kurt­osis | JB-Stat. | Std. err. | Avg. bias | CovP (90) in % |
|  | # |  |  | true | est. |  |  |  |  |  |  |  |
|  | (1) |  | (2) | (3) | (4) | (5) | (6) | (7) | (8) | (9) | (10) | (11) |
| ATE | 1 | Basic | 1.20 | - | - | 2.87 | 0.0 | 3.1 | 0.3 | 1.21 | 0.01 | 74 |
| GATE | 2 |  | 1.23 | - | - | 3.10 | 0.0 | 3.1 | 0.3 | 1.25 | 0.02 | 74 |
| GATE | 32 |  | 1.14 | 0.35 | 0.20 | 2.98 | 0.0 | 3.1 | 0.6 | 1.28 | 0.72 | 79 |
| IATE | 5000 |  | 1.28 | 1.71 | 1.39 | 6.38 | 0.0 | 3.1 | 3.6 | 2.00 | 0.34 | 87 |
| ATE | 1 | OneF. | 1.18 | - | - | 2.83 | -0.1 | 2.8 | 3.8 | 1.20 | 0.06 | 75 |
| GATE | 2 | VarT | 1.25 | - | - | 3.24 | -0.1 | 2.8 | 3.7 | 1.21 | 0.08 | 73 |
| GATE | 32 |  | 1.11 | 0.35 | 0.08 | 2.81 | -0.1 | 2.8 | 3.5 | 1.22 | 0.10 | 78 |
| IATE | 5000 |  | 1.53 | 1.71 | 0.67 | 5.88 | -0.1 | 3.0 | 2.4 | 1.58 | 0.50 | 83 |
| ATE | 1 | OneF. | 1.18 | - | - | 2.77 | 0.1 | 2.9 | 0.9 | 1.17 | 0.04 | 76 |
| GATE | 2 | MCE | 1.24 | - | - | 3.09 | 0.0 | 2.9 | 1.1 | 1.19 | 0.07 | 74 |
| GATE | 32 |  | 1.13 | 0.35 | 0.13 | 2.83 | 0.1 | 3.0 | 1.5 | 1.22 | 0.11 | 80 |
| IATE | 5000 |  | 1.41 | 1.71 | 0.80 | 5.48 | 0.0 | 3.1 | 2.8 | 1.65 | 0.54 | 87 |
| ATE | 1 | OneF. | 0.80 | - | - | 2.33 | -0.1 | 2.9 | 2.2 | 1.30 | 0.07 | 86 |
| GATE | 2 | MCE. | 0.86 | - | - | 2.69 | -0.1 | 2.7 | 2.2 | 1.32 | 0.08 | 84 |
| GATE | 32 | LC-2 | 0.74 | 0.35 | 0.10 | 2.46 | -0.1 | 2.9 | 1.8 | 1.36 | 0.08 | 86 |
| IATE | 5000 |  | 1.24 | 1.71 | 0.58 | 5.54 | 0.0 | 3.0 | 2.0 | 1.77 | 0.18 | 82 |
| ATE | 1 | OneF. | 0.56 | - | - | 2.09 | -0.1 | 3.0 | 1.9 | 1.33 | 0.23 | 92 |
| GATE | 2 | MCE. | 0.64 | - | - | 2.44 | -0.1 | 3.0 | 1.8 | 1.34 | 0.24 | 89 |
| GATE | 32 | Pen | 0.50 | 0.35 | 0.11 | 2.15 | -0.1 | 2.9 | 1.8 | 1.35 | 0.25 | 92 |
| IATE | 5000 |  | 1.37 | 1.71 | 1.05 | 6.70 | -0.1 | 3.0 | 4.1 | 1.94 | 0.67 | 89 |
| ATE | 1 | OneF. | 0.42 | - | - | 2.58 | 0.0 | 2.9 | 0.3 | 1.55 | 0.04 | 90 |
| GATE | 2 | MCE. | 0.51 | - | - | 2.93 | 0.0 | 3.0 | 0.3 | 1.56 | 0.05 | 87 |
| GATE | 32 | Pen | 0.37 | 0.35 | 0.09 | 2.70 | 0.0 | 3.0 | 0.5 | 1.58 | 0.04 | 90 |
| IATE | 5000 | LC-2 | 1.26 | 1.71 | 0.76 | 7.40 | 0.0 | 3.1 | 3.2 | 2.20 | 0.15 | 84 |
| ATE | 1 | OneF. | 0.36 | - | - | 1.93 | 0.1 | 3.1 | 2.7 | 1.34 | 0.05 | 90 |
| GATE | 2 | MCE. | 0.51 | - | - | 2.27 | 0.1 | 3.1 | 2.8 | 1.35 | 0.05 | 87 |
| GATE | 32 | Pen | 0.32 | 0.35 | 0.08 | 2.03 | 0.1 | 3.1 | 2.4 | 1.37 | 0.05 | 90 |
| IATE | 5000 | LC-5 | 1.22 | 1.71 | 0.69 | 6.15 | 0.0 | 3.2 | 6.0 | 1.91 | 0.12 | 82 |

Note: For GATE and IATE the *average bias* is the absolute value of the bias for the specific group (GATE) / observation (IATE) averaged over all groups / observation (each group / observation receives the same weight). *CovP (90%)* denotes the (average) probability that the true value is part of the 90% confidence interval. The simulation errors of the mean MSEs are around 0.1

*Table C.10: Simulation results for N=4,000, earnings dependent effect, and main estimators*

| | | | True & estimated effects | | | Estimation error of effects (averages) | | | | | Estimation of std. error | |
|---|---|---|---|---|---|---|---|---|---|---|---|---|
| | Groups | Est. | Avg. bias | X-sectional std. dev. | | MSE | Skew­ness | Kurt­osis | JB-Stat. | Std. err. | Avg. bias | CovP (90) in % |
| | # | | | true | est. | | | | | | | |
| | (1) | | (2) | (3) | (4) | (5) | (6) | (7) | (8) | (9) | (10) | (11) |
| ATE | 1 | Basic | 0.94 | - | - | 1.20 | -0.3 | 3.2 | 4.8 | 0.56 | 0.07 | 54 |
| GATE | 2 | | 0.96 | - | - | 1.27 | -0.3 | 3.1 | 3.8 | 0.59 | 0.08 | 58 |
| GATE | 32 | | 0.91 | 0.35 | 0.28 | 1.30 | -0.2 | 3.1 | 4.0 | 0.67 | 0.12 | 69 |
| IATE | 5000 | | 1.09 | 1.71 | 1.82 | 4.36 | -0.1 | 3.0 | 2.2 | 1.61 | 0.23 | 85 |
| ATE | 1 | OneF. | 1.06 | - | - | 1.55 | 0.3 | 3.0 | 5.2 | 0.61 | 0.08 | 56 |
| GATE | 2 | VarT | 1.14 | - | - | 1.86 | 0.3 | 3.0 | 3.8 | 0.63 | 0.09 | 54 |
| GATE | 32 | | 1.02 | 0.35 | 0.13 | 1.53 | 0.3 | 3.0 | 3.5 | 0.65 | 0.11 | 64 |
| IATE | 5000 | | 1.40 | 1.71 | 0.93 | 4.24 | 0.0 | 3.3 | 11.2 | 1.15 | 0.48 | 79 |
| ATE | 1 | OneF. | 0.87 | - | - | 1.03 | -0.2 | 2.9 | 1.2 | 0.53 | 0.10 | 62 |
| GATE | 2 | MCE | 0.90 | - | - | 1.17 | -0.2 | 2.9 | 1.6 | 0.55 | 0.12 | 65 |
| GATE | 32 | | 0.83 | 0.35 | 0.20 | 1.08 | -0.1 | 2.8 | 1.5 | 0.60 | 0.17 | 74 |
| IATE | 5000 | | 1.03 | 1.71 | 1.22 | 2.78 | 0.0 | 3.0 | 2.0 | 1.13 | 0.54 | 90 |
| ATE | 1 | OneF. | 0.55 | - | - | 0.74 | 0.0 | 3.2 | 0.3 | 0.66 | 0.03 | 82 |
| GATE | 2 | MCE. | 0.60 | - | - | 0.96 | 0.0 | 3.2 | 0.4 | 0.68 | 0.03 | 76 |
| GATE | 32 | LC-2 | 0.50 | 0.35 | 0.13 | 0.81 | 0.0 | 3.2 | 1.9 | 0.72 | 0.04 | 84 |
| IATE | 5000 | | 0.97 | 1.71 | 0.78 | 3.06 | 0.1 | 3.1 | 3.7 | 1.22 | 0.13 | 80 |
| ATE | 1 | OneF. | 0.21 | - | - | 0.75 | -0.3 | 2.9 | 3.3 | 0.84 | 0.13 | 93 |
| GATE | 2 | MCE. | 0.41 | - | - | 0.96 | -0.3 | 2.9 | 3.1 | 0.85 | 0.14 | 89 |
| GATE | 32 | Pen | 0.21 | 0.35 | 0.15 | 0.83 | -0.3 | 2.9 | 3.3 | 0.87 | 0.16 | 93 |
| IATE | 5000 | | 1.10 | 1.71 | 1.05 | 4.33 | -0.1 | 3.0 | 2.2 | 1.54 | 0.56 | 89 |
| ATE | 1 | OneF. | 0.16 | - | - | 0.73 | 0.0 | 3.0 | 0.0 | 0.85 | 0.05 | 90 |
| GATE | 2 | MCE. | 0.45 | - | - | 0.98 | 0.0 | 3.0 | 0.0 | 0.86 | 0.05 | 87 |
| GATE | 32 | Pen | 0.19 | 0.35 | 0.12 | 0.83 | 0.0 | 3.0 | 0.6 | 0.87 | 0.06 | 90 |
| IATE | 5000 | LC-2 | 1.07 | 1.71 | 0.82 | 4.58 | 0.0 | 3.0 | 2.3 | 1.62 | 0.16 | 82 |
| ATE | 1 | OneF. | 0.12 | - | - | 0.54 | -0.2 | 3.0 | 1.1 | 0.73 | 0.05 | 92 |
| GATE | 2 | MCE. | 0.45 | - | - | 0.78 | -0.2 | 3.0 | 1.1 | 0.73 | 0.05 | 86 |
| GATE | 32 | Pen | 0.19 | 0.35 | 0.13 | 0.63 | -0.2 | 3.0 | 1.6 | 0.75 | 0.06 | 91 |
| IATE | 5000 | LC-5 | 1.06 | 1.71 | 0.74 | 4.04 | 0.0 | 3.1 | 2.5 | 1.43 | 0.10 | 79 |

Note: For GATE and IATE the *average bias* is the absolute value of the bias for the specific group (GATE) / observation (IATE) averaged over all groups / observation (each group / observation receives the same weight). *CovP (90%)* denotes the (average) probability that the true value is part of the 90% confidence interval. The simulation errors of the mean MSEs are around 0.07.

*Table C.11: Simulation results for N=8,000, earnings dependent effect, and main estimators*

| | | | True & estimated effects | | | Estimation error of effects (averages) | | | | | Estimation of std. error | |
|---|---|---|---|---|---|---|---|---|---|---|---|---|
| | Groups # | Est. | Avg. bias | X-sectional std. dev. true | est. | MSE | Skewness | Kurtosis | JB-Stat. | Std. err. | Avg. bias | CovP (90) in % |
| | (1) | | (2) | (3) | (4) | (5) | (6) | (7) | (8) | (9) | (10) | (11) |
| ATE | 1 | Basic | 0.77 | - | - | 0.74 | -0.4 | 2.9 | 2.8 | 0.38 | 0.08 | 46 |
| GATE | 2 | | 0.78 | - | - | 0.77 | -0.4 | 2.9 | 2.7 | 0.40 | 0.09 | 50 |
| GATE | 32 | | 0.75 | 0.35 | 0.30 | 0.86 | 0.0 | 2.9 | 1.4 | 0.53 | 0.10 | 68 |
| IATE | 5000 | | 0.99 | 1.71 | 1.97 | 3.63 | 0.0 | 3.0 | 2.3 | 1.46 | 0.21 | 85 |
| ATE | 1 | OneF. | 0.69 | - | - | 0.73 | 0.4 | 3.1 | 2.9 | 0.51 | 0.12 | 78 |
| GATE | 2 | VarT | 0.75 | - | - | 1.00 | 0.3 | 3.1 | 2.6 | 0.52 | 0.13 | 66 |
| GATE | 32 | | 0.63 | 0.35 | 0.14 | 0.74 | 0.3 | 3.1 | 2.7 | 0.53 | 0.14 | 80 |
| IATE | 5000 | | 1.32 | 1.71 | 1.02 | 4.01 | 0.0 | 2.9 | 1.8 | 1.16 | 0.63 | 83 |
| ATE | 1 | OneF. | 0.73 | - | - | 0.68 | -0.2 | 3.0 | 1.2 | 0.39 | 0.06 | 51 |
| GATE | 2 | MCE | 0.76 | - | - | 0.77 | -0.2 | 3.0 | 1.0 | 0.41 | 0.08 | 54 |
| GATE | 32 | | 0.68 | 0.35 | 0.19 | 0.71 | -0.2 | 3.2 | 2.0 | 0.46 | 0.16 | 72 |
| IATE | 5000 | | 0.87 | 1.71 | 1.36 | 2.12 | 0.0 | 3.0 | 1.9 | 1.00 | 0.48 | 90 |
| ATE | 1 | OneF. | 0.51 | - | - | 0.50 | -0.1 | 2.8 | 0.4 | 0.50 | 0.00 | 73 |
| GATE | 2 | MCE. | 0.55 | - | - | 0.66 | -0.1 | 2.8 | 0.5 | 0.51 | 0.01 | 68 |
| GATE | 32 | LC-2 | 0.47 | 0.35 | 0.19 | 0.56 | -0.1 | 2.9 | 1.6 | 0.56 | 0.01 | 77 |
| IATE | 5000 | | 0.82 | 1.71 | 0.97 | 2.30 | 0.0 | 2.9 | 1.7 | 1.07 | 0.11 | 80 |
| ATE | 1 | OneF. | 0.06 | - | - | 0.28 | -0.3 | 3.7 | 4.5 | 0.53 | 0.24 | 98 |
| GATE | 2 | MCE. | 0.31 | - | - | 0.40 | -0.3 | 3.7 | 5.1 | 0.54 | 0.24 | 96 |
| GATE | 32 | Pen | 0.15 | 0.35 | 0.18 | 0.35 | -0.2 | 3.5 | 3.4 | 0.56 | 0.28 | 97 |
| IATE | 5000 | | 0.84 | 1.71 | 1.01 | 2.88 | 0.0 | 3.0 | 1.9 | 1.27 | 0.68 | 93 |
| ATE | 1 | OneF. | 0.00 | - | - | 0.44 | 0.0 | 2.7 | 0.6 | 0.67 | 0.01 | 90 |
| GATE | 2 | MCE. | 0.40 | - | - | 0.62 | 0.0 | 2.7 | 0.6 | 0.69 | 0.01 | 82 |
| GATE | 32 | Pen | 0.17 | 0.35 | 0.16 | 0.53 | 0.0 | 2.7 | 0.9 | 0.71 | 0.02 | 89 |
| IATE | 5000 | LC-2 | 0.94 | 1.71 | 0.72 | 3.75 | -0.1 | 2.9 | 1.7 | 1.57 | 0.10 | 82 |
| ATE | 1 | OneF. | 0.04 | - | - | 0.31 | 0.0 | 2.5 | 0.8 | 0.55 | 0.02 | 90 |
| GATE | 2 | MCE. | 0.40 | - | - | 0.48 | 0.0 | 2.6 | 0.6 | 0.56 | 0.02 | 80 |
| GATE | 32 | Pen | 0.17 | 0.35 | 0.14 | 0.39 | 0.0 | 2.7 | 0.8 | 0.58 | 0.02 | 89 |
| IATE | 5000 | LC-5 | 0.98 | 1.71 | 0.79 | 3.21 | 0.0 | 3.0 | 2.3 | 1.23 | 0.10 | 78 |

Note: For GATE and IATE the *average bias* is the absolute value of the bias for the specific group (GATE) / observation (IATE) averaged over all groups / observation (each group / observation receives the same weight). *CovP (90%)* denotes the (average) probability that the true value is part of the 90% confidence interval. The simulation errors of the mean MSEs are around 0.08.

## Appendix C.3 (Online): Further estimators

In this section, we present additional results for three further estimators using the DGPs with selectivity and the four specification of the IATEs. The first estimator is the *Basic* estimator in a configuration that might be considered standard: It uses the full sample for training (instead of the splitted half), but when building each tree, the respective subsample is splitted and 50% is used for building the tree, and the other 50% is used for computing the effects. This procedure has been termed 'honesty' in the literature (e.g., Wager and Athey, 2018).

*Table C.12: Simulation results for additional estimators: Main DGP with selectivity*

| | | | True & estimated effects | | | Estimation error of effects (averages) | | | | | Estimation of std. error | |
|---|---|---|---|---|---|---|---|---|---|---|---|---|
| | Groups | Est. | Avg. bias | X-sectional std. dev. | | MSE | Skewness | Kurtosis | JB-Stat. | Std. err. | Avg. bias | CovP (90) in % |
| | # | | | true | est. | | | | | | | |
| | (1) | | (2) | (3) | (4) | (5) | (6) | (7) | (8) | (9) | (10) | (11) |
| | | | | | | N = 1,000 | | | | | | |
| ATE | 1 | Basic. | 1.78 | - | - | 3.87 | 0.0 | 2.8 | 1.2 | 0.84 | 0.02 | 33 |
| GATE | 2 | One | 1.78 | - | - | 3.92 | 0.0 | 2.9 | 0.5 | 0.87 | 0.02 | 37 |
| GATE | 32 | Sam | 1.74 | 0.17 | 0.11 | 3.93 | 0.0 | 3.0 | 1.3 | 0.97 | 0.01 | 44 |
| IATE | 5000 | | 1.80 | 1.72 | 1.08 | 8.00 | 0.0 | 3.1 | 3.2 | 3.27 | -0.13 | 67 |
| ATE | 1 | OneF | 1.91 | - | - | 5.03 | 0.1 | 2.7 | 3.6 | 1.18 | 0.04 | 50 |
| GATE | 2 | | 1.90 | - | - | 5.04 | 0.0 | 2.7 | 3.2 | 1.20 | 0.06 | 52 |
| GATE | 32 | | 1.88 | 0.17 | 0.07 | 5.12 | 0.0 | 2.8 | 3.9 | 1.24 | 0.13 | 57 |
| IATE | 5000 | | 1.94 | 1.72 | 0.59 | 8.64 | 0.0 | 3.0 | 1.8 | 1.71 | 0.63 | 76 |
| ATE | 1 | OneF. | 1.50 | - | - | 4.06 | 0.0 | 3.1 | 0.20 | 1.34 | 0.09 | 71 |
| GATE | 2 | VarT. | 1.50 | - | - | 4.09 | 0.0 | 3.1 | 0.21 | 1.35 | 0.10 | 72 |
| GATE | 32 | Pen | 1.49 | 0.17 | 0.16 | 4.06 | 0.0 | 3.1 | 0.53 | 1.36 | 0.12 | 73 |
| IATE | 5000 | | 1.51 | 1.72 | 1.26 | 6.40 | -0.1 | 3.5 | 42.5* | 1.89 | 0.53 | 89 |
| | | | | | | N = 4,000 | | | | | | |
| ATE | 1 | Basic. | 1.45 | - | - | 2.30 | 0.2 | 2.8 | 1.3 | 0.43 | 0.03 | 15 |
| GATE | 2 | One | 1.46 | - | - | 2.34 | 0.1 | 2.9 | 1.1 | 0.45 | 0.03 | 19 |
| GATE | 32 | Sam | 1.42 | 0.17 | 0.13 | 2.41 | 0.1 | 2.9 | 1.7 | 0.60 | -0.02 | 35 |
| IATE | 5000 | | 1.56 | 1.72 | 1.38 | 6.06 | 0.0 | 3.0 | 2.3 | 1.58 | -0.23 | 76 |
| ATE | 1 | OneF | 1.60 | - | - | 2.90 | -0.2 | 3.2 | 2.5 | 0.58 | 0.07 | 15 |
| GATE | 2 | | 1.60 | - | - | 2.90 | -0.2 | 3.2 | 2.1 | 0.60 | 0.08 | 19 |
| GATE | 32 | | 1.58 | 0.17 | 0.08 | 2.97 | -0.1 | 3.2 | 1.7 | 0.67 | 0.16 | 35 |
| IATE | 5000 | | 1.66 | 1.72 | 0.74 | 5.91 | 0.0 | 3.0 | 2.2 | 1.25 | 0.58 | 76 |
| ATE | 1 | OneF. | 1.21 | - | - | 1.93 | 0.2 | 3.6 | 5.6 | 0.70 | 0.12 | 58 |
| GATE | 2 | VarT. | 1.20 | - | - | 1.94 | 0.2 | 3.6 | 5.0 | 0.71 | 0.13 | 61 |
| GATE | 32 | Pen | 1.19 | 0.17 | 0.13 | 1.62 | 0.2 | 3.6 | 5.6 | 0.71 | 0.14 | 63 |
| IATE | 5000 | | 1.24 | 1.72 | 1.61 | 3.78 | 0.0 | 2.9 | 1.7 | 1.34 | 0.62 | 88 |
| | | | | | | N = 8,000 | | | | | | |
| ATE | 1 | Basic. | 1.34 | - | - | 1.92 | 0.2 | 2.7 | 1.5 | 0.37 | -0.04 | 0 |
| GATE | 2 | One | 1.35 | - | - | 1.96 | 0.2 | 2.9 | 0.9 | 0.39 | -0.04 | 3 |
| GATE | 32 | Sam | 1.30 | 0.17 | 0.14 | 1.97 | 0.1 | 3.0 | 1.5 | 0.49 | -0.04 | 14 |
| IATE | 5000 | | 1.47 | 1.72 | 1.42 | 4.84 | 0.0 | 2.9 | 1.8 | 1.31 | -0.18 | 59 |
| ATE | 1 | OneF | 0.92 | - | - | 1.08 | 0.1 | 3.4 | 0.8 | 0.48 | 0.02 | 43 |
| GATE | 2 | | 0.92 | - | - | 1.09 | 0.0 | 3.3 | 0.7 | 0.50 | 0.02 | 45 |
| GATE | 32 | | 0.90 | 0.17 | 0.06 | 1.09 | 0.0 | 2.0 | 2.0 | 0.57 | 0.05 | 52 |
| IATE | 5000 | | 1.18 | 1.72 | 0.62 | 3.56 | 0.0 | 1.1 | 1.9 | 1.19 | 0.13 | 70 |
| ATE | 1 | OneF. | 1.05 | - | - | 1.32 | 0.2 | 2.9 | 0.6 | 0.46 | 0.16 | 46 |
| GATE | 2 | VarT. | 1.04 | - | - | 1.32 | 0.2 | 2.9 | 0.6 | 0.47 | 0.16 | 52 |
| GATE | 32 | Pen | 1.04 | 0.17 | 0.14 | 1.32 | 0.1 | 2.9 | 0.7 | 0.48 | 0.17 | 55 |
| IATE | 5000 | | 1.10 | 1.72 | 1.77 | 3.07 | 0.0 | 2.9 | 1.7 | 1.18 | 0.61 | 89 |

Note: For GATE and IATE the *average bias* is the absolute value of the bias for the specific group (GATE) / observation (IATE) averaged over all groups / observation (each group / observation receives the same weight). *CovP (90%)* denotes the (average) probability that the true value is part of the 90% confidence interval. *OneF.VarT.Penalty:* Baseline penalty multiplied by 100.

The second estimator is *OneF*. It is identical to *OneF.MCE* when setting MCE to zero. Therefore, it has some small computational advantages compared to *OneF.MCE* and *OneF.MCE.Penalty*. The third estimator is *OneF.VarT* with and added penalty function. To economize on computation costs. For the latter two estimators we do report results for *N=8,000* in selected DGPs.

*Table C.13: Simulation results for additional estimators: DGP with no effect and selectivity*

| | Groups # | Est. | Avg. bias | X-sectional std. dev. true | X-sectional std. dev. est. | MSE | Skew­ness | Kurt­osis | JB-Stat. | Std. err. | Avg. bias | CovP (90) in % |
|---|---|---|---|---|---|---|---|---|---|---|---|---|
| | (1) | | (2) | (3) | (4) | (5) | (6) | (7) | (8) | (9) | (10) | (11) |
| | | | | | | **N = 1,000** | | | | | | |
| ATE | 1 | Basic. | 1.01 | - | - | 1.68 | 0.0 | 3.2 | 0.7 | 0.81 | 0.05 | 69 |
| GATE | 2 | One | 1.02 | - | - | 1.75 | 0.0 | 3.2 | 1.6 | 0.84 | 0.04 | 69 |
| GATE | 32 | Sam | 0.97 | 0 | 0.21 | 1.89 | 0.0 | 3.1 | 0.9 | 0.96 | 0.01 | 74 |
| IATE | 5000 | | 1.06 | 0 | 0.93 | 4.71 | 0.0 | 3.0 | 2.5 | 1.79 | -0.09 | 79 |
| ATE | 1 | OneF | 1.08 | - | - | 2.34 | -0.1 | 2.9 | 2.3 | 1.07 | 0.12 | 78 |
| GATE | 2 | | 1.09 | - | - | 2.38 | -0.1 | 2.9 | 2.0 | 1.09 | 0.14 | 79 |
| GATE | 32 | | 1.08 | 0 | 0.07 | 2.49 | -0.1 | 2.9 | 1.6 | 1.15 | 0.21 | 83 |
| IATE | 5000 | | 1.09 | 0 | 0.43 | 4.04 | -0.1 | 3.0 | 2.8 | 1.63 | 0.68 | 94 |
| ATE | 1 | OneF. | 0.96 | - | - | 2.46 | -0.1 | 3.1 | 2.6 | 1.24 | 0.12 | 84 |
| GATE | 2 | VarT. | 0.96 | - | - | 2.48 | -0.1 | 3.1 | 2.8 | 1.25 | 0.13 | 84 |
| GATE | 32 | Pen | 0.96 | 0 | 0.04 | 2.51 | -0.1 | 3.1 | 2.9 | 1.26 | 0.14 | 85 |
| IATE | 5000 | | 0.97 | 0 | 0.56 | 4.07 | -0.1 | 3.0 | 11.3 | 1.68 | 0.57 | 93 |
| | | | | | | **N = 4,000** | | | | | | |
| ATE | 1 | Basic. | 0.85 | - | - | 0.89 | 0.3 | 3.0 | 4.0 | 0.41 | 0.04 | 39 |
| GATE | 2 | One | 0.86 | - | - | 0.94 | 0.3 | 3.1 | 3.6 | 0.44 | 0.03 | 43 |
| GATE | 32 | Sam | 0.79 | 0 | 0.21 | 1.04 | 0.1 | 3.1 | 2.3 | 0.61 | -0.02 | 59 |
| IATE | 5000 | | 1.05 | 0 | 0.93 | 3.89 | 0.0 | 3.0 | 2.4 | 1.51 | -0.18 | 73 |
| ATE | 1 | OneF | 0.91 | - | - | 1.16 | -0.3 | 3.1 | 3.6 | 0.57 | 0.05 | 56 |
| GATE | 2 | | 0.90 | - | - | 1.18 | -0.3 | 3.0 | 3.7 | 0.59 | 0.06 | 56 |
| GATE | 32 | | 0.91 | - | - | 1.24 | -0.2 | 3.0 | 2.6 | 0.65 | 0.17 | 73 |
| IATE | 5000 | | 0.91 | - | 0.51 | 2.59 | -0.0 | 2.9 | 1.7 | 1.22 | 0.58 | 94 |
| ATE | 1 | OneF. | 0.86 | - | - | 1.22 | 0.0 | 3.0 | 0.0 | 0.69 | 0.13 | 77 |
| GATE | 2 | VarT. | 0.86 | - | - | 1.23 | 0.0 | 3.0 | 0.3 | 0.70 | 0.14 | 77 |
| GATE | 32 | Pen | 0.85 | 0 | 0.05 | 1.23 | 0.0 | 3.0 | 0.4 | 0.71 | 0.15 | 79 |
| IATE | 5000 | | 1.00 | 0 | 0.71 | 2.96 | 0.0 | 3.0 | 1.3 | 1.30 | 0.62 | 92 |

Note: For GATE and IATE the *average bias* is the absolute value of the bias for the specific group (GATE) / observation (IATE) averaged over all groups / observation (each group / observation receives the same weight). *CovP (90%)* denotes the (average) probability that the true value is part of the 90% confidence interval. *OneF.VarT.Penalty:* Baseline penalty multiplied by 100.

*Table C.14: Simulation results for additional estimators: DGP with strong effect and selectivity*

| | Groups # | Est. | Avg. bias | X-sectional std. dev. true | X-sectional std. dev. est. | MSE | Skewness | Kurtosis | JB-Stat. | Std. err. | Avg. bias | CovP (90) in % |
|---|---|---|---|---|---|---|---|---|---|---|---|---|
| | (1) | | (2) | (3) | (4) | (5) | (6) | (7) | (8) | (9) | (10) | (11) |
| | | | | | | **N = 1,000** | | | | | | |
| ATE | 1 | Basic. | 2.94 | - | - | 9.49 | 0.1 | 3.2 | 2.5 | 0.93 | -0.03 | 33 |
| GATE | 2 | One | 2.94 | - | - | 9.54 | 0.1 | 3.1 | 1.6 | 0.95 | -0.03 | 37 |
| GATE | 32 | Sam | 2.90 | 0.67 | 0.45 | 9.59 | 0.1 | 3.2 | 6.4 | 1.03 | -0.03 | 44 |
| IATE | 5000 | | 3.03 | 6.86 | 4.24 | 24.10 | 0.0 | 3.0 | 2.6 | 2.05 | -0.18 | 67 |
| ATE | 1 | OneF | 3.20 | - | - | 12.01 | 0.0 | 2.9 | 0.5 | 1.32 | 0.12 | 27 |
| GATE | 2 | | 3.19 | - | - | 11.93 | 0.0 | 2.9 | 0.5 | 1.33 | 0.14 | 29 |
| GATE | 32 | | 3.15 | 0.66 | 0.27 | 11.91 | 0.0 | 2.9 | 0.5 | 1.36 | 0.18 | 34 |
| IATE | 5000 | | 3.98 | 6.87 | 2.75 | 34.22 | 0.0 | 3.0 | 2.1 | 2.04 | 0.54 | 61 |
| ATE | 1 | OneF. | 1.79 | - | - | 5.64 | 0.0 | 2.8 | 2.2 | 1.56 | 0.21 | 74 |
| GATE | 2 | VarT. | 1.79 | - | - | 5.62 | 0.0 | 2.8 | 2.2 | 1.56 | 0.21 | 75 |
| GATE | 32 | Pen | 1.75 | 0.66 | 0.48 | 5.58 | 0.0 | 2.8 | 2.2 | 1.57 | 0.23 | 76 |
| IATE | 5000 | | 2.38 | 6.87 | 5.14 | 14.62 | 0.0 | 3.0 | 2.4 | 2.44 | 0.63 | 82 |
| | | | | | | **N = 4,000** | | | | | | |
| ATE | 1 | Basic. | 2.10 | - | - | 4.65 | 0.0 | 3.1 | 0.1 | 0.49 | -0.01 | 1 |
| GATE | 2 | One | 2.11 | - | - | 4.72 | 0.0 | 3.1 | 0.6 | 0.51 | -0.01 | 1 |
| GATE | 32 | Sam | 2.10 | 0.67 | 0.56 | 4.81 | 0.0 | 3.0 | 1.5 | 0.61 | -0.01 | 4 |
| IATE | 5000 | | 2.31 | 6.86 | 5.27 | 13.74 | 0.0 | 3.0 | 2.1 | 1.62 | -0.18 | 57 |
| ATE | 1 | OneF | 2.02 | - | - | 4.61 | -0.1 | 2.9 | 0.7 | 0.71 | 0.14 | 18 |
| GATE | 2 | | 2.02 | - | - | 4.58 | -0.1 | 2.9 | 0.7 | 0.72 | 0.15 | 21 |
| GATE | 32 | | 1.98 | 0.66 | 0.42 | 4.61 | -0.1 | 2.9 | 1.4 | 0.76 | 0.19 | 29 |
| IATE | 5000 | | 2.56 | 6.87 | 4.13 | 16.44 | 0.0 | 2.9 | 1.8 | 1.49 | 0.52 | 67 |
| ATE | 1 | OneF. | 0.88 | - | - | 1.64 | 0.3 | 2.9 | 4.0 | 0.93 | 0.07 | 82 |
| GATE | 2 | VarT. | 0.87 | - | - | 1.64 | 0.3 | 2.9 | 4.0 | 0.93 | 0.09 | 82 |
| GATE | 32 | Pen | 0.85 | 0.66 | 0.56 | 1.64 | 0.3 | 2.9 | 3.7 | 0.95 | 0.11 | 82 |
| IATE | 5000 | | 1.37 | 6.87 | 5.91 | 6.61 | 0.2 | 3.2 | 4.8 | 1.83 | 0.46 | 89 |

Note: For GATE and IATE the *average bias* is the absolute value of the bias for the specific group (GATE) / observation (IATE) averaged over all groups / observation (each group / observation receives the same weight). *CovP (90%)* denotes the (average) probability that the true value is part of the 90% confidence interval. *OneF.VarT.Penalty:* Baseline penalty multiplied by 100.

*Table C.15: Simulation results for additional estimators: DGP with earnings dependent effect and selectivity*

|  | Groups # | Est. | True & estimated effects | | | Estimation error of effects (averages) | | | | | Estimation of std. error | |
|---|---|---|---|---|---|---|---|---|---|---|---|---|
|  |  |  | Avg. bias | X-sectional std. dev. | | MSE | Skewness | Kurtosis | JB-Stat. | Std. err. | Avg. bias | CovP (90) in % |
|  |  |  |  | true | est. |  |  |  |  |  |  |  |
|  | (1) |  | (2) | (3) | (4) | (5) | (6) | (7) | (8) | (9) | (10) | (11) |
| \multicolumn{12}{c}{N = 1,000} | | | | | | | | | | | |
| ATE | 1 | Basic. | 1.10 | - | - | 1.86 | 0.0 | 3.0 | 0.1 | 0.80 | -0.08 | 65 |
| GATE | 2 | One | 1.14 | - | - | 2.04 | 0.0 | 2.9 | 0.3 | 0.84 | -0.07 | 65 |
| GATE | 32 | Sam | 1.06 | 0.35 | 0.22 | 1.99 | 0.0 | 3.0 | 0.4 | 0.91 | -0.05 | 70 |
| IATE | 5000 |  | 1.20 | 1.71 | 1.45 | 4.83 | 0.0 | 3.0 | 2.1 | 1.65 | -0.10 | 76 |
| ATE | 1 | OneF | 1.09 | - | - | 2.44 | -0.1 | 2.9 | 1.2 | 1.11 | 0.10 | 78 |
| GATE | 2 |  | 1.14 | - | - | 2.70 | -0.1 | 2.9 | 1.2 | 1.14 | 0.12 | 80 |
| GATE | 32 |  | 1.05 | 0.35 | 0.15 | 2.54 | 0.0 | 2.9 | 0.6 | 1.18 | 0.19 | 83 |
| IATE | 5000 |  | 1.28 | 1.71 | 1.02 | 5.20 | 0.0 | 3.1 | 2.5 | 1.71 | 0.66 | 91 |
| ATE | 1 | OneF. | 0.97 | - | - | 2.47 | -0.1 | 2.7 | 5.9 | 1.23 | 0.14 | 83 |
| GATE | 2 | VarT. | 1.04 | - | - | 2.88 | -0.1 | 2.7 | 5.6 | 1.25 | 0.15 | 81 |
| GATE | 32 | Pen | 0.90 | 0.35 | 0.09 | 2.47 | -0.1 | 2.7 | 5.9 | 1.26 | 0.16 | 85 |
| IATE | 5000 |  | 1.47 | 1.71 | 0.78 | 6.17 | -0.1 | 3.1 | 7.5 | 1.71 | 0.58 | 85 |
| \multicolumn{12}{c}{N = 4,000} | | | | | | | | | | | |
| ATE | 1 | Basic. | 0.88 | - | - | 0.92 | -0.3 | 3.1 | 2.8 | 0.44 | 0.06 | 34 |
| GATE | 2 | One | 0.89 | - | - | 0.99 | -0.2 | 3.1 | 6.1 | 0.47 | 0.04 | 38 |
| GATE | 32 | Sam | 0.85 | 0.35 | 0.30 | 1.06 | -0.2 | 3.1 | 4.3 | 0.56 | 0.01 | 52 |
| IATE | 5000 |  | 1.05 | 1.71 | 1.84 | 3.79 | 0.0 | 3.0 | 2.1 | 1.25 | -0.20 | 71 |
| ATE | 1 | OneF | 0.81 | - | - | 0.99 | 0.0 | 3.6 | 3.8 | 0.57 | 0.05 | 68 |
| GATE | 2 |  | 0.84 | - | - | 1.11 | 0.0 | 3.5 | 3.2 | 0.60 | 0.07 | 66 |
| GATE | 32 |  | 0.77 | 0.35 | 0.21 | 1.06 | -0.1 | 3.4 | 3.4 | 0.66 | 0.16 | 80 |
| IATE | 5000 |  | 0.96 | 1.71 | 1.33 | 2.94 | 0.0 | 3.1 | 2.2 | 1.25 | 0.59 | 92 |
| ATE | 1 | OneF. | 0.75 | - | - | 1.00 | -0.3 | 2.9 | 4.1 | 0.82 | 0.16 | 79 |
| GATE | 2 | VarT. | 0.81 | - | - | 1.32 | -0.3 | 3.0 | 4.5 | 0.84 | 0.16 | 74 |
| GATE | 32 | Pen | 0.68 | 0.35 | 0.12 | 1.01 | -0.3 | 2.9 | 3.5 | 0.86 | 0.17 | 83 |
| IATE | 5000 |  | 1.40 | 1.71 | 1.04 | 4.63 | 0.0 | 3.0 | 1.6 | 1.93 | 0.65 | 83 |

Note: For GATE and IATE the *average bias* is the absolute value of the bias for the specific group (GATE) / observation (IATE) averaged over all groups / observation (each group / observation receives the same weight). *CovP (90%)* denotes the (average) probability that the true value is part of the 90% confidence interval. *OneF.VarT.Penalty:* Baseline penalty multiplied by 100.

# Appendix D (online): Simulation results for the experimental case

In this section, we consider the case without selection bias, i.e., where treatment is fully randomized with a propensity score of 0.5 for all observations. To save computation time, the cases of *N = 8,000* and LC-5 are omitted.

*Table D.1: Simulation results for N=1,000, main DGP, no selectivity, and main estimators*

|      | Groups # | Est. | True & estimated effects | | | Estimation error of effects (averages) | | | | | Estimation of std. error | |
|------|----------|------|--------|--------|--------|------|------|------|------|------|------|------|
|      |          |      | Avg. bias | X-sectional std. dev. true | X-sectional std. dev. est. | MSE | Skewness | Kurtosis | JB-Stat. | Std. err. | Avg. bias | CovP (90) in % |
|      | (1)      |      | (2)    | (3)    | (4)    | (5)  | (6)  | (7)  | (8)  | (9)  | (10) | (11) |
| ATE  | 1    | Basic    | -0.01 | -    | -    | 1.23 | -0.2 | 3.1 | 7.3 | 1.11 | 0.07  | 92 |
| GATE | 2    |          | 0.03  | -    | -    | 1.29 | -0.2 | 3.1 | 7.1 | 1.13 | 0.09  | 92 |
| GATE | 32   |          | 0.07  | 0.17 | 0.12 | 1.42 | -0.2 | 3.1 | 5.1 | 1.19 | 0.12  | 93 |
| IATE | 5000 |          | 0.82  | 1.72 | 0.83 | 4.49 | -0.1 | 3.1 | 5.1 | 1.89 | 0.43  | 92 |
| ATE  | 1    | OneF. | -0.02 | -    | -    | 1.28 | 0.1  | 2.9 | 2.7 | 1.13 | 0.05  | 92 |
| GATE | 2    | VarT  | 0.05  | -    | -    | 1.31 | 0.1  | 2.9 | 2.5 | 1.15 | 0.07  | 92 |
| GATE | 32   |       | 0.10  | 0.17 | 0.06 | 1.39 | 0.1  | 2.9 | 2.0 | 1.17 | 0.09  | 92 |
| IATE | 5000 |       | 1.10  | 1.72 | 0.51 | 3.98 | 0.1  | 2.9 | 2.2 | 1.56 | 0.54  | 91 |
| ATE  | 1    | OneF. | -0.05 | -    | -    | 1.41 | -0.1 | 3.1 | 1.4 | 1.19 | -0.01 | 91 |
| GATE | 2    | MCE   | 0.06  | -    | -    | 1.45 | -0.1 | 3.1 | 1.3 | 1.20 | 0.01  | 91 |
| GATE | 32   |       | 0.10  | 0.17 | 0.06 | 1.54 | -0.1 | 3.1 | 1.1 | 1.23 | 0.07  | 92 |
| IATE | 5000 |       | 1.04  | 1.72 | 0.57 | 4.19 | 0.0  | 3.0 | 1.9 | 1.67 | 0.57  | 92 |
| ATE  | 1    | OneF. | -0.05 | -    | -    | 1.68 | 0.0  | 2.8 | 2.8 | 1.29 | 0.06  | 92 |
| GATE | 2    | MCE.  | 0.06  | -    | -    | 1.72 | 0.0  | 2.8 | 2.1 | 1.31 | 0.07  | 92 |
| GATE | 32   | LC-2  | 0.11  | 0.17 | 0.05 | 1.82 | 0.0  | 2.9 | 1.5 | 1.34 | 0.08  | 92 |
| IATE | 5000 |       | 1.08  | 1.72 | 0.53 | 4.83 | 0.0  | 3.0 | 1.7 | 1.82 | 0.17  | 86 |
| ATE  | 1    | OneF. | 0.02  | -    | -    | 1.28 | -0.1 | 2.8 | 4.2 | 1.13 | 0.05  | 91 |
| GATE | 2    | MCE.  | 0.05  | -    | -    | 1.32 | -0.1 | 2.8 | 3.6 | 1.15 | 0.07  | 91 |
| GATE | 32   | Pen   | 0.10  | 0.17 | 0.06 | 1.41 | -0.1 | 2.9 | 4.0 | 1.18 | 0.12  | 93 |
| IATE | 5000 |       | 1.06  | 1.72 | 0.54 | 4.03 | 0.0  | 3.0 | 2.2 | 1.60 | 0.58  | 92 |
| ATE  | 1    | OneF. | -0.08 | -    | -    | 1.85 | -0.1 | 3.0 | 1.0 | 1.36 | 0.00  | 90 |
| GATE | 2    | MCE.  | 0.08  | -    | -    | 1.89 | -0.1 | 3.0 | 0.9 | 1.37 | 0.00  | 90 |
| GATE | 32   | Pen   | 0.12  | 0.17 | 0.05 | 2.01 | -0.1 | 2.9 | 1.3 | 1.42 | 0.01  | 89 |
| IATE | 5000 | LC-2  | 1.13  | 1.72 | 0.48 | 5.05 | 0.0  | 3.0 | 1.9 | 1.97 | 0.12  | 85 |

Note: For GATE and IATE the *average bias* is the absolute value of the bias for the specific group (GATE) / observation (IATE) averaged over all groups / observation (each group / observation receives the same weight). *CovP (90%)* denotes the (average) probability that the true value is part of the 90% confidence interval. The simulation errors of the mean MSEs are around 0.06.

*Table D.2: Simulation results for N=4,000, main DGP, no selectivity, and main estimators*

|  | Groups # | Est. | True & estimated effects | | | Estimation error of effects (averages) | | | | | Estimation of std. error | |
|---|---|---|---|---|---|---|---|---|---|---|---|---|
|  |  |  | Avg. bias | X-sectional std. dev. true | X-sectional std. dev. est. | MSE | Skewness | Kurtosis | JB-Stat. | Std. err. | Avg. bias | CovP (90) in % |
|  | (1) |  | (2) | (3) | (4) | (5) | (6) | (7) | (8) | (9) | (10) | (11) |
| ATE | 1 | Basic | 0.06 | - | - | 0.29 | 0.2 | 3.0 | 1.9 | 0.53 | 0.06 | 94 |
| GATE | 2 |  | 0.06 | - | - | 0.31 | 0.2 | 3.0 | 2.0 | 0.55 | 0.07 | 94 |
| GATE | 32 |  | 0.07 | 0.17 | 0.15 | 0.43 | 0.1 | 3.0 | 1.0 | 0.65 | 0.10 | 95 |
| IATE | 5000 |  | 0.44 | 1.72 | 1.34 | 2.81 | 0.0 | 3.0 | 2.2 | 1.56 | 0.28 | 92 |
| ATE | 1 | OneF. | -0.1 | - | - | 0.34 | -0.1 | 2.7 | 1.5 | 0.58 | 0.01 | 90 |
| GATE | 2 | VarT | 0.03 | - | - | 0.36 | -0.2 | 2.7 | 1.7 | 0.60 | 0.03 | 91 |
| GATE | 32 |  | 0.08 | 0.17 | 0.08 | 0.39 | -0.1 | 2.7 | 2.6 | 0.62 | 0.05 | 92 |
| IATE | 5000 |  | 0.82 | 1.72 | 0.84 | 2.10 | 0.0 | 3.0 | 1.9 | 1.10 | 0.51 | 93 |
| ATE | 1 | OneF. | -0.01 | - | - | 0.32 | 0.1 | 2.9 | 0.3 | 0.57 | 0.02 | 92 |
| GATE | 2 | MCE | 0.05 | - | - | 0.34 | 0.1 | 3.0 | 0.6 | 0.58 | 0.03 | 92 |
| GATE | 32 |  | 0.06 | 0.17 | 0.09 | 0.41 | 0.1 | 2.9 | 0.9 | 0.63 | 0.12 | 94 |
| IATE | 5000 |  | 0.72 | 1.72 | 0.93 | 2.15 | 0.0 | 3.0 | 2.1 | 1.19 | 0.54 | 94 |
| ATE | 1 | OneF. | 0.02 | - | - | 0.46 | 0.1 | 2.9 | 0.3 | 0.68 | -0.01 | 91 |
| GATE | 2 | MCE. | 0.04 | - | - | 0.48 | 0.1 | 2.9 | 0.8 | 0.69 | -0.01 | 90 |
| GATE | 32 | LC-2 | 0.09 | 0.17 | 0.07 | 0.55 | 0.0 | 2.9 | 0.6 | 0.73 | 0.00 | 91 |
| IATE | 5000 |  | 0.88 | 1.72 | 0.76 | 2.64 | 0.0 | 3.0 | 2.2 | 1.27 | 0.13 | 84 |
| ATE | 1 | OneF. | 0.03 | - | - | 0.32 | 0.0 | 2.8 | 0.4 | 0.57 | 0.02 | 90 |
| GATE | 2 | MCE. | 0.04 | - | - | 0.34 | 0.0 | 2.9 | 0.6 | 0.58 | 0.04 | 91 |
| GATE | 32 | Pen | 0.08 | 0.17 | 0.08 | 0.41 | -0.1 | 2.9 | 0.9 | 0.63 | 0.11 | 93 |
| IATE | 5000 |  | 0.78 | 1.72 | 0.86 | 2.17 | 0.0 | 3.0 | 2.0 | 1.15 | 0.52 | 93 |
| ATE | 1 | OneF. | 0.00 | - | - | 0.44 | -0.1 | 2.8 | 0.7 | 0.67 | 0.01 | 88 |
| GATE | 2 | MCE. | 0.03 | - | - | 0.46 | -0.1 | 2.8 | 0.7 | 0.69 | 0.01 | 89 |
| GATE | 32 | Pen | 0.08 | 0.16 | 0.08 | 0.54 | -0.1 | 2.9 | 1.0 | 0.74 | 0.02 | 89 |
| IATE | 5000 | LC-2 | 0.90 | 1.72 | 0.76 | 2.68 | 0.0 | 3.0 | 2.3 | 1.42 | 0.14 | 83 |

Note: For GATE and IATE the *average bias* is the absolute value of the bias for the specific group (GATE) / observation (IATE) averaged over all groups / observation (each group / observation receives the same weight). *CovP (90%)* denotes the (average) probability that the true value is part of the 90% confidence interval. The simulation errors of the mean MSEs are around 0.06.

*Table D.3: Simulation results for N=1,000, no effect, no selectivity, and main estimators*

| | Groups # | Est. | True & estimated effects | | | Estimation error of effects (averages) | | | | | Estimation of std. error | |
|---|---|---|---|---|---|---|---|---|---|---|---|---|
| | | | Avg. bias | X-sectional std. dev. | | MSE | Skewness | Kurtosis | JB-Stat. | Std. err. | Avg. bias | CovP (90) in % |
| | | | | true | est. | | | | | | | |
| | (1) | | (2) | (3) | (4) | (5) | (6) | (7) | (8) | (9) | (10) | (11) |
| ATE | 1 | Basic | 0.02 | - | - | 1.27 | -0.1 | 3.0 | 1.8 | 1.13 | 0.04 | 91 |
| GATE | 2 | | 0.02 | - | - | 1.32 | -0.1 | 3.0 | 1.7 | 1.15 | 0.05 | 91 |
| GATE | 32 | | 0.02 | 0 | 0.01 | 1.47 | -0.1 | 3.0 | 2.8 | 1.21 | 0.09 | 92 |
| IATE | 5000 | | 0.04 | 0 | 0.04 | 3.36 | -0.1 | 3.1 | 3.1 | 1.83 | 0.45 | 96 |
| ATE | 1 | OneF. | 0.00 | - | - | 1.26 | 0.0 | 3.1 | 0.7 | 1.12 | 0.04 | 92 |
| GATE | 2 | VarT | 0.00 | - | - | 1.29 | 0.0 | 3.1 | 1.1 | 1.14 | 0.07 | 92 |
| GATE | 32 | | 0.02 | 0 | 0.01 | 1.34 | 0.0 | 3.1 | 0.9 | 1.16 | 0.09 | 92 |
| IATE | 5000 | | 0.02 | 0 | 0.03 | 2.36 | 0.0 | 3.0 | 1.9 | 1.53 | 0.53 | 97 |
| ATE | 1 | OneF. | -0.04 | - | - | 1.24 | 0.0 | 3.2 | 2.3 | 1.11 | 0.05 | 92 |
| GATE | 2 | MCE | 0.04 | - | - | 1.28 | 0.0 | 3.2 | 2.0 | 1.13 | 0.07 | 92 |
| GATE | 32 | | 0.04 | 0 | 0.01 | 1.38 | 0.0 | 3.2 | 2.8 | 1.18 | 0.13 | 93 |
| IATE | 5000 | | 0.04 | 0 | 0.03 | 2.56 | 0.0 | 3.1 | 2.0 | 1.59 | 0.60 | 97 |
| ATE | 1 | OneF. | -0.08 | - | - | 1.73 | 0.1 | 3.1 | 2.3 | 1.31 | 0.03 | 91 |
| GATE | 2 | MCE. | 0.08 | - | - | 1.78 | 0.1 | 3.2 | 2.7 | 1.33 | 0.04 | 91 |
| GATE | 32 | LC-2 | 0.08 | 0 | 0.00 | 1.88 | 0.1 | 3.1 | 2.6 | 1.37 | 0.05 | 92 |
| IATE | 5000 | | 0.08 | 0 | 0.04 | 3.31 | 0.0 | 3.8 | 3.8 | 1.82 | 0.16 | 93 |
| ATE | 1 | OneF. | 0.03 | - | - | 1.32 | 0.0 | 2.9 | 0.9 | 1.15 | 0.01 | 91 |
| GATE | 2 | MCE. | 0.03 | - | - | 1.35 | 0.0 | 2.9 | 1.0 | 1.16 | 0.03 | 91 |
| GATE | 32 | Pen | 0.03 | 0 | 0.05 | 1.43 | 0.0 | 2.9 | 0.6 | 1.20 | 0.09 | 93 |
| IATE | 5000 | | 0.03 | 0 | 0.03 | 2.54 | 0.0 | 2.9 | 1.9 | 1.59 | 0.57 | 97 |
| ATE | 1 | OneF. | 0.00 | - | - | 1.73 | -0.1 | 3.0 | 3.0 | 1.32 | 0.03 | 91 |
| GATE | 2 | MCE. | 0.01 | - | - | 1.78 | -0.1 | 3.0 | 2.8 | 1.34 | 0.03 | 91 |
| GATE | 32 | Pen | 0.01 | 0 | 0.01 | 1.89 | -0.1 | 3.0 | 3.1 | 1.37 | 0.04 | 90 |
| IATE | 5000 | LC-2 | 0.03 | 0 | 0.04 | 3.24 | -0.1 | 3.0 | 2.6 | 1.80 | 0.15 | 93 |

Note: For GATE and IATE the *average bias* is the absolute value of the bias for the specific group (GATE) / observation (IATE) averaged over all groups / observation (each group / observation receives the same weight). *CovP (90%)* denotes the (average) probability that the true value is part of the 90% confidence interval. The simulation errors of the mean MSEs are around 0.06.

*Table D.4: Simulation results for N=4,000, no effect, no selectivity, and main estimators*

| | Groups | Est. | True & estimated effects | | | Estimation error of effects (averages) | | | | | Estimation of std. error | |
|---|---|---|---|---|---|---|---|---|---|---|---|---|
| | | | Avg. bias | X-sectional std. dev. | | MSE | Skewness | Kurtosis | JB-Stat. | Std. err. | Avg. bias | CovP (90) in % |
| | # | | | true | est. | | | | | | | |
| | (1) | | (2) | (3) | (4) | (5) | (6) | (7) | (8) | (9) | (10) | (11) |
| ATE | 1 | Basic | -0.01 | - | - | 0.28 | 0.1 | 2.8 | 1.4 | 0.53 | 0.05 | 93 |
| GATE | 2 | | 0.01 | - | - | 0.30 | 0.1 | 2.8 | 1.4 | 0.55 | 0.07 | 92 |
| GATE | 32 | | 0.02 | 0 | 0.01 | 0.47 | 0.1 | 2.7 | 1.9 | 0.67 | 0.08 | 94 |
| IATE | 5000 | | 0.07 | 0 | 0.09 | 2.28 | 0.0 | 3.0 | 2.1 | 1.50 | 0.29 | 94 |
| ATE | 1 | OneF. | -0.02 | - | - | 0.28 | 0.2 | 3.1 | 2.0 | 0.58 | 0.05 | 94 |
| GATE | 2 | VarT | 0.02 | - | - | 0.30 | 0.2 | 3.0 | 1.5 | 0.62 | 0.08 | 94 |
| GATE | 32 | | 0.02 | 0 | 0.01 | 0.32 | 0.2 | 3.1 | 1.8 | 0.66 | 0.10 | 95 |
| IATE | 5000 | | 0.05 | 0 | 0.05 | 1.06 | 0.1 | 3.0 | 1.9 | 1.55 | 0.53 | 98 |
| ATE | 1 | OneF. | 0.07 | - | - | 0.30 | -0.2 | 2.7 | 2.0 | 0.54 | 0.04 | 94 |
| GATE | 2 | MCE | 0.07 | - | - | 0.32 | -0.2 | 2.8 | 2.1 | 0.56 | 0.06 | 94 |
| GATE | 32 | | 0.07 | 0 | 0.02 | 0.40 | -0.2 | 2.9 | 2.4 | 0.63 | 0.13 | 95 |
| IATE | 5000 | | 0.08 | 0 | 0.06 | 1.26 | 0.0 | 3.0 | 1.8 | 1.11 | 0.57 | 98 |
| ATE | 1 | OneF. | 0.07 | - | | 0.44 | 0.1 | 2.7 | 1.1 | 0.66 | 0.00 | 89 |
| GATE | 2 | MCE. | 0.07 | - | | 0.46 | 0.1 | 2.7 | 1.2 | 0.68 | 0.01 | 90 |
| GATE | 32 | LC-2 | 0.08 | 0 | 0.02 | 0.52 | 0.1 | 3.0 | 1.2 | 0.72 | 0.02 | 91 |
| IATE | 5000 | | 0.08 | 0 | 0.07 | 1.57 | 0.0 | 3.0 | 2.3 | 1.25 | 0.13 | 93 |
| ATE | 1 | OneF. | -0.02 | - | - | 0.28 | 0.1 | 2.7 | 0.9 | 0.53 | 0.05 | 93 |
| GATE | 2 | MCE. | 0.02 | - | - | 0.30 | 0.1 | 2.8 | 1.0 | 0.55 | 0.07 | 93 |
| GATE | 32 | Pen | 0.02 | 0 | 0.01 | 0.37 | 0.1 | 2.8 | 0.8 | 0.61 | 0.14 | 96 |
| IATE | 5000 | | 0.05 | 0 | 0.06 | 1.16 | -0.1 | 3.0 | 2.9 | 1.07 | 0.57 | 98 |
| ATE | 1 | OneF. | 0.01 | - | - | 0.38 | 0.0 | 3.0 | 0.0 | 0.62 | 0.05 | 92 |
| GATE | 2 | MCE. | 0.01 | - | - | 0.40 | 0.0 | 3.0 | 0.7 | 0.63 | 0.05 | 92 |
| GATE | 32 | Pen | 0.02 | 0 | 0.01 | 0.47 | 0.0 | 3.0 | 0.9 | 0.68 | 0.06 | 93 |
| IATE | 5000 | LC-2 | 0.05 | 0 | 0.06 | 1.57 | 0.0 | 3.1 | 3.4 | 1.25 | 0.15 | 94 |

Note: For GATE and IATE the *average bias* is the absolute value of the bias for the specific group (GATE) / observation (IATE) averaged over all groups / observation (each group / observation receives the same weight). *CovP (90%)* denotes the (average) probability that the true value is part of the 90% confidence interval. The simulation errors of the mean MSEs are around 0.03.

*Table D.5: Simulation results for N=1,000, strong effect, no selectivity, and main estimators*

| | Groups # | Est. | Avg. bias | X-sectional std. dev. | | MSE | Skewness | Kurtosis | JB-Stat. | Std. err. | Avg. bias | CovP (90) in % |
| | | | | true | est. | | | | | | | |
|---|---|---|---|---|---|---|---|---|---|---|---|---|
| | (1) | | (2) | (3) | (4) | (5) | (6) | (7) | (8) | (9) | (10) | (11) |
| ATE | 1 | Basic | 0.04 | - | - | 1.27 | 0.1 | 3.1 | 1.8 | 1.12 | 0.15 | 93 |
| GATE | 2 | | 0.04 | - | - | 1.31 | 0.1 | 3.1 | 2.0 | 1.14 | 0.17 | 93 |
| GATE | 32 | | 0.12 | 0.67 | 0.62 | 1.45 | 0.1 | 3.1 | 3.1 | 1.19 | 0.21 | 94 |
| IATE | 5000 | | 1.27 | 6.87 | 5.93 | 8.04 | 0.0 | 3.1 | 3.1 | 2.20 | 0.41 | 86 |
| ATE | 1 | OneF. | -0.09 | - | - | 1.30 | -0.1 | 3.3 | 4.3 | 1.14 | 0.14 | 93 |
| GATE | 2 | VarT | 0.10 | - | - | 1.34 | -0.1 | 3.3 | 4.2 | 1.15 | 0.16 | 94 |
| GATE | 32 | | 0.23 | 0.67 | 0.43 | 1.46 | -0.1 | 3.2 | 3.4 | 1.18 | 0.19 | 93 |
| IATE | 5000 | | 2.23 | 6.87 | 4.50 | 11.25 | 0.0 | 3.0 | 1.8 | 2.00 | 0.50 | 77 |
| ATE | 1 | OneF. | -0.02 | - | - | 1.22 | 0.0 | 2.9 | 0.7 | 1.11 | 0.18 | 95 |
| GATE | 2 | MCE | 0.09 | - | - | 1.25 | 0.0 | 2.9 | 0.9 | 1.12 | 0.19 | 95 |
| GATE | 32 | | 0.21 | 0.67 | 0.45 | 1.39 | 0.0 | 2.9 | 0.8 | 1.15 | 0.23 | 95 |
| IATE | 5000 | | 2.01 | 6.87 | 4.67 | 10.77 | 0.0 | 2.9 | 2.3 | 2.08 | 0.51 | 79 |
| ATE | 1 | OneF. | -0.02 | - | - | 1.83 | 0.0 | 3.0 | 0.4 | 1.35 | 0.06 | 91 |
| GATE | 2 | MCE. | 0.15 | - | - | 1.91 | 0.0 | 3.0 | 0.3 | 1.37 | 0.06 | 91 |
| GATE | 32 | LC-2 | 0.31 | 0.67 | 0.32 | 2.09 | 0.1 | 3.0 | 0.9 | 1.40 | 0.07 | 90 |
| IATE | 5000 | | 3.25 | 6.87 | 3.38 | 18.63 | 0.0 | 3.2 | 8.9 | 2.12 | 0.03 | 54 |
| ATE | 1 | OneF. | 0.00 | - | - | 1.33 | 0.0 | 2.9 | 0.8 | 1.28 | 0.13 | 94 |
| GATE | 2 | MCE. | 0.09 | - | - | 1.37 | 0.0 | 2.9 | 1.1 | 1.31 | 0.14 | 93 |
| GATE | 32 | Pen | 0.21 | 0.67 | 0.45 | 1.49 | 0.0 | 2.9 | 0.8 | 1.37 | 0.18 | 94 |
| IATE | 5000 | | 1.97 | 6.87 | 4.47 | 10.59 | 0.0 | 2.9 | 3.1 | 2.54 | 0.46 | 79 |
| ATE | 1 | OneF. | 0.00 | - | - | 1.94 | -0.1 | 3.0 | 0.60 | 1.39 | 0.02 | 91 |
| GATE | 2 | MCE. | 0.14 | - | - | 2.00 | -0.1 | 3.0 | 0.68 | 1.41 | 0.02 | 91 |
| GATE | 32 | Pen | 0.32 | 0.67 | 0.31 | 2.18 | -0.1 | 3.0 | 0.57 | 1.46 | 0.03 | 90 |
| IATE | 5000 | LC-2 | 3.29 | 6.87 | 3.34 | 18.85 | 0.0 | 3.1 | 6.18 | 2.10 | 0.02 | 53 |

Note: For GATE and IATE the *average bias* is the absolute value of the bias for the specific group (GATE) / observation (IATE) averaged over all groups / observation (each group / observation receives the same weight). *CovP (90%)* denotes the (average) probability that the true value is part of the 90% confidence interval. The simulation errors of the mean MSEs are around 0.15.

*Table D.6: Simulation results for N=4,000, strong effect, no selectivity, and main estimators*

| | Groups # | Est. | True & estimated effects | | | Estimation error of effects (averages) | | | | | Estimation of std. error | |
|---|---|---|---|---|---|---|---|---|---|---|---|---|
| | | | Avg. bias | X-sectional std. dev. | | MSE | Skew ness | Kurt- osis | JB- Stat. | Std. err. | Avg. bias | CovP (90) in % |
| | | | | true | est. | | | | | | | |
| | (1) | | (2) | (3) | (4) | (5) | (6) | (7) | (8) | (9) | (10) | (11) |
| ATE | 1 | Basic | 0.01 | - | - | 0.29 | -0.1 | 2.6 | 1.6 | 0.54 | 0.10 | 95 |
| GATE | 2 | | 0.06 | - | - | 0.32 | -0.1 | 2.6 | 1.9 | 0.56 | 0.11 | 97 |
| GATE | 32 | | 0.16 | 0.67 | 0.74 | 0.48 | -0.1 | 2.7 | 1.7 | 0.66 | 0.13 | 94 |
| IATE | 5000 | | 1.10 | 6.87 | 6.53 | 5.24 | 0.0 | 3.0 | 2.2 | 1.64 | 0.33 | 84 |
| ATE | 1 | OneF. | 0.00 | - | - | 0.24 | -0.1 | 2.9 | 0.5 | 0.50 | 0.14 | 98 |
| GATE | 2 | VarT | 0.05 | - | - | 0.27 | -0.1 | 2.8 | 0.6 | 0.51 | 0.15 | 98 |
| GATE | 32 | | 0.13 | 0.67 | 0.56 | 0.32 | 0.0 | 2.8 | 1.0 | 0.55 | 0.18 | 97 |
| IATE | 5000 | | 1.23 | 6.87 | 5.77 | 4.62 | 0.0 | 2.9 | 1.9 | 1.35 | 0.50 | 86 |
| ATE | 1 | OneF. | 0.05 | - | - | 0.29 | -0.1 | 3.1 | 0.7 | 0.54 | 0.10 | 94 |
| GATE | 2 | MCE | 0.05 | - | - | 0.31 | -0.1 | 3.1 | 0.8 | 0.55 | 0.11 | 94 |
| GATE | 32 | | 0.13 | 0.67 | 0.56 | 0.39 | -0.1 | 3.1 | 2.0 | 0.60 | 0.17 | 95 |
| IATE | 5000 | | 1.11 | 6.87 | 5.93 | 4.36 | 0.0 | 3.0 | 1.9 | 1.35 | 0.60 | 89 |
| ATE | 1 | OneF. | 0.03 | - | - | 0.48 | -0.1 | 2.8 | 0.4 | 0.70 | 0.00 | 90 |
| GATE | 2 | MCE. | 0.16 | - | - | 0.53 | -0.1 | 2.9 | 0.9 | 0.72 | 0.01 | 90 |
| GATE | 32 | LC-2 | 0.22 | 0.67 | 0.44 | 0.61 | -0.1 | 2.9 | 0.7 | 0.75 | 0.01 | 88 |
| IATE | 5000 | | 2.21 | 6.87 | 4.78 | 9.40 | 0.0 | 3.1 | 5.0 | 1.52 | 0.04 | 58 |
| ATE | 1 | OneF. | 0.06 | - | - | 0.29 | 0.0 | 2.8 | 0.3 | 0.53 | 0.10 | 95 |
| GATE | 2 | MCE. | 0.06 | - | - | 0.30 | 0.0 | 2.8 | 0.4 | 0.54 | 0.12 | 95 |
| GATE | 32 | Pen | 0.13 | 0.67 | 0.57 | 0.36 | 0.0 | 2.7 | 1.3 | 0.58 | 0.18 | 96 |
| IATE | 5000 | | 1.14 | 6.87 | 5.89 | 4.44 | 0.0 | 3.0 | 1.7 | 1.34 | 0.55 | 87 |
| ATE | 1 | OneF. | -0.05 | - | - | 0.41 | -0.1 | 3.0 | 0.6 | 0.64 | 0.05 | 91 |
| GATE | 2 | MCE. | 0.12 | - | - | 0.45 | -0.1 | 3.0 | 1.1 | 0.66 | 0.06 | 92 |
| GATE | 32 | Pen | 0.23 | 0.67 | 0.45 | 0.55 | -0.1 | 3.0 | 0.8 | 0.69 | 0.06 | 91 |
| IATE | 5000 | LC-2 | 2.12 | 6.87 | 4.90 | 8.93 | -0.0 | 3.0 | 3.6 | 1.51 | 0.05 | 60 |

Note: For GATE and IATE the *average bias* is the absolute value of the bias for the specific group (GATE) / observation (IATE) averaged over all groups / observation (each group / observation receives the same weight). *CovP (90%)* denotes the (average) probability that the true value is part of the 90% confidence interval. The simulation errors of the mean MSEs are around 0.06.

*Table D.7: Simulation results for N=1,000, earnings dependent effect, no selectivity, and main estimators*

|  | Groups | Est. | True & estimated effects | | | Estimation error of effects (averages) | | | | | Estimation of std. error | |
|---|---|---|---|---|---|---|---|---|---|---|---|---|
|  |  |  | Avg. bias | X-sectional std. dev. | | MSE | Skewness | Kurtosis | JB-Stat. | Std. err. | Avg. bias | CovP (90) in % |
|  | # |  |  | true | est. |  |  |  |  |  |  |  |
|  | (1) |  | (2) | (3) | (4) | (5) | (6) | (7) | (8) | (9) | (10) | (11) |
| ATE | 1 | Basic | -0.08 | - | - | 1.27 | 0.0 | 3.3 | 4.2 | 1.13 | 0.06 | 91 |
| GATE | 2 |  | 0.19 | - | - | 1.39 | 0.0 | 3.3 | 4.7 | 1.16 | 0.07 | 91 |
| GATE | 32 |  | 0.14 | 0.35 | 0.24 | 1.46 | 0.0 | 3.3 | 3.8 | 1.20 | 0.13 | 92 |
| IATE | 5000 |  | 0.50 | 1.71 | 1.02 | 4.47 | 0.0 | 3.1 | 3.1 | 1.94 | 0.41 | 93 |
| ATE | 1 | OneF. | 0.03 | - | - | 1.26 | 0.1 | 3.0 | 0.6 | 1.12 | 0.06 | 92 |
| GATE | 2 | VarT | 0.42 | - | - | 1.48 | 0.1 | 3.0 | 0.8 | 1.14 | 0.09 | 90 |
| GATE | 32 |  | 0.20 | 0.35 | 0.09 | 1.41 | 0.1 | 3.0 | 0.7 | 1.16 | 0.12 | 92 |
| IATE | 5000 |  | 1.01 | 1.71 | 0.40 | 4.24 | 0.0 | 3.0 | 2.0 | 1.54 | 0.56 | 90 |
| ATE | 1 | OneF. | -0.03 | - | - | 1.20 | 0.1 | 2.9 | 1.1 | 1.09 | 0.09 | 92 |
| GATE | 2 | MCE | 0.35 | - | - | 1.37 | 0.1 | 2.9 | 1.1 | 1.12 | 0.11 | 92 |
| GATE | 32 |  | 0.19 | 0.35 | 0.14 | 1.39 | 0.1 | 2.9 | 1.3 | 1.16 | 0.16 | 94 |
| IATE | 5000 |  | 0.82 | 1.71 | 0.61 | 4.01 | 0.0 | 3.0 | 2.2 | 1.62 | 0.63 | 93 |
| ATE | 1 | OneF. | -0.05 | - | - | 1.85 | 0.0 | 2.7 | 3.9 | 1.36 | 0.00 | 90 |
| GATE | 2 | MCE. | 0.42 | - | - | 2.07 | 0.0 | 2.7 | 3.4 | 1.38 | 0.00 | 88 |
| GATE | 32 | LC-2 | 0.22 | 0.16 | 0.11 | 2.08 | 0.0 | 2.7 | 3.6 | 1.42 | 0.01 | 90 |
| IATE | 5000 |  | 0.97 | 1.71 | 0.45 | 5.24 | 0.0 | 3.0 | 2.1 | 1.87 | 0.13 | 85 |
| ATE | 1 | OneF. | 0.06 | - | - | 1.34 | 0.0 | 2.9 | 0.3 | 1.16 | 0.03 | 91 |
| GATE | 2 | MCE. | 0.36 | - | - | 1.52 | 0.0 | 3.0 | 0.3 | 1.18 | 0.05 | 90 |
| GATE | 32 | Pen | 0.15 | 0.35 | 0.15 | 1.50 | 0.0 | 2.9 | 0.6 | 1.21 | 0.10 | 92 |
| IATE | 5000 |  | 0.83 | 1.71 | 0.62 | 4.06 | 0.0 | 3.0 | 2.7 | 1.64 | 0.56 | 92 |
| ATE | 1 | OneF. | 0.01 | - | - | 1.68 | 0.0 | 2.8 | 1.6 | 1.30 | 0.06 | 92 |
| GATE | 2 | MCE. | 0.40 | - | - | 1.89 | 0.0 | 2.8 | 1.5 | 1.32 | 0.06 | 90 |
| GATE | 32 | Pen. | 0.19 | 0.16 | 0.12 | 1.89 | 0.0 | 2.9 | 1.6 | 1.35 | 0.07 | 91 |
| IATE | 5000 | LC-2 | 0.96 | 1.71 | 0.47 | 4.92 | 0.0 | 2.2 | 2.2 | 1.80 | 0.17 | 86 |

Note: For GATE and IATE the *average bias* is the absolute value of the bias for the specific group (GATE) / observation (IATE) averaged over all groups / observation (each group / observation receives the same weight). *CovP (90%)* denotes the (average) probability that the true value is part of the 90% confidence interval. The simulation errors of the mean MSEs are around 0.06

*Table D.8: Simulation results for N=4,000, earnings dependent effect, no selectivity, and main estimators*

|  | Groups # | Est. | True & estimated effects | | | Estimation error of effects (averages) | | | | | Estimation of std. error | |
|---|---|---|---|---|---|---|---|---|---|---|---|---|
|  |  |  | Avg. bias | X-sectional std. dev. | | MSE | Skew ness | Kurt- osis | JB- Stat. | Std. err. | Avg. bias | CovP (90) in % |
|  |  |  |  | true | est. |  |  |  |  |  |  |  |
|  | (1) |  | (2) | (3) | (4) | (5) | (6) | (7) | (8) | (9) | (10) | (11) |
| ATE | 1 | Basic | 0.02 | - | - | 0.29 | -0.1 | 2.9 | 0.5 | 0.59 | 0.05 | 92 |
| GATE | 2 |  | 0.03 | - | - | 0.34 | -0.5 | 2.9 | 1.1 | 0.64 | 0.06 | 92 |
| GATE | 32 |  | 0.04 | 0.35 | 0.34 | 0.43 | -0.5 | 2.9 | 0.7 | 0.76 | 0.11 | 95 |
| IATE | 5000 |  | 0.40 | 1.71 | 1.50 | 2.78 | 0.0 | 3.0 | 1.6 | 1.85 | 0.28 | 92 |
| ATE | 1 | OneF. | 0.00 | - | - | 0.31 | 0.0 | 2.8 | 0.3 | 0.56 | 0.03 | 91 |
| GATE | 2 | VarT | 0.33 | - | - | 0.45 | 0.0 | 2.9 | 1.3 | 0.58 | 0.05 | 87 |
| GATE | 32 |  | 0.16 | 0.35 | 0.16 | 0.40 | 0.0 | 2.8 | 0.8 | 0.60 | 0.08 | 93 |
| IATE | 5000 |  | 0.80 | 1.71 | 0.66 | 2.39 | 0.0 | 3.0 | 2.3 | 1.07 | 0.50 | 90 |
| ATE | 1 | OneF. | 0.05 | - | - | 0.31 | -0.2 | 3.2 | 3.1 | 0.56 | 0.03 | 92 |
| GATE | 2 | MCE | 0.24 | - | - | 0.40 | -0.2 | 3.3 | 2.9 | 0.58 | 0.06 | 90 |
| GATE | 32 |  | 0.10 | 0.35 | 0.23 | 0.42 | -0.2 | 3.2 | 2.6 | 0.63 | 0.13 | 94 |
| IATE | 5000 |  | 0.57 | 1.71 | 0.95 | 2.15 | -0.1 | 3.0 | 2.5 | 1.18 | 0.55 | 94 |
| ATE | 1 | OneF. | -0.01 | - | - | 0.45 | 0.0 | 3.0 | 0.0 | 0.67 | 0.00 | 91 |
| GATE | 2 | MCE. | 0.33 | - | - | 0.58 | 0.0 | 3.1 | 0.2 | 0.70 | 0.01 | 86 |
| GATE | 32 | LC-2 | 0.17 | 0.35 | 0.15 | 0.57 | 0.0 | 2.9 | 0.9 | 0.73 | 0.01 | 89 |
| IATE | 5000 |  | 0.80 | 1.71 | 0.66 | 2.85 | 0.0 | 3.1 | 1.3 | 1.26 | 0.13 | 83 |
| ATE | 1 | OneF. | 0.01 | - | - | 0.29 | -0.1 | 2.8 | 1.2 | 0.59 | 0.05 | 93 |
| GATE | 2 | MCE. | 0.23 | - | - | 0.37 | -0.2 | 2.9 | 2.0 | 0.64 | 0.07 | 91 |
| GATE | 32 | Pen | 0.10 | 0.35 | 0.23 | 0.39 | -0.1 | 2.8 | 1.0 | 0.75 | 0.14 | 95 |
| IATE | 5000 |  | 0.54 | 1.71 | 0.97 | 1.98 | -0.1 | 3.0 | 2.3 | 1.68 | 0.56 | 95 |
| ATE | 1 | OneF. | -0.05 | - | - | 0.40 | -0.1 | 3.3 | 1.2 | 0.63 | 0.04 | 92 |
| GATE | 2 | MCE. | 0.32 | - | - | 0.53 | 0.0 | 3.3 | 1.7 | 0.65 | 0.04 | 89 |
| GATE | 32 | Pen | 0.17 | 0.16 | 0.17 | 0.53 | -0.1 | 3.3 | 1.9 | 0.70 | 0.05 | 92 |
| IATE | 5000 | LC-2 | 0.77 | 1.72 | 0.71 | 2.73 | 0.0 | 3.1 | 2.8 | 1.26 | 0.15 | 84 |

Note: For GATE and IATE the *average bias* is the absolute value of the bias for the specific group (GATE) / observation (IATE) averaged over all groups / observation (each group / observation receives the same weight). *CovP (90%)* denotes the (average) probability that the true value is part of the 90% confidence interval. The simulation errors of the mean MSEs are around 0.04.